%% file: main.tex
\documentclass[a4paper,noarxiv,onecolumn]{quantumarticle}
\usepackage[utf8]{inputenc}
\usepackage[english]{babel}
\usepackage[T1]{fontenc}

\usepackage{lipsum}
\usepackage{mdframed}
\usepackage{booktabs}
\usepackage{longtable}

\usepackage{adjustbox}
\usepackage{quantikz}
\usepackage{amsmath}
\usepackage{amssymb}
\usepackage{tikz}
\usetikzlibrary{tikzmark}
\usepackage{caption}
\usepackage{subcaption}
\usepackage{multicol}
\usepackage{multirow}
\usepackage{nicefrac}
\usepackage{braket}
\usepackage[linesnumbered,ruled,vlined]{algorithm2e}
\usepackage{diagbox}
\usepackage{tabularx}
\usepackage{placeins}
\usepackage{nicematrix}
\usepackage{natbib}
\usepackage{comment}
\usepackage{array}

\newcommand\Tstrut{\rule{0pt}{2.6ex}}         
\newcommand\Bstrut{\rule[-0.9ex]{0pt}{0pt}}   

\author{David Winderl}
\affiliation{%
CIT Department of Computer Science, Technical University of Munich, Garching, Germany
}
\email{david.winderl@tum.de}
\author{Qunsheng Huang}
\affiliation{%
CIT Department of Computer Science, Technical University of Munich, Garching, Germany
}
\email{keefe.huang@tum.de}
\author{Arianne Meijer--van de Griend}
\affiliation{%
Department of Computer Science, University of Helsinki, Helsinki, Finland
}
\email{ariannemeijer@gmail.com}
\author{Richie Yeung}
\affiliation{%
Department of Computer Science, University of Oxford, Oxford,
United Kingdom
}
\email{richie.yeung@cs.ox.ac.uk}

\title{Architecture-Aware Synthesis of Stabilizer Circuits from Clifford Tableaus}

\begin{document}

\begin{abstract}
    Traditionally, Clifford tableaus are used for classical simulation of quantum circuits. However, we repurpose this data structure for quantum compilation to optimize the number of gates executed on specific hardware.
    This is required for current NISQ computers, but will also be applicable for compiling error correction codes on fault-tolerant devices.
    
    We propose a new algorithm that synthesizes stabilizer circuits from Clifford tableaus while respecting the connectivity graph of a given quantum device. 
    We do this by only synthesizing CNOT gates between connected qubits on the quantum chip.
    
    For deep circuits, our method outperforms other state-of-the-art synthesis techniques, when compiled for a specific hardware. Additionally, we show that our synthesized circuits show higher fidelity and reduce the overall execution times when executed on real IBM devices. 

    Our proposed algorithm can be used to compile quantum circuits by noting which single-qubit non-Clifford gates are executed with respect to which stabilizer state and generate those stabilizers on multiple qubits in parallel using our algorithm.
\end{abstract}

\maketitle

\section{Introduction}
\input{struct/intro}
\section{Preliminaries}
\input{struct/preliminaries}
\input{struct/relatedwork}
\section{Methods}
\input{struct/methods}
\section{Evaluations}
\input{struct/evaluation}

\FloatBarrier
\section{Conclusion}
\input{struct/conclusion}

\subsection*{Acknowledgements}
This research was partially funded by BayQS, UKRI, Simon Harrison, and Business Finland projects FrameQ and EM4QS.

\clearpage
\newpage
\setcitestyle{numbers} 

\input{main.bbl}

\clearpage
\newpage
\appendix
\input{struct/appendix}
\end{document}

%% file: struct/intro.tex
In the current noisy intermediate-scale quantum (NISQ) era of quantum computing, the effects of decoherence and imperfect physical quantum gates mean that deep circuits typically accrue significant errors before computation and measurement can be completed. Hence, optimizing circuits to improve overall fidelity is important to obtain meaningful results from quantum algorithms.

However, current quantum computing architectures typically limit the set of possible multi-qubit interactions, increasing the circuit's size as additional gates are required to route the interactions along limited connectivity. Typically, the device's connectivity constraints are represented using a simple graph where physical qubits are represented by nodes in the graph, and edges between nodes indicate that the represented physical qubits can interact directly. We henceforth refer to this graph as the architecture or connectivity graph of the given quantum system.

The problem of making a quantum circuit fit the connectivity constraints of a particular device is called the qubit routing problem. Traditionally, this problem was solved from the perspective that if the architecture does not allow a gate in the circuit, additional SWAP gates would be introduced to swap qubits to the correct registers. These strategies are still dominant in existing compilers, such as the Qiskit transpiler~\cite{qiskit}.

Recent advances in architecture-aware synthesis techniques solve this problem through the perspective that gates in the circuit are wrong. Whenever the quantum circuit was created, incompatible gates were chosen. Instead, a new circuit should be synthesized in an architecture-aware fashion. This approach to routing a circuit makes it easier to make global optimizations of the circuit, such that the gate overhead of architecture-aware synthesis should be lower than the gate overhead of swapping qubits.

These architecture-aware synthesis techniques typically use parity maps or phase polynomials~\cite{Amy2013} to represent pieces of the circuit that only contain CNOT gates~\cite{Nash_2020,kissinger2019cnot,Wu_2023,vandegriend2023dynamic}, or CNOT gates with $R_Z(\alpha)$-gates~\cite{Amy_2018, Nash_2020,degriend2020architectureaware,vandaele2022phase}, respectively. Then, these intermediate representations are stitched together to form a universal synthesis method~\cite{gheorghiu2022reducing,Martiel_2022,gogioso2022annealing,winderl2023recursively,vandegriend2023towards}.

This paper proposes an efficient classical architecture-aware synthesis algorithm that uses a different class of simulable quantum circuits, namely those only containing Clifford gates. These circuits are often called stabilizer circuits or Clifford circuits, and they can be efficiently represented using a binary matrix called a Clifford tableau. Although these circuits are considered classical in nature, the architecture-aware synthesis of these circuits can be used in the compilation of universal quantum programs by cutting the circuit on the single-qubit non-Clifford gates and re-synthesizing the stabilizer circuits between them.

Synthesis of stabilizer circuits from Clifford tableaus is a highly-explored field, and extensive work has been invested into tableau synthesis algorithms~\cite{Aaronson_2004, Dehaene_2003, maslov_2018, Bravyi_2021}. We adapted the algorithm from \citet{berg2021simple} to be architecture-aware using strategies from the RowCol algorithm for architecture-aware CNOT synthesis~\cite{Wu_2023}.

As a proof of concept, we evaluate our algorithm against the existing state-of-the-art Clifford tableau synthesis algorithms~\cite{Bravyi_2021,berg2021simple,maslov_2018,peham2023depthoptimal,Duncan_2020}.
When targeting different IBM quantum computers, we demonstrate a significant reduction of CNOT gates for circuits with sufficient depth. Our experimental data shows an asymptotic upper bound on CNOT gates for input circuits with sufficient depth for all synthesis methods, where the circuits were routed using SABRE (Qiskit impl.)~\cite{li2019tackling} if the algorithm did not consider the connectivity constraints. 

Conversely, we show that directly transpiling the given circuit without synthesis does not incur an upper bound. Instead, the CNOT gate count increases linearly with input circuit size. 

Additionally, we executed the optimized circuits on real quantum devices available on the IBM quantum platform, demonstrating higher fidelity and shorter runtime compared to circuits synthesized with the two methods we used as a baseline. Although this is expected from running a smaller circuit, we believe this is a nice addition to see how large the effect is in practice.

%% file: struct/preliminaries.tex
In this section, we describe two important mathematical constructs used in the proposed algorithm: Steiner trees and Clifford tableaus.

\subsection{Steiner Tree}
In graph theory, a Steiner Tree of a graph, $G$, and a subset of vertices $V$ of $G$, $T$, called Terminals, is defined as the minimum tree over $G$ containing at least $T$. In a sense, this can be seen as a generalization of a minimum path where there are many start/end points, namely, the Terminals $T$. Similarly, it can be seen as an incomplete minimum spanning tree over $G$ defined by the Terminals $T$. The nodes in the Steiner tree that were not originally in $T$ are called Steiner nodes. 

The Steiner tree problem is known to be an NP-hard problem~\cite{karp2010reducibility}. However, efficient approximate algorithms do exist. For our experiments, we use \citet{kou_1981}.

\subsection{Clifford Tableau}
The $q$-qubit Pauli group $\mathcal{P}_q$ consists of the $4^q$ $q$-element tensor products of Pauli matrices: $X = \begin{bsmallmatrix}
0 & 1\\
1 & 0
\end{bsmallmatrix}$, $Y=\begin{bsmallmatrix}
0 & -i\\
i & 0
\end{bsmallmatrix}$, $Z=\begin{bsmallmatrix}
0 & 1\\
1 & 0, 
\end{bsmallmatrix}$, and $I_2$ with multiplicative factors $\pm{1}$ and $\pm{i}$ for a total of $4 \cdot 4^q$ elements.
Noting that $\mathcal{P}_q$ spans all $2^q \times 2^q$ complex matrices, we can characterize any unitary operation by its action on $P_q$. As we limit our considerations to Clifford operations, which, by definition, normalize $\mathcal{P}_q$ through conjugation, we can fully characterize any Clifford operation by its action on each element of $\mathcal{P}_q$.

We then define a basis consisting of $2q$ elements of $\mathcal{P}_q$ written as tensor products of Pauli operators as follows\footnote{These are typically abbreviated as a Pauli string. So, for instance, $Z \otimes I \otimes I$ can be written as $ZII$.}:
\begin{align*}
p_1 &= X_1 \otimes I_2 \otimes I_3 \otimes  ...  \otimes  I_q \quad & p_{q+1} &= Z_1  \otimes I_2 \otimes  I_3 \otimes   ... \otimes   I_q\\
p_2 &= I_1  \otimes  X_2 \otimes  I_3  \otimes  ...  \otimes  I_q \quad & p_{q+2} &= I_1  \otimes  Z_2  \otimes  I_3   \otimes ...  \otimes  I_q\\
&... &... \\
p_{q} &= I_1  \otimes  I_2  \otimes  I_3  \otimes ...  \otimes  X_q \quad & p_{2q} &= I_1 \otimes  I_2  \otimes I_3  \otimes ...   \otimes Z_q,
\end{align*}
the products of which can represent any element of $\mathcal{P}_q$ up to a phase of $\pm{1}$ or $\pm{i}$ \cite{Grier_2022}. The first $q$ elements of this basis are destabilizer generators for the stabilizer state $\ket{0}^{\otimes q}$ and the last $q$ elements are the corresponding stabilizer generators.

 Keeping track of how Clifford operators modify this basis directly leads us to Aaronson's canonical formatting of the Clifford tableau, where each basis element $p_j$ is encoded into a $2q +1 $ large vector $r \in GF(2)$~\footnote{$GF(2)$ is the unique field with two elements with its additive and multiplicative identities} corresponding to the $j$th row of said tableau, by the following formula:
\begin{equation}\label{eq.encoding_ct}
    f(r^j) = (-1)^{r^j_{2q+1}}\bigotimes_{i=1}^n (Z^{r^j_{i}} X^{r^j_{i+q}})
\end{equation}
The underlying principle in \autoref{eq.encoding_ct} is the following relation among Paulis: $Y = -iZX$. Hence for a qubit $i$, a Pauli $Y$ can be encoded by setting the bits $r_i$ and $r_{i+q}$ to one, a Pauli $X$ by setting $r_{i+q}$ to one, a Pauli $Z$ can be encoded by setting $r_{i}$ to one and finally a $I$ can be encoded by leaving both $r_i$ and $r_{i+q}$ as zero. At last, one can undo the possible change of signs occurring due to the prefactor of $-i$; therefore, each row possesses a sign column $r_{2q+1}$, which can be flipped to undo the change of sign. 

 Then, ignoring the final column storing the sign bit, this encoding effectively blocks the Clifford tableau into four $q \times q$ sub-matrices as follows, with the positions of the bits representing the $j$th destabilizer and stabilizer on qubit $i$ indicated:
$$
\left[
\begin{array}{c c c | c c c }
            & \vdots &          &        & \vdots    &        \\
     \cdots & r^j_i  & \cdots   & \cdots & r^j_{i+q} & \cdots \\
            & \vdots &          &        & \vdots    &        \\
    \hline
            & \vdots     &          &        & \vdots        &        \\
     \cdots & r^{j+q}_i  & \cdots   & \cdots & r^{j+q}_{i+q} & \cdots \\
            & \vdots     &          &        & \vdots        &        \\
     
\end{array}
\right]
\rightarrow
\left[
\begin{array}{ c | c }
X_x & Z_x
\Bstrut\\\hline\Tstrut
X_z & Z_z
\end{array}
\right] 
$$

The labels $X$ and $Z$ indicate the applied Pauli gate and the subscript $x$ and $z$ correspond to the destabilizer and stabilizer respectively. We further note that destabilizers and stabilizers are ordered row-wise in pairs, such that the $j$th destabilizer ($j$th row in the Clifford tableau) anti-commutes with the $j$th stabilizer ($j+n$th row in the Clifford tableau) and commute with all other rows in the tableau. This property is detailed in \cite{Aaronson_2004} and is preserved under conjugation by Clifford operations. Utilizing this setup, any $n$-qubit Clifford circuit can be represented by $2n(2n+1)$ bits and allows updates by adding $H$, $S$, and CNOT gates to the end (appending) or the front (prepending)~\cite{gottesman1997stabilizer, Aaronson_2004}. 

The specific actions of appending and prepending $H$, $S$, and CNOT gates in the Clifford tableau formulation are laid out in \cite{Gidney_2021,gottesman1997stabilizer} and visually specified in \autoref{fig:append_op} sans the effects on the sign-flip vector.
Then, recognizing the similarities between the Clifford tableau representation and the ubiquitous parity map representation of CNOT-composed circuits used in circuit optimization by \cite{Duncan_2020,kissinger2019cnot,Nash_2020}, we obtain an easily manipulable representation for Clifford circuit synthesis that allows reuse of existing machinery.
\begin{figure}[ht]
    \centering
    \begin{subfigure}{0.3\textwidth}
        \centering
        \begin{tikzpicture}[scale=0.3]
            \fill[white] (0,0) rectangle (10,10);
            \draw[black, thick] (0,0) rectangle (10,10);
            \draw[black, thick] (5,0) -- (5,10);
            
            \fill[gray] (6,0) rectangle (7,10);
            \draw[<->, thick, bend right] (1.5, -0.2) to node[midway, below]{swap} (6.5, -0.2);
            \node[anchor=south] at (6.5,10.2) {$z_{i}$};
            
            \fill[gray] (1,0) rectangle (2,10);
    
            \node[anchor=south] at (1.5,10.2) {$x_{i}$};
        \end{tikzpicture} 
        \caption{Append: $H_{i}$}
    \end{subfigure}
    \hfill
    \begin{subfigure}{0.3\textwidth}
        \centering
        \begin{tikzpicture}[scale=0.3]
            \fill[white] (0,0) rectangle (10,10);
            \draw[black, thick] (0,0) rectangle (10,10);
            \draw[black, thick] (5,0) -- (5,10);
            
            \fill[gray] (6,0) rectangle (7,10);
            \node[anchor=south] at (6.5,10.2) {$z_{i}$};
            
            \fill[gray] (1,0) rectangle (2,10);
    
            \draw[->, thick, bend right] (1.5,-0.2) to node[pos=1, below]{$\oplus$} (6.5, -0.2);
            \node[anchor=south] at (1.5,10.2) {$x_{i}$};
        \end{tikzpicture}   
        \caption{Append: $S_{i}$}   
    \end{subfigure}
    \hfill
    \begin{subfigure}{0.3\textwidth}
        \centering
        \begin{tikzpicture}[scale=0.3]
            \fill[white] (0,0) rectangle (10,10);
            \draw[black, thick] (0,0) rectangle (10,10);
            \draw[black, thick] (5,0) -- (5,10);
            
            \fill[gray] (8,0) rectangle (9,10);
            \fill[gray] (6,0) rectangle (7,10);
            \draw[->, thick, bend left] (8.5, -0.2) to node[pos=1, below]{$\oplus$} (6.5, -0.2);
            \node[anchor=south] at (8.5,10.2) {$z_{t}$};
            \node[anchor=south] at (6.5,10.2) {$z_{c}$};
            
            \fill[gray] (3,0) rectangle (4,10);
            \fill[gray] (1,0) rectangle (2,10);
    
            \draw[<-, thick, bend left] (3.5,-0.2) to node[pos=0, below]{$\oplus$} (1.5, -0.2);
            \node[anchor=south] at (3.5,10.2) {$x_{t}$};
            \node[anchor=south] at (1.5,10.2) {$x_{c}$};
        \end{tikzpicture}
        \caption{Append: $CNOT_{c, t}$}
    \end{subfigure}
    \caption{Action of appending Clifford gates $H$, $S$ and CNOT to an $n$ qubit Clifford tableau.}
    \label{fig:append_op}
\end{figure}
\begin{figure}
\centering
    \begin{subfigure}{0.3\textwidth}
        \centering
        \begin{tikzpicture}[scale=0.29]
            \fill[white] (0,0) rectangle (10,10);
            \draw[black, thick] (0,0) rectangle (10,10);
            \draw[black, thick] (0,5) -- (10,5);
    
            \fill[gray] (0,6) rectangle (10,7);
            \draw[<->, thick, bend left] (-0.2,1.5) to node[midway, sloped, above]{swap} (-0.2,6.5);
            \node[anchor=west] at (10.2,6.5) {$i$};
            
            \fill[gray] (0,1) rectangle (10,2);
    
            \node[anchor=west] at (10.2,1.5) {$i+n$};
        \end{tikzpicture}
        \caption{Prepend: $H_i$}
    \end{subfigure}
    \hfill
    \begin{subfigure}{0.3\textwidth}
        \centering
        \begin{tikzpicture}[scale=0.29]
            \fill[white] (0,0) rectangle (10,10);
            \draw[black, thick] (0,0) rectangle (10,10);
            \draw[black, thick] (0,5) -- (10,5);
            \fill[gray] (0,6) rectangle (10,7);
            \draw[->, thick, bend left] (-0.2,1.5) to node[pos=1, left]{$\oplus$} (-0.2,6.5);
            \node[anchor=west] at (10.2,6.5) {$i$};
            
            \fill[gray] (0,1) rectangle (10,2);
            \draw (0,0) rectangle (10,2);
    
            \node[anchor=west] at (10.2,1.5) {$i+n$};
        \end{tikzpicture}
        \caption{Prepend: $S_i$}
    \end{subfigure}
    \hfill
    \begin{subfigure}{0.3\textwidth}
        \centering
        \begin{tikzpicture}[scale=0.29]
            \fill[white] (0,0) rectangle (10,10);
            \draw[black, thick] (0,0) rectangle (10,10);
            \draw[black, thick] (0,5) -- (10,5);
            
            \fill[gray] (0,8) rectangle (10,9);
            \fill[gray] (0,6) rectangle (10,7);
            \draw[->, thick, bend left] (-0.2,6.5) to node[pos=1, left]{$\oplus$} (-0.2,8.5);
            \node[anchor=west] at (10.2,8.5) {$c$};
            \node[anchor=west] at (10.2,6.5) {$t$};
            
            \fill[gray] (0,3) rectangle (10,4);
            \fill[gray] (0,1) rectangle (10,2);
    
            \draw[<-, thick, bend left] (-0.2,1.5) to node[pos=0, left]{$\oplus$} (-0.2,3.5);
            \node[anchor=west] at (10.2,3.5) {$c+n$};
            \node[anchor=west] at (10.2,1.5) {$t+n$};
            
        \end{tikzpicture}
        \caption{Prepend: $\text{CNOT}_{c, t}$}
    \end{subfigure}

\caption{Action of prepending Clifford gates $H$, $S$ and CNOT to an $n$ qubit  Clifford tableau.}
\label{fig:prepending_op}
\end{figure}

%% file: struct/relatedwork.tex
\section{Related work}
This paper proposes a novel algorithm for the architecture-aware synthesis of stabilizer circuits from Clifford tableaus. This section gives an overview of existing Clifford tableau synthesis algorithms and how they relate to each other and the proposed algorithm.

Most approaches to synthesizing circuits from Clifford tableaus rely on normal forms. An initial synthesis algorithm was proposed by \citet{Aaronson_2004} using the normal form: $H-C-P-C-P-C-H-P-C-P-C$, where $H$ represented a region consisting of Hadamard gates, $P$ a region of Phase gates and $C$ a region of CNOT gates. \citet{Dehaene_2003} formulated a courser-grained 5-layer form, which was later improved by \citet{maslov_2018} and \citet{nest2009classical}, respectively. \citet{Duncan_2020} have implemented an improved $H-S-CZ-CX-H-CZ-S-H$ normal form in the pyzx library~\cite{kissinger2020Pyzx}. 



We particularly point the reader to the algorithm by \citet{Bravyi_2021}, the state-of-the-art baseline algorithm that performed the best in our benchmarks out of all baseline algorithms. Their method uses a combination of template matching, designed explicitly for Clifford circuits, to level out CNOT and SWAP gates. They propose a canonical decomposition of any Clifford circuit in the form of a modified Bruhat~\cite{maslov_2018} decomposition of $-CX-CZ-P-H-P-CZ-CX-$ and perform gate reduction by first shifting gates into the initial $-CX-CZ-P-$ layers before optimizing the $CX$ layer via a parity matrix representation and the $CZ$ layer using a phase polynomial form. Similar to the current work, the focus of their optimization lies in reducing the number of entangling gates, which are $CX$ and $CZ$ in their normal form.  Currently, their algorithm is implemented as the standard when synthesizing a Clifford tableau in the qiskit software stack~\cite{qiskit}.

These algorithms are heuristical and do not promise a minimal circuit implementation. Alternatively, it is possible to find such a minimal implementation using an SAT-solver~\cite{peham2023depthoptimal}. Although such a method is not feasible in practical use in compilers, it does show how close the heuristic methods are to finding an optimal decomposition of the Clifford Tableau.

Unlike the proposed algorithm, these algorithms assume that two-qubit gates are available between arbitrary qubits, an assumption that is violated by many current and near-term quantum computers. Alternatively, synthesis algorithms that target connectivity graphs with a particular structure, such as a line~\cite{Maslov_2023,debrugière2023shallower,maslov_2018} or a grid~\cite{debrugière2023shallower} structure, do exist. This can be valuable when compiling for devices that happen to have such a structure, but when using them for devices that do not have such a structure, such as the ones used in our experiments, they still require an inefficient qubit-routing step.
Instead, we propose an algorithm that can decompose the Clifford Tableau to arbitrary graphs.

The general structure of our method relies on the synthesis method proposed by \citet{berg2021simple}. To the best of our knowledge, this method is the first one that does not use the normal forms of synthesis but relies on the sanitization (clearing of elements in the context of \citet{berg2021simple}) and removal of interactions (sweeping in the context of \citet{berg2021simple}). This algorithm is currently implemented in the \texttt{stim} library~\cite{Gidney_2021}—a highly efficient library for simulating Clifford circuits. In the next section, we give an overview of the proposed algorithm. When executed using a fully connected quantum architecture without dynamically mapping the qubits, this algorithm is the same as the one proposed by \citet{berg2021simple}. Additionally, we prove its correctness which had not yet been shown previously in literature.

%% file: struct/methods.tex
In the following, we describe the proposed algorithm for the architecture-aware synthesis of Clifford tableaus. As such, it heavily borrows from an existing algorithm for simulating Clifford circuits~\cite{berg2021simple} that is implemented in \texttt{stim}~\cite{Gidney_2021}.
Similar to this algorithm, our proposed algorithm also consists of the following steps:
\begin{enumerate}
    \item Pick a pivot qubit in the Clifford tableau
    \item Sanitize the qubit w.r.t. destabilizer
    \item Remove destabilizer interactions with other qubits
    \item Sanitize the qubit w.r.t. stabilizer
    \item Remove stabilizer interactions with other qubits
    \item Remove the qubit from the problem and go to (1) while there are still qubits left
    \item Sanitize the tableau signs
\end{enumerate}
Contrarily to the previous algorithm, the proposed algorithm will only synthesize Clifford gates allowed by a target architecture as specified by a given connectivity graph during this process.

By construction, the synthesis process is the inverse of constructing the Clifford tableau. As such, if we apply gates to a Clifford tableau $C$ until the tableau becomes an identity matrix, the gates will correspond to the Clifford tableau $C^\dagger$. Hence, we start the procedure by inverting the tableau such that we generate the circuit corresponding to $(C^\dagger)^\dagger = C$. This inversion process is well-known in literature and involves the term-wise modification of the transpose of the Clifford tableau. Alternatively, by nature of the inversibility of quantum computing, it is possible to generate a circuit from $C$ directly, reversing the order of operations of the solution, and applying the gate-wise inverse.

\subsection{The architecture-aware Clifford synthesis algorithm}\label{sec:description_algorithm}
The algorithm works by applying Clifford gates to the tableau to transform it into an identity matrix. 
$$
\left[
\begin{array}{c | c}
X_x & Z_x
\Bstrut\\\hline\Tstrut
X_z & Z_z
\end{array}
\right]
\rightsquigarrow
\left[
\begin{array}{c|c}
I & 0
\Bstrut\\\hline\Tstrut
0 & I
\end{array}
\right]
$$

We do this by iteratively picking a qubit and turning its corresponding rows and columns into identity rows and columns; this equivalently means that the qubit no longer interacts with the other qubits and can be removed from the Clifford tableau. We continue this process until the tableau is the identity; in other words, it is empty.

First, we pick a qubit to remove from the tableau; we call this the pivot qubit. In principle, any qubit is a suitable pivot as long as it does not disconnect the connectivity graph upon removal (i.e., it is a non-cutting vertex). Thus, we will assume in the following that the pivot qubit is any non-cutting vertex in the graph, and we will later provide a heuristic for picking a specific pivot. 

Given a tableau representing a $q$-qubit Clifford circuit, the pivot qubit (the $p^{th}$ qubit in this example) can be removed if its corresponding rows, $r_p$ and $r_{q+p}$, and columns, $c_p$ and $c_{q+p}$, in the tableau, have a single non-zero entry along the diagonal of the tableau. In other words, if it describes the stabilizer state $I\dots IZ_pI \dots I$ and destabilizer state $I\dots IX_pI \dots I$.

We remove the undesired non-zero entries from row $r_p$ in two phases. First, we remove the non-zero entries from the second half of row $r_p$, which are the elements in the $Z_x$ block matrix. Suppose the row has a $1$ in column $c_{q+i}$; we then apply a Hadamard gate on qubit $i$ if the row has a $0$ in column $c_i$ and an S gate otherwise. This will make the entry in $c_{q+i}$ a $0$ and the one in $c_i$ a $1$, thus sanitizing row $r_p$ when applying this for each non-zero entry in the second half of $r_p$. This process is equivalent to converting each element of the corresponding destabilizer to a Pauli string consisting only of Xs and Is.

Schematically, $r_p$ changes at $c_i$ and $c_{q+i}$ when sanitizing as follows
\begin{align*}
    &\textbf{if } c_i=0: \\
    &\quad \left[ \begin{array}{ccc | ccc}
        & c_i &  &  & c_{q+i} &  \\
     ... & 0 & ... & ... & 1 & ... \\
\end{array} \right]  \xrightarrow{H_i} \left[ \begin{array}{ccc | ccc}
        & c_i &  &  & c_{q+i} &  \\
     ... & 1 & ... & ... & 0 & ... \\
\end{array} \right] \\ \\
    &\textbf{if } c_i=1:\\
    & \quad \left[ \begin{array}{ccc | ccc}
        & c_i &  &  & c_{q+i} &  \\
     ... & 1 & ... & ... & 1 & ... \\
\end{array} \right] \xrightarrow{S_i} \left[ \begin{array}{ccc | ccc}
        & c_i &  &  & c_{q+i} &  \\
     ... & 1 & ... & ... & 0 & ... \\
\end{array} \right]
\\
\end{align*}
Then, we want to remove the undesired non-zero entries from the first half of row $r_p$. We do this by applying CNOT gates. Since the CNOT gates only act within the respective halves of the row, the second half or $r_p$ will remain zero. When applying the CNOT gates, we want to turn row $r_p$ into an identity row, but we are restricted to the connectivity constraints of the given connectivity graph. We can do this like architecture-aware CNOT synthesis~\cite{Nash_2020,kissinger2019cnot,Wu_2023,vandegriend2023dynamic} by building a Steiner tree over the interacting qubits in the graph. Then, we can apply CNOT gates to turn the entries corresponding to Steiner nodes into a $1$ and use those nodes to make all entries in the Steiner tree a $0$ except for the root (i.e., the pivot qubit). This results in row $r_p$ being an identity row in the tableau.

\begin{figure}[p]
    \begin{center}
    \begin{tabular}{ c c c c c c }
    &
     \includegraphics[]{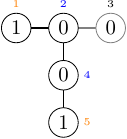} &&
     \includegraphics[]{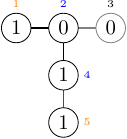} &&
     \includegraphics[]{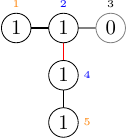} \\ 
    &
     $[\:1\:0\:0\:0\:1\:|\:0\:0\:0\:0\:0\:]$ &$\xrightarrow{CNOT_{5,4}}$ & $[\:1\:1\:0\:0\:1\:|\:0\:0\:0\:0\:0\:]$ &$\xrightarrow{CNOT_{4,2}}$ & $[\:1\:1\:0\:1\:1\:|\:0\:0\:0\:0\:0\:]$\\\\
    &
     \includegraphics[]{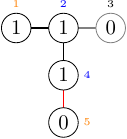} &&
     \includegraphics[]{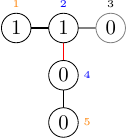} &&
     \includegraphics[]{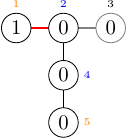} \\ 
    $\xrightarrow{CNOT_{4,5}}$ &
     $[\:1\:1\:0\:1\:0\:|\:0\:0\:0\:0\:0\:]$ &  $\xrightarrow{CNOT_{2,4}}$ & $[\:1\:1\:0\:0\:0\:|\:0\:0\:0\:0\:0\:]$ & $\xrightarrow{CNOT_{1,2}}$ & $[\:1\:0\:0\:0\:0\:|\:0\:0\:0\:0\:0\:]$\\
    \end{tabular}
    \end{center}
    \caption{An example of eliminating the interaction between non-adjacent qubits 1 and 5 on the IBM Quito device. The Steiner tree is indicated with bold edges ($q_1$, $q_2$, $q_4$, $q_5$) and the operation of the CNOT is indicated with red.}
    \label{fig:steiner-example}
\end{figure}

For example, suppose we have the 5 qubit IBM Quito device and the pivot row $r_p=r_1$ looks as follows: $\left[ 1\:0\:0\:0\:1|\:0\:0\:0\:0\:0\:\right]$. Then, we cannot directly add $c_1$ to $c_5$ with a CNOT since those qubits are not connected. Instead, we build a Steiner tree over the connectivity graph given qubit $q_1$ and $q_5$ and use it to define which CNOTs needs to be placed, as shown in \autoref{fig:steiner-example}. 
Hence, the pivot row $r_p=r_1$ is successfully turned into an identity row.

Moving on towards the stabilizer row $r_{q+p}$, which we will decompose similarly to the destabilizer. We can look at the first half of $r_{q+p}$, which is part of the $X_z$ block matrix in the tableau. Here, we want to apply a Hadamard gate on qubit $i$ when row $r_{q+p}$ has a $1$ in column $c_{i}$ and a $0$ in column $c_{q+i}$ or an S gate followed by a Hadamard gate when there is a $1$ in both column $c_{i}$ and $c_{q+i}$. Applying these gates for each qubit will sanitize row $r_{q+p}$.

Schematically, $r_{q+p}$ changes at $c_i$ and $c_{q+i}$ when sanitizing as follows:
\begin{align*}
    &\textbf{if } c_{q+i}=0:  \\
    &\quad \left[ \begin{array}{ccc | ccc}
        & c_i &  &  & c_{q+i} &  \\
     ... & 1 & ... & ... & 0 & ... \\
\end{array} \right] 
 \xrightarrow{H_i} \left[ \begin{array}{ccc | ccc}
        & c_i &  &  & c_{q+i} &  \\
     ... & 0 & ... & ... & 1 & ... \\
\end{array} \right]  \\
& \textbf{if } c_{q+i}=1: \\
& \quad \left[ \begin{array}{ccc | ccc}
        & c_i &  &  & c_{q+i} &  \\
     ... & 1 & ... & ... & 1 & ... \\
\end{array} \right] 
 \xrightarrow{S_i} \left[ \begin{array}{ccc | ccc}
        & c_i &  &  & c_{q+i} &  \\
     ... & 1 & ... & ... & 0 & ... \\
\end{array} \right] 
 \xrightarrow{H_i} \left[ \begin{array}{ccc | ccc}
        & c_i &  &  & c_{q+i} &  \\
     ... & 0 & ... & ... & 1 & ... \\
\end{array} \right] 
\end{align*}

However, we must be careful when this process applies to the pivot qubit $i=p$. By construction, row $r_p$ corresponds to a destabilizer of the form $I...X...I$, where only $X$ is on the pivot qubit $p$. Since row $r_{q+p}$ is a stabilizer, $r_p$ and $r_{q+p}$ need to be anti-commuting. Consequently, the $p^{th}$ entry of the stabilizer corresponding to $r_{q+p}$ is either $Z$ or $Y$. Thus, we know that row $r_{q+p}$ only has a $1$ in column $c_p$ if and only if there is also a $1$ in column $c_{q+p}$ (i.e., if it is a $Y$). Then, we could create an S gate followed by a Hadamard gate to sanitize row $r_{q+p}$. However, when doing that on the pivot qubit, we move the $1$ in row $r_p$ to column $c_{q+p}$. So we must first apply an H-, then an S-, followed by an H-Gate. That way, row $r_p$ remains an identity row, and row $r_{q+p}$ is sanitized. Additionally, row $r_{q+p}$ has a $1$ in column $c_{q+p}$, or else the row would commute with row $r_p$.

Schematically,$r_p$ and $r_{q+p}$ change at $c_p$ and $c_{q+p}$ when sanitizing as follows:
\begin{align*}
\left[ \begin{array}{ccc | ccc}
        & c_p &  &  & c_{q+p} &  \\
     ... & 1 & ... & ... & 0 & ... \\
      & \vdots &  &  & \vdots &  \\
     ... & 1 & ... & ... & 1 & ... \\
\end{array} \right]
& \xrightarrow{H_p} \left[ \begin{array}{ccc | ccc}
        & c_p &  &  & c_{q+p} &  \\
     ... & 0 & ... & ... & 1 & ... \\
      & \vdots &  &  & \vdots &  \\
     ... & 1 & ... & ... & 1 & ... \\
\end{array} \right] 
 \xrightarrow{S_p} \left[ \begin{array}{ccc | ccc}
        & c_p &  &  & c_{q+p} &  \\
     ... & 0 & ... & ... & 1 & ... \\
      & \vdots &  &  & \vdots &  \\
     ... & 1 & ... & ... & 0 & ... \\
\end{array} \right] \\
& \xrightarrow{H_p} \left[ \begin{array}{ccc | ccc}
        & c_p &  &  & c_{q+p} &  \\
     ... & 1 & ... & ... & 0 & ... \\
      & \vdots &  &  & \vdots &  \\
     ... & 0 & ... & ... & 1 & ... \\
\end{array} \right] \\
\end{align*}

After sanitization, we want to remove the remaining non-zero entries in row $r_{q+p}$ using CNOT gates. For this, we use the exact same strategy as for row $r_p$, but all CNOTs will be applied in the opposite direction because they are acting on the stabilizer, which is a Pauli string consisting solely of Is and Zs. To avoid introducing new interactions on the corresponding destabilizer state, we need to ensure that we never add any column to the pivot column, $c_{q+p}$. It is only needed to do this when row $r_{q+p}$ has a $0$ in column $c_{q+p}$. However, we have seen from the sanitization process that this is never the case because the stabilizer corresponding to $r_{q+p}$ needs to anti-commute with $r_p$. So after sanitization, the $p^{th}$ entry of the stabilizer corresponding to row $r_{q+p}$ is always a $Z$, and thus row $r_{q+p}$ will always have a $1$ in column $c_{q+p}$.

When rows $r_p$ and $r_{q+p}$ are identity rows in the Clifford tableau, we need to make the corresponding columns $c_p$ and $c_{q+p}$ into identity. Luckily, because of the nature of the Clifford tableau, this is already the case. In the previous steps of the algorithm, we have turned row $r_p$ and row $r_{q+p}$ into two rows that correspond to a destabilizer and a stabilizer that only have $I$ except for on the pivot qubit $p$ where it has an $X$ and $Z$, respectively. Additionally, we know that both these rows must commute with all other rows in the Clifford tableau. Thus, all other rows correspond to destabilizers or stabilizers with an $I$ on the pivot qubit $p$. Hence, the columns $c_p$ and $c_{q+p}$ have zeroes everywhere except on the diagonal, and we can safely remove the pivot qubit from the tableau since it no longer interacts with any other qubits.

We can continue this process of sanitization of the destabilizer, removal of interactions in the destabilizer, sanitization of the stabilizer, and removal of interactions in the stabilizer by choosing a new pivot. After $q-1$ many iterations, a single qubit will be left, which can be synthesized trivially. A visual example of steps (2)-(5) on a 3-qubit Clifford tableau is shown in Appendix~\ref{apenx:tableau_example}.

Lastly, the Clifford tableau keeps track of the sign of the stabilizers because $Z\ket{1} = -\ket{1}$ and $X\ket{-} = -\ket{-}$. If there is a sign on a qubit after synthesis, this must be undone. We do this by applying the corresponding gate once more. For non-zero entries in the first $q$ rows of the sign column, this means applying $X_i = H_iS_iS_iH_i$, and for the non-zero entries in the bottom $q$ rows of the sign column, this means applying $Z_i = S_iS_i$.

The pseudocode for the proposed algorithm can be found in Algorithm \ref{ch4:alg:clifford_synth}.

\subsection{Heuristic for choosing the pivot qubit}
\label{subsec:pivot_choice}
As previously stated, the pivot qubit can be any non-cutting vertex in the connectivity graph. However, the choice of pivot will influence which gates will be synthesized. Since our primary goal is synthesizing the Clifford tableau with as few CNOTs as possible, we want to choose the row with the smallest Steiner tree as this reduces the number of CNOTs required to traverse the Steiner trees in steps (3) and (5) of the proposed algorithm. However, identifying the minimum Steiner tree is NP-hard~\cite{karp2010reducibility}. Instead, we use the sum of the shortest paths as an approximation for the size of the minimum Steiner tree.
Additionally, we also opted to pre-compute the shortest paths between each qubit using the Floyd-Warshall algorithm~\cite{floyd1962algorithm, warshall1962theorem} to reduce the computational overhead. This provides a lookup table for distances: $d$. Thus, we can use these distances to approximate the cost of the Steiner tree for the pivot row $r$ as follows:
\begin{equation}
	s(r) = \sum_i d_{r,j}\left[\mathbb{I}(T_{r,i} \neq 0 \vee T_{r,i+q} \neq 0) + \mathbb{I}(T_{r+q,i} \neq 0 \vee T_{r+q,i+q} \neq 0)\right]
\end{equation}
where $T$ is the Clifford tableau and $\mathbb{I}: \mathbb{B} \rightarrow \{0, 1\}$ describes the indicator function converting a boolean expression to its corresponding pseudo-boolean integer. Intuitively, we are counting the number of non-identity interactions for each row and estimating the distance of removing those. Hence, for our heuristic, we want to pick a non-cutting vertex $r$ for the graph $G$ such that $s(r)$ is minimized.

Our initial investigation uses a simple greedy heuristic, where we choose the pivot qubit with the lowest cost at each iteration. However, similar to other NP-hard problems with a super-polynomial search space, greedy methods do not guarantee a good solution. Hence, we conducted a secondary investigation to identify the sensitivity of our method to pivot qubit selection. 
We implemented two stochastic approaches for qubit selection: random sampling, and a randomized greedy approach. In the random sampling approach, we choose a valid pivot qubit at random, disregarding cost. In the randomized greedy approach, we apply a weighted selection with the probability of choosing a pivot qubit $r_i$ related to $\textnormal{P}(r=r_i) = \frac{\textnormal{e}^{-s(r_i)}}{\sum_j{\textnormal{e}^{-s(r_j)}}}$, then choosing the best final circuit after $20$ iterations of the full synthesis algorithm. We note that this is a very rudimentary investigation into pivot selection as only one cost function was considered and no tuning of the random greedy algorithm was performed. 

\subsection{Qubit placement heuristic}\label{subsec:mapping}
The above algorithm uses the benefit of applying routing throughout the Clifford synthesis, such that all gates are synthesized with the target architecture in mind. Thus, avoiding the need to add SWAP gates during a later routing step. 
Although this can greatly improve the CNOT count of the final circuit, it is unable to place the qubits. We remedy this by the following heuristic.

We start the synthesis without a mapping from logical qubit to vertex on the connectivity graph. When considering a pivot, we pick from the qubits that have not yet been mapped or have been mapped (in a later step) to a vertex that is not cutting the remaining graph. Then, we follow the procedure as before, but when building the Steiner tree, we map the required qubits that have not yet been mapped to the graph as close to the other qubits as possible. Note that during sanitization, we might add single qubit gates to qubits that have not yet been mapped. In this case, we can buffer them and place them as soon as they are mapped.

\begin{algorithm}[h! bt]
	\caption{Architecture-aware Clifford tableau Synthesis}\label{ch4:alg:clifford_synth}
	\SetKwFunction{Synthesis}{tableau\_synth}
	\SetKwFunction{RemoveInteractionsZ}{remove\_interactions\_z}
	\SetKwFunction{RemoveInteractionsX}{remove\_interactions\_x}
	\SetKwFunction{SanitizeZ}{sanitize\_z}
	\SetKwFunction{SanitizeX}{sanitize\_x}
	\SetKwFunction{SanitizeSigns}{sanitize\_signs}
	\SetKwProg{Fn}{Function}{:}{}
	
	\Fn{\Synthesis{\texttt{G}, \texttt{tableau}}}{
		$\texttt{tableau} \gets \texttt{tableau}^{-1}$  \tcp*{Invert the tableau}
		$ qc \gets$ A Quantum Circuit, which all operations are appended to\;
        $ mapping \gets $ Empty dictionary\;
        $ q \gets $ Number of qubits\;
		\While{G has nodes}{
            \tcp{1. Pick a pivot qubit in the Clifford tableau}
			$p \gets$ choose an unmapped qubit or a non-cutting vertex from G according to heuristic $s(r)$ \tcp*{Map qubit if needed}
            \tcp{2. Sanitize $p$ w.r.t. destabilizer}
            \For{$i\in 1...q$ \textbf{where} $\texttt{tableau}_{p,q+i} =1 \And \texttt{tableau}_{p,i} = 0$}{$qc \gets H_i$}
            \For{$i\in 1...q$ \textbf{where} $\texttt{tableau}_{p,q+i} =1 \And \texttt{tableau}_{p,i} = 1$}{$qc \gets S_i$}
            \tcp{3. Remove interactions w.r.t. destabilizer}
			$nodes \gets$ Non zero entries of $\texttt{tableau}_p$\;
            $T \gets $ Steiner tree over $G$ containing $nodes$ and $p$ \tcp*{Map qubits if needed}
            \For{$parent, child \in bottom_{up}(T)$ \textbf{where} $parent \notin nodes$}{$qc \gets CNOT_{child, parent}$}
            \For{$parent, child \in bottom_{up}(T)$}{$qc \gets CNOT_{parent, child}$}
            \tcp{4. Sanitize $p$ w.r.t. the stabilizer}
            \If{$\texttt{tableau}_{q+p,p} =1$}{$qc \gets H_pS_pH_p$}
            \For{$i\in 1...q$ \textbf{where} $\texttt{tableau}_{q+p,q+i} =0 \And \texttt{tableau}_{q+p,i} = 1$}{$qc \gets H_i$}
            \For{$i\in 1...q$ \textbf{where} $\texttt{tableau}_{q+p,q+i} =1 \And \texttt{tableau}_{q+p,i} = 1$}{$qc \gets S_i$\; $qc \gets H_i$\; }
            
            \tcp{5. Remove interactions w.r.t. stabilizer}
			$nodes \gets$ Non zero entries of $\texttt{tableau}_{q+p}$\;
            $T \gets $ Steiner tree over $G$ containing $nodes$ \tcp*{Map qubits if needed}
            \For{$parent, child \in bottom_up(T)$ \textbf{where} $parent \notin nodes$}{$qc \gets CNOT_{parent, child}$}
            \For{$parent, child \in bottom_up(T)$}{$qc \gets CNOT_{child, parent}$}
			Remove $p$ from G \tcp*{Step 6.}
		}
        \tcp{7. Sanitize the signs}
        \For{$i\in 1...q ; \texttt{tableau}_{i, 2q}\neq 1$}{$qc \gets H_iS_iS_iH_i$\;}
        \For{$i\in 1...q ; \texttt{tableau}_{q+i, 2q}\neq 1$}{$qc \gets S_iS_i$\;}
		\KwRet{qc, mapping} \tcp*[r]{Return the quantum circuit synthesized}
	}
\end{algorithm}
\clearpage

\subsection{Runtime analysis}
We analyze the time complexity of our algorithm as follows. Recalling \autoref{sec:description_algorithm}, we can note that one has to iterate for all $q$ elements of the diagonal to convert the stabilizer and destabilizer to their original form. Per iteration, it is required to approximate a minimal Steiner tree, which was done in our work by \citet{kou_1981}. This requires a computational overhead of $\mathcal{O}(q^3)$, assuming that the number of edges is smaller than the number of nodes. Afterward, we perform a breadth-first traversal through the tree twice. This traversal has a smaller complexity than the building of the tree. Sanitization can be done with a linear overhead. We hence end up with a runtime of:
\begin{equation}
    \mathcal{O}(q (q^3 + q)) = \mathcal{O}(q^4)
\end{equation}

\subsection{Gate complexity analysis}
Additionally, since our proposed algorithm is used for synthesizing and optimizing stabilizer circuits, we analyze the number of gates that would be generated using our algorithm. We analyze the gate complexity of our algorithm by focusing on the three gates generating the Clifford group: H, S, and CNOT.

The algorithm generates at most $q$ H gates for the sanitization of the destabilizer and $q+1$ H gates for the sanitization of the stabilizer. Additionally, $2$ H gates per qubit might be generated to sanitize the signs. Thus, the maximum number of H gates that our algorithm can generate is $\mathcal{O}(q^2)$.

The maximum number of generated S gates follows analogously to the upper bound of the H gates, except that it could generate $q$ S gates to sanitize the stabilizer instead of $q+1$. Thus, the maximum number of S gates our algorithm generates is $\mathcal{O}(q^2)$.

Lastly, for the CNOT gates, our algorithm generates two Steiner trees for each qubit, one for the stabilizer and one for the destabilizer. In the worst case, it is possible that the Steiner tree covers the full connectivity graph, and only the leaves and root of the tree are terminal nodes, resulting in a maximum number of Steiner nodes. Then, our algorithm generates a CNOT for each Steiner node ($\mathcal{O}(q)$) and then one for all nodes in the Steiner tree, except the root ($\mathcal{O}(q)$). Thus, the maximum number of CNOTs generated by our algorithm is: 
\begin{equation}
    \mathcal{O}(\sum_{i=0}^q 4(q-i)) = \mathcal{O}(2q^2-2q) = \mathcal{O}(q^2)
\end{equation}

If we compare this result with the layered normal forms, we see that we can generate many more single qubit gates. However, we generate all CNOTs in a single Gaussian-Elimination-like pass. Although this does not seem very different from the CNOT complexity of the previously discussed normal forms, it does matter practically whether you generate double or triple the amount of CNOT gates. Each CNOT layer in the different normal forms requires Gaussian Elimination to synthesize. This can be done in $\mathcal{O}(\frac{q^2}{\log (q)})$ in the unconstrained case~\cite{patel2008optimal} and $\mathcal{O}(q^2)$ or $\mathcal{O}(\frac{q^2}{log \delta})$ architecture-aware~\cite{Wu_2023}, where $\delta$ is a connectivity graph dependent number. The original normal form from \citet{Aaronson_2004} requires $5$ CNOT layers. The normal form \citet{Bravyi_2021} has $2$ CNOT layers and the one from \citet{Duncan_2020} only $1$ CNOT layer, but these also contain $2$ CZ layers, for which no architecture-aware synthesis algorithm exists. 

However, in practice, circuits generated from these normal forms do not obey the target connectivity constraints and need to be routed. This routing then adds SWAP gates to optimally move the qubits, but that takes $3$ CNOTs per SWAP gate, which can be $\mathcal{O}(d)$ per CNOT where $d$ is the maximum distance of the connectivity graph. 

We want to point out to the reader that we do not expect our algorithm to outperform these normal-form methods in the unconstrained case. Nevertheless, finding a similar algorithm as \citet{patel2008optimal} might further fuel the asymptotic reduction of CNOTs. 

Given that, for NISQ devices, two-qubit gates generally have the lowest fidelity~\cite{cowtan2020generic}, reducing the number of generated CNOT gates is extremely valuable as is shown experimentally in the next section.

%% file: struct/evaluation.tex
To determine the efficacy of the proposed method, we conducted experiments on different architectures available in the IBM Software stack\footnote{Our implementation can be found on Github: \url{https://github.com/daehiff/pauliopt/tree/clifford_synthesis}. We used qiskit version \texttt{0.39.0}, qiskit-aer \texttt{0.11.0}. qiskit-ibm-runtime \texttt{0.8.0} and qiskit-ibm-provider \texttt{0.19.2}. The architectures \textit{ibm\_ithaca} and \textit{ibm\_brisbane} where obtained from the qiskit API: \href{https://api-qcon.quantum-computing.ibm.com/api/users/backends}{https://api-qcon.quantum-computing.ibm.com/api/users/backends} (version September 2023)}. More specifically, we targeted the backends: \textit{quito} (5 qubits),  \textit{nairobi} (7 qubits), \textit{guadalupe} (16 qubits), \textit{mumbai} (27 qubits), \textit{ithaca} (65 qubits) and \textit{brisbane} (127 qubits). An outline of the used connectivity graphs can be found in \autoref{apndx:arch_study}.

Per architecture, we created 20 unique circuits by adjusting the number of gates within the set {H, S, CX} and selecting each gate with equal probability. For each architecture, we utilized increasing input gate counts until asymptotic behavior was observed for most methods, seen in \autoref{fig:cx_comparison_smaller_arch} and \autoref{fig:cx_comparison_larger_arch}. When generating each random circuit, for each input gate, we sampled the gate type as well as position in the architecture uniformly from the indicated gate set and available qubits; this implies that single-qubit gates are more likely to be present in the circuit. 

Since our algorithm relies on a simple heuristic for picking the pivot qubit, we investigate the quality of that heuristic with respect to two randomized picking strategies, as explained before. From this investigation, we show that the heuristic is better than random pivot picking without requiring the multiple compilations needed for the greedy sampling. In the later experiments, we use the simple heuristic as proposed. Additionally, in all the experiments both the transpiler and the proposed algorithm allow for finding a good qubit mapping onto the connectivity graph. For the proposed algorithm, we used the method outlined in \autoref{subsec:mapping}. We did this to avoid poor performance due to a poor choice of qubit mapping on the constrained architectures.

We compared our algorithm against five baselines, which we all transpiled using the default \texttt{transpile} method from Qiskit~\citet{qiskit} parameterized by SABRE swap~\cite{li2019tackling} when the assumed target connectivity graph did not match the actual connectivity graph. In particular, we used the default transpilation of the original random circuit (Transpile), the method of \citet{Bravyi_2021} as implemented in Qiskit~\cite{qiskit}, the method of \citet{berg2021simple} as implemented in Stim~\cite{Gidney_2021}, the method of \citet{Duncan_2020} as implemented in PyZX~\cite{kissinger2020Pyzx}, and the method of \citet{Maslov_2023} as implemented in Qiskit v1.0~\cite{qiskit}.
We evaluate our algorithm against these baselines with respect to gate count and gate depth of the resulting circuits in both the constrained and unconstrained case.

To investigate how close to optimal these heuristic methods are, we compare our method, \citet{Maslov_2023}, and \citet{Bravyi_2021} against the depth optimal synthesis approach of \citet{peham2023depthoptimal}\footnote{Note that the runtime of gate count optimal version of this algorithm was too high for comparison for 3-qubit architectures, so we were advised to use the depth-optimal variant by the authors.} on small architectures: The 3-, 4-, and 5-qubit line and complete architectures, and the 5-qubit IBM \textit{quito} device. 
Moreover, we compare both the depth and CNOT count in both the constrained and unconstrained cases to show the effect of routing on a depth-optimal circuit.

Lastly, we give a simple demo to show how the more optimized transpiled circuits result in a faster runtime and a higher circuit fidelity when running on actual IBM hardware (\texttt{quito}). For this, we compared our algorithms against \citet{Bravyi_2021} and \citet{berg2021simple}.

\subsection{Evaluation of choosing pivot qubit}
\begin{figure}[bt]
	\centering
     \vspace{-1em}
    \begin{subfigure}{\textwidth}
    \centering
    \includegraphics[width=0.8\textwidth]{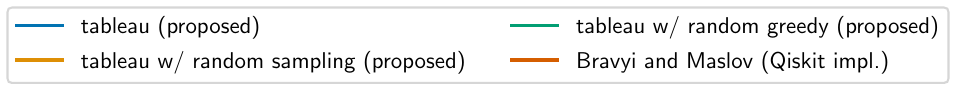}
    \end{subfigure}
    \vspace{-1em}
    \begin{subfigure}{\textwidth}
        \begin{subfigure}{0.47\textwidth}
    	    \centering
    		\includegraphics[width=\textwidth]{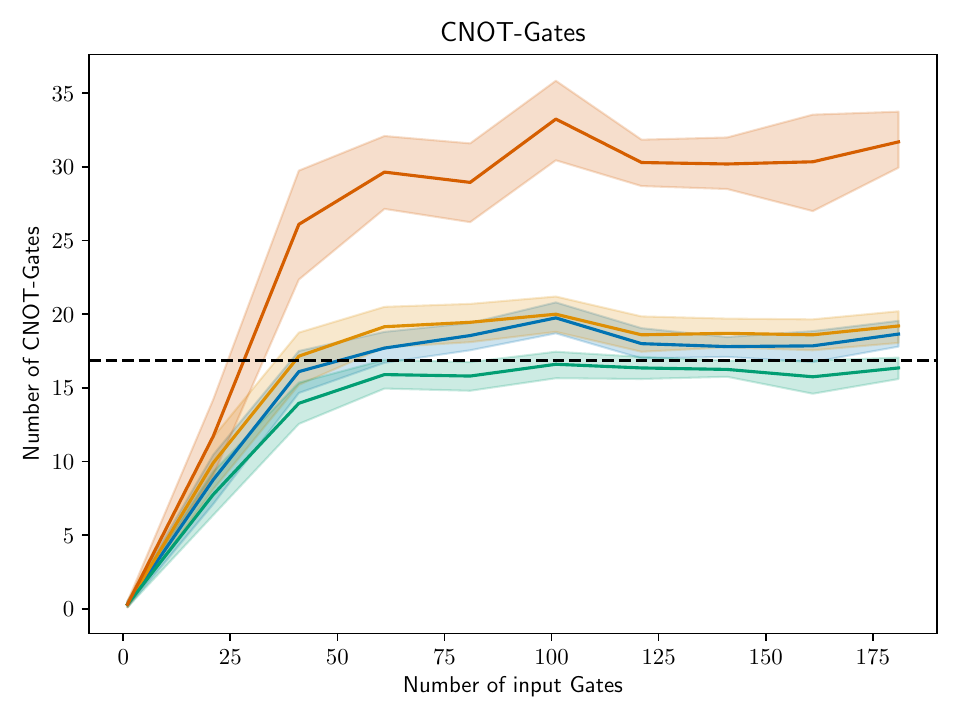}
    		\caption{Quito (5 Qubits)}
    	\end{subfigure}
    \hfill
        \begin{subfigure}{0.47\textwidth}
    	    \centering
            \includegraphics[width=\textwidth]{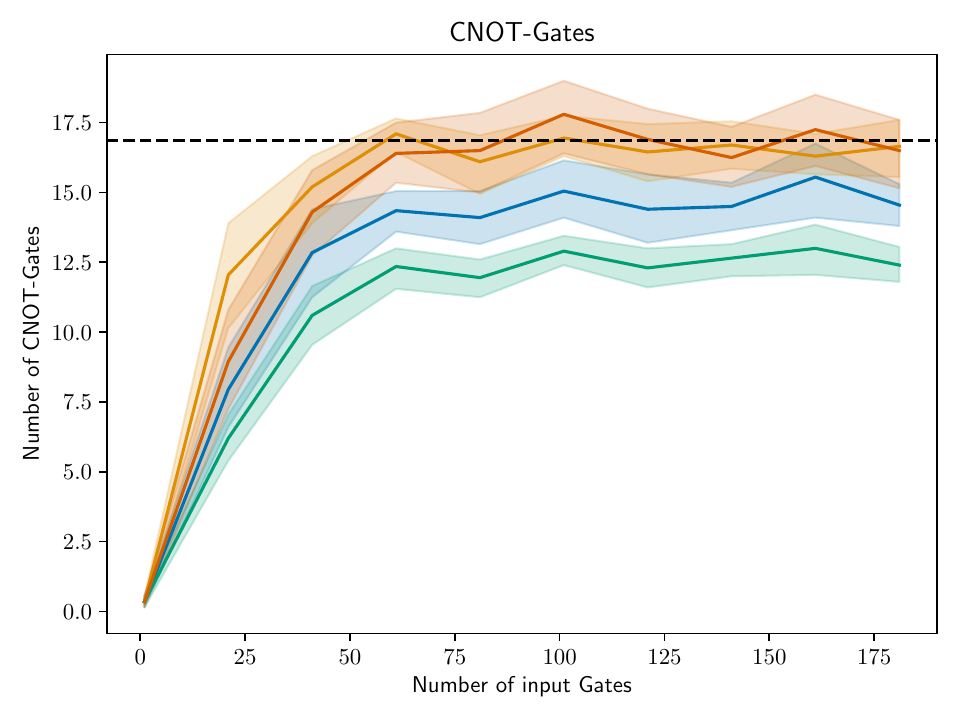}
    		\caption{Complete (5 Qubits)}
    	\end{subfigure}
    \hfill
     	\begin{subfigure}{0.47\textwidth}
    	    \centering
    		\includegraphics[width=\textwidth]{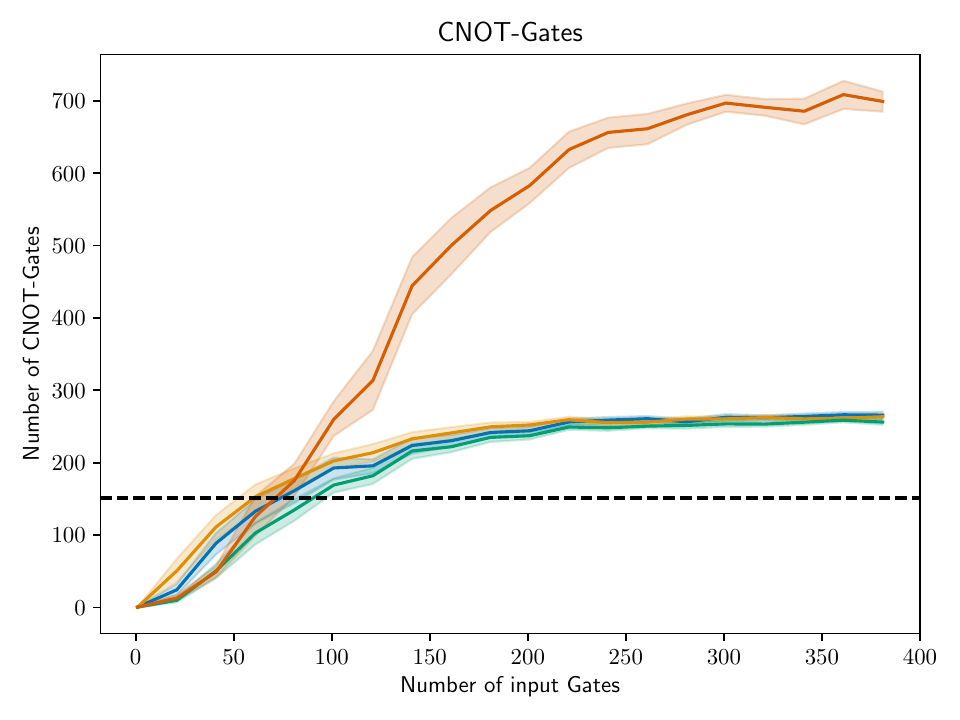}
    		\caption{Guadalupe (16 Qubits)}
    	\end{subfigure}
     \hfill
     	\begin{subfigure}{0.47\textwidth}
    	    \centering
    		\includegraphics[width=\textwidth]{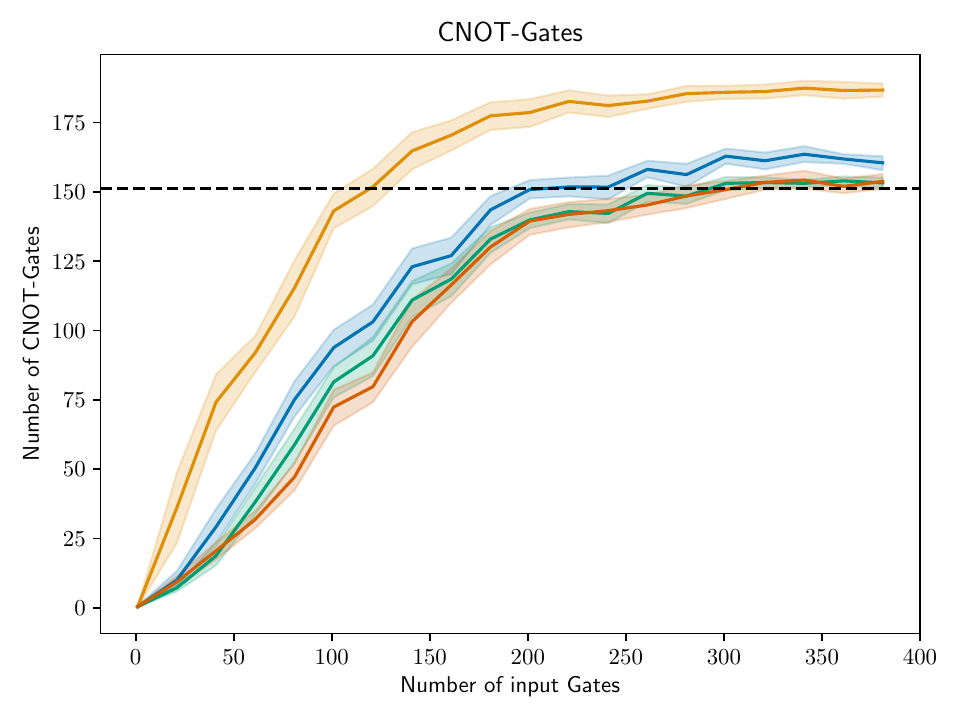}
    		\caption{Complete (16 Qubits)}
    	\end{subfigure}
     \hfill
    	\begin{subfigure}{0.47\textwidth}
    	    \centering
    		\includegraphics[width=\textwidth]{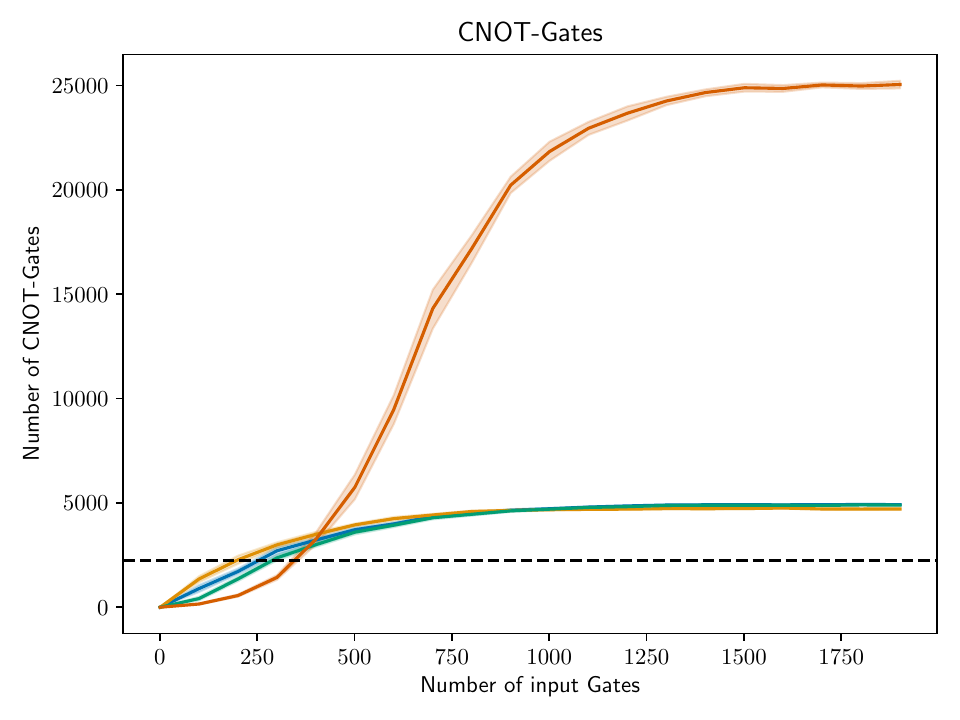}
    		\caption{Ithica (65 Qubits)}
    	\end{subfigure}
     \hfill
        \begin{subfigure}{0.47\textwidth}
    	    \centering
    		\includegraphics[width=\textwidth]{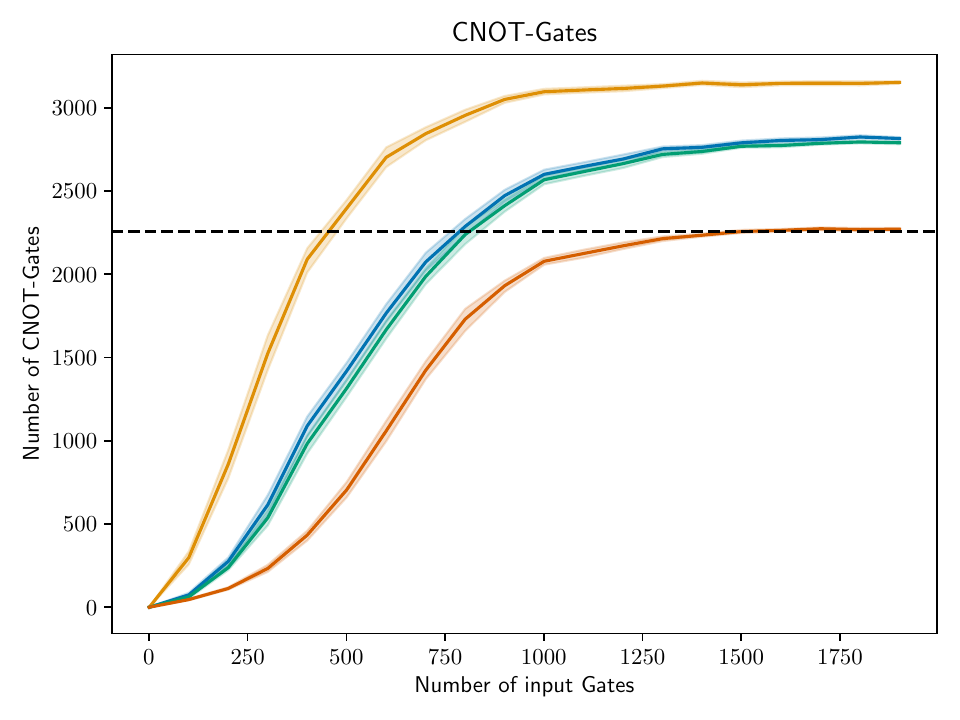}
    		\caption{Complete (65 Qubits)}
    	\end{subfigure}
    \end{subfigure}
	\caption{CNOT count for architectures on 5, 16 and 65 qubits when investigating the effect of pivot qubit selection. Specifically, the backends Quito, Guadalupe, and Ithaca were used. We additionally reported the CNOT count of the complete architectures of the same size.}\label{fig:cx_comparison_pivot}
    \vspace{3em}
\end{figure}

\begin{figure}[bt]
	\centering
    \vspace{-1em}
    \begin{subfigure}{\textwidth}
    \centering
    \includegraphics[width=0.8\textwidth]{img_pivot/legend_pivot.pdf}
    \end{subfigure}
    \vspace{-1em}
    \begin{subfigure}{\textwidth}
        \begin{subfigure}{0.47\textwidth}
    	    \centering
    		\includegraphics[width=\textwidth]{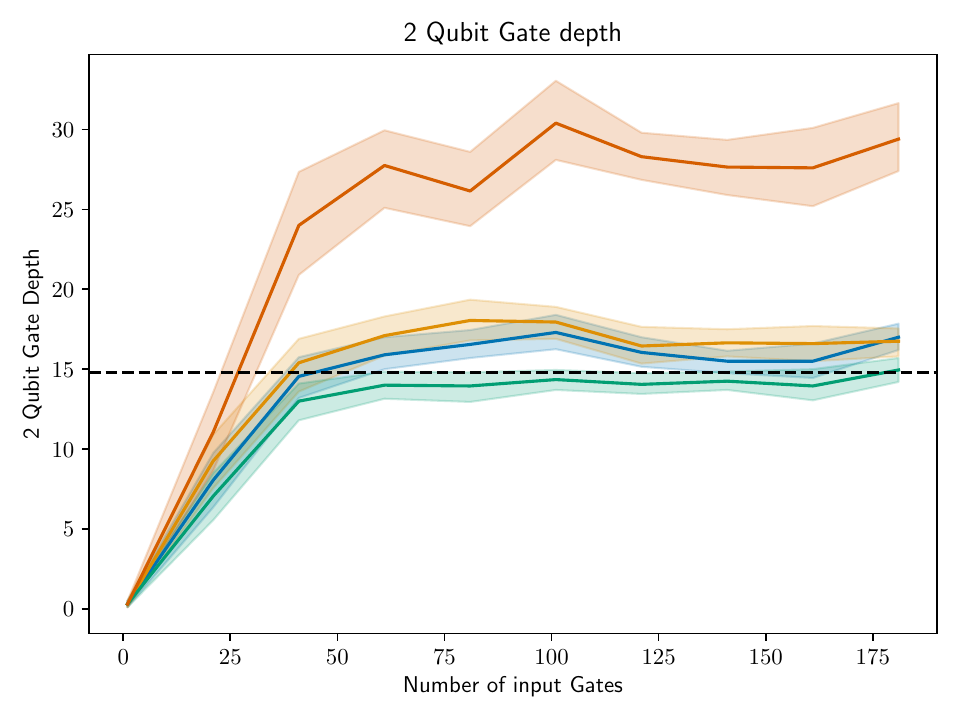}
    		\caption{Quito (5 Qubits)}
    	\end{subfigure}
    \hfill
        \begin{subfigure}{0.47\textwidth}
    	    \centering
            \includegraphics[width=\textwidth]{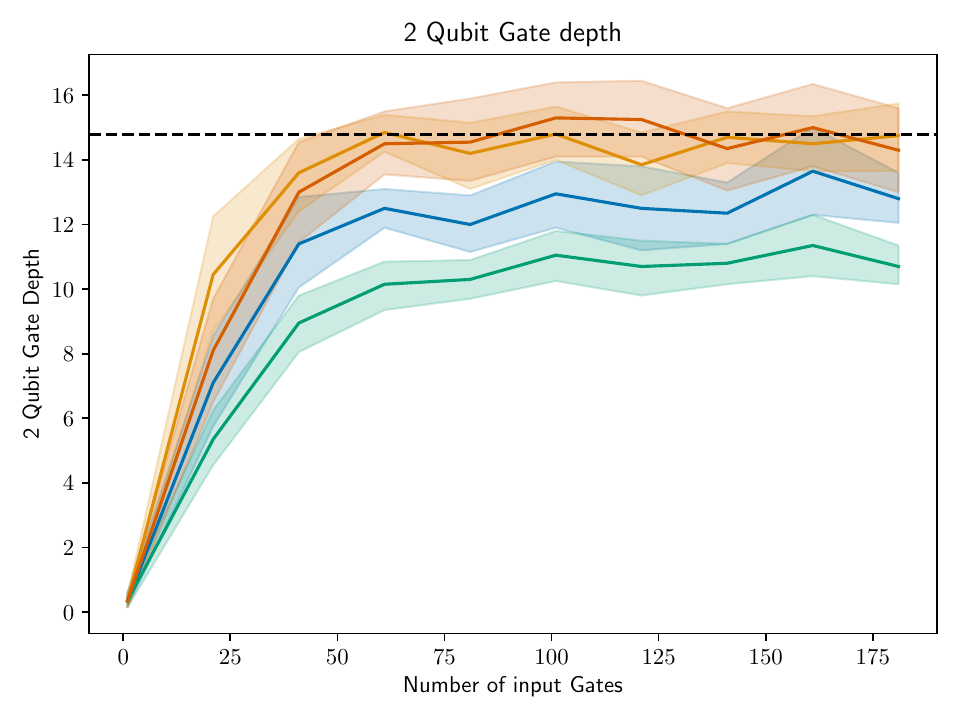}
    		\caption{Complete (5 Qubits)}
    	\end{subfigure}
    \hfill
     	\begin{subfigure}{0.47\textwidth}
    	    \centering
    		\includegraphics[width=\textwidth]{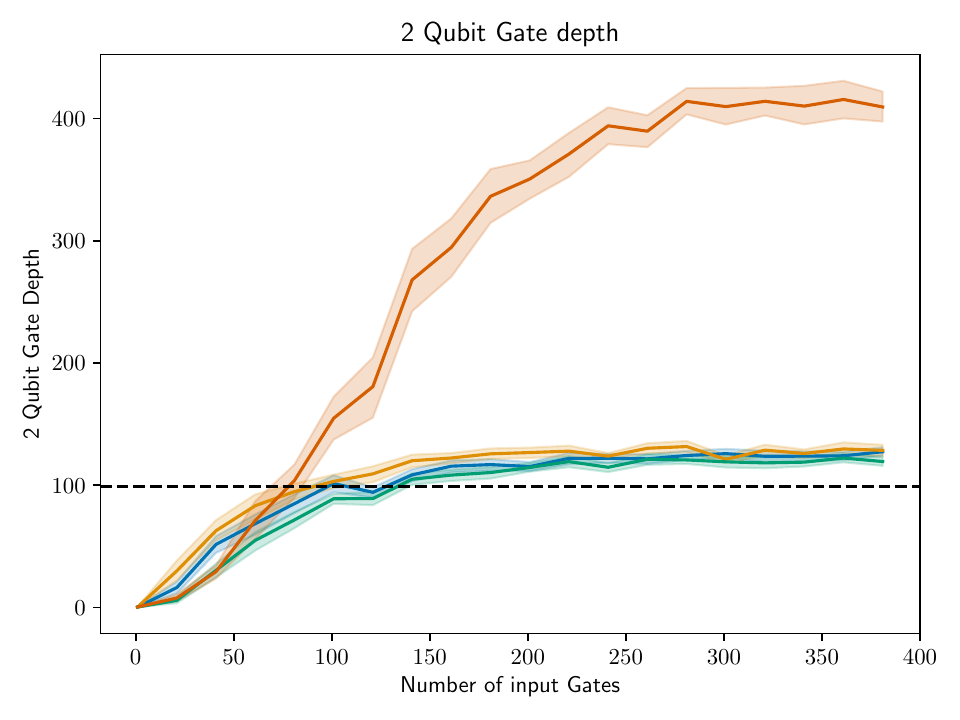}
    		\caption{Guadalupe (16 Qubits)}
    	\end{subfigure}
     \hfill
     	\begin{subfigure}{0.47\textwidth}
    	    \centering
    		\includegraphics[width=\textwidth]{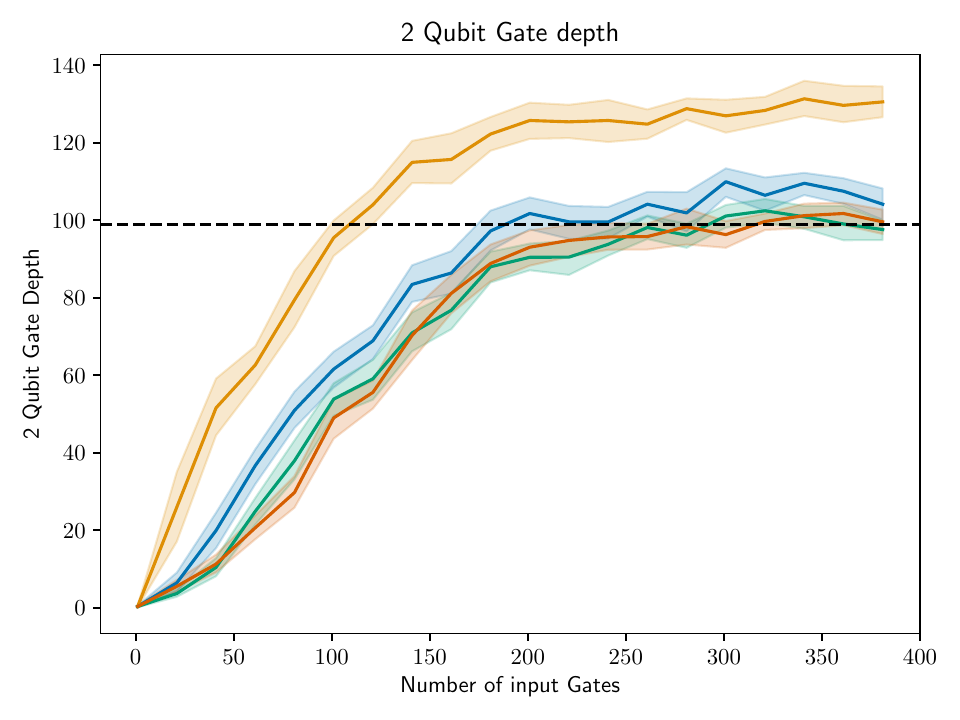}
    		\caption{Complete (16 Qubits)}
    	\end{subfigure}
     \hfill
    	\begin{subfigure}{0.47\textwidth}
    	    \centering
    		\includegraphics[width=\textwidth]{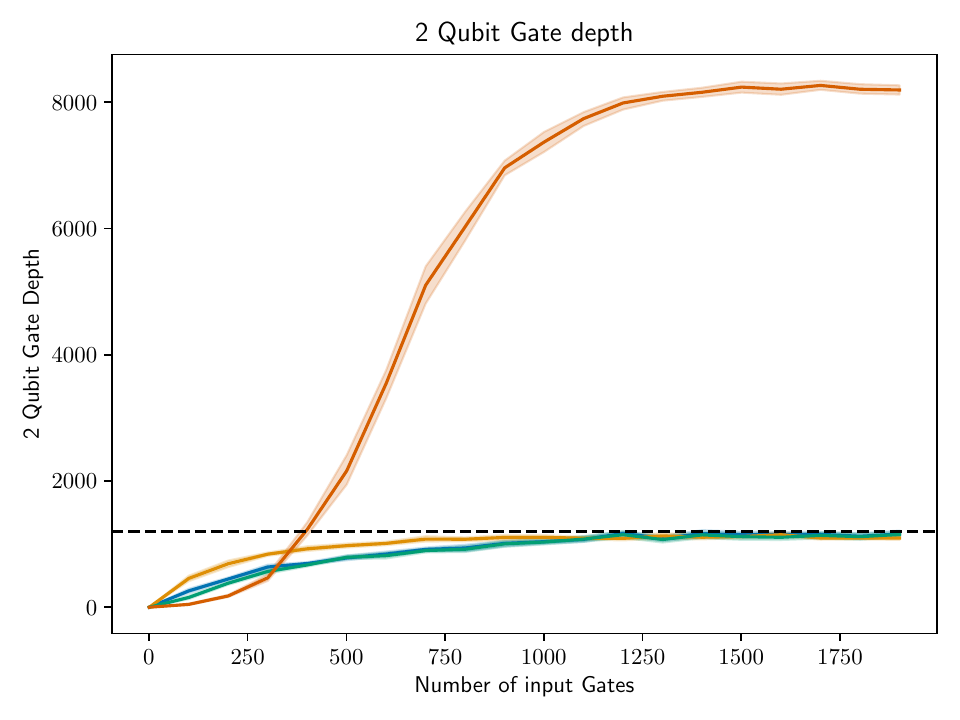}
    		\caption{Ithica (65 Qubits)}
    	\end{subfigure}
     \hfill
        \begin{subfigure}{0.47\textwidth}
    	    \centering
    		\includegraphics[width=\textwidth]{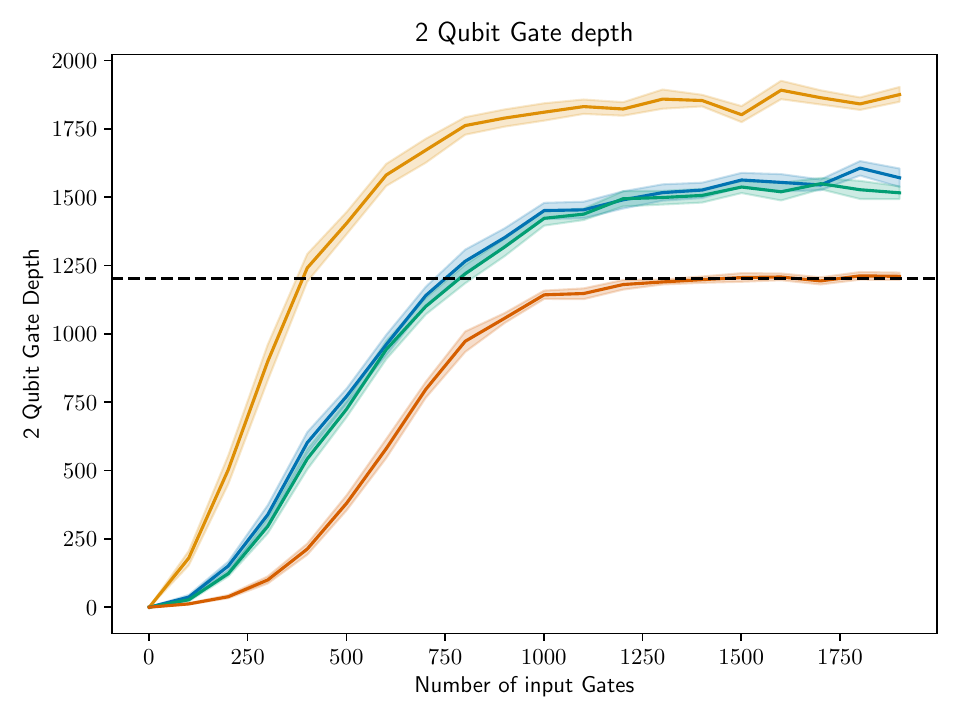}
    		\caption{Complete (65 Qubits)}
    	\end{subfigure}
    \end{subfigure}
	\caption{CNOT depth for architectures on 5, 16, and 65 qubits when investigating the effect of pivot qubit selection. Specifically, the backends Quito, Guadalupe, and Ithaca were used. We additionally reported the CNOT count of the complete architectures of the same size.}\label{fig:depth_comparison_pivot}
    \vspace{3em}
\end{figure}

We compared the results of using a random selection and a randomized greedy algorithm as outlined in \autoref{subsec:pivot_choice}. As seen in \autoref{fig:cx_comparison_pivot} and \autoref{fig:depth_comparison_pivot}, for smaller architectures, such as Quito, Nairobi, and Guadalupe, qubit selection with a randomized greedy algorithm allows us to remain competitive with \citet{Bravyi_2021} at low gate counts on both complete and restricted architectures. Beyond 16 qubits, \citet{Bravyi_2021} still outperforms our algorithm on complete architectures and for low gate counts in restricted architectures. The approach could be tuned for further improvements. However, the benefits of this approach appear negligible for larger architectures and the approach itself does not scale in this context as it involves running the entire synthesis for multiple iterations.

Randomly selecting a pivot qubit unsurprisingly leads to worse asymptotic behavior on complete architectures. However, for restricted architectures, there is only minor impact on two-qubit gate depth and almost no discernable difference in CNOT gate count. This makes sense as the search space for pivot qubit selection sees a combinatorial explosion with increasing architecture size for complete architectures. Taking the results of the depth investigation into account, we conjecture that our simple greedy heuristic struggles to find a reasonably good pivot qubit selection with no connectivity restraints. However, the highly restricted architecture of the IBM quantum devices reduces the search space sufficiently so it produces a reasonably good solution.

Overall, the randomized greedy approach does not provide a significant improvement considering its cost, though experimenting with pivot selection heuristics remains an open topic. Additionally, one could further investigate the sensitivity of synthesis performance with respect to pivot selection over a range of graph-theoretic shapes in the connectivity graph. We suspect that the pivot qubit selection heuristic will need to be specifically designed for target architectures for the best results.

\FloatBarrier
\subsection{Evaluation of the Gate Count}
We focus the evaluation of our algorithm on the CNOT counts because CNOTs typically have a lower fidelity than single qubit gates on quantum hardware. For completeness, the single-qubit gate counts obtained in our evaluation can be found in \autoref{apenx:eval_single_qubit}.

\begin{figure}[bt]
    \centering
    \vspace{-1em}
    \begin{subfigure}{\textwidth}
    \centering
    \includegraphics[width=0.8\textwidth]{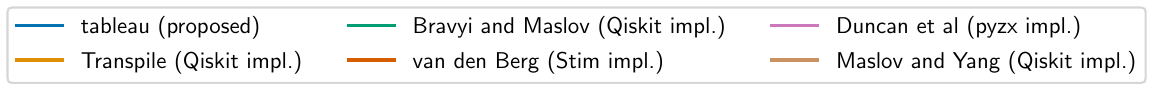}
    \end{subfigure}
    \vspace{-1em}
    \begin{subfigure}{\textwidth}
        \begin{subfigure}{0.47\textwidth}
    	    \centering
    		\includegraphics[width=\textwidth]{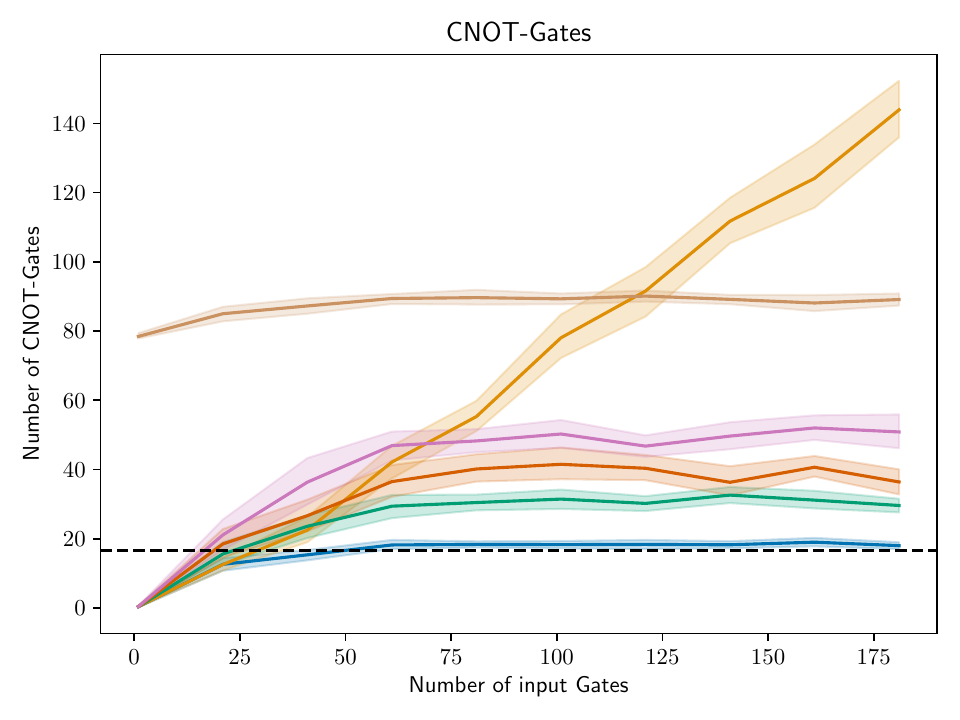}
    		\caption{Quito (5 Qubits)}
    	\end{subfigure}
    \hfill
        \begin{subfigure}{0.47\textwidth}
    	    \centering
            \includegraphics[width=\textwidth]{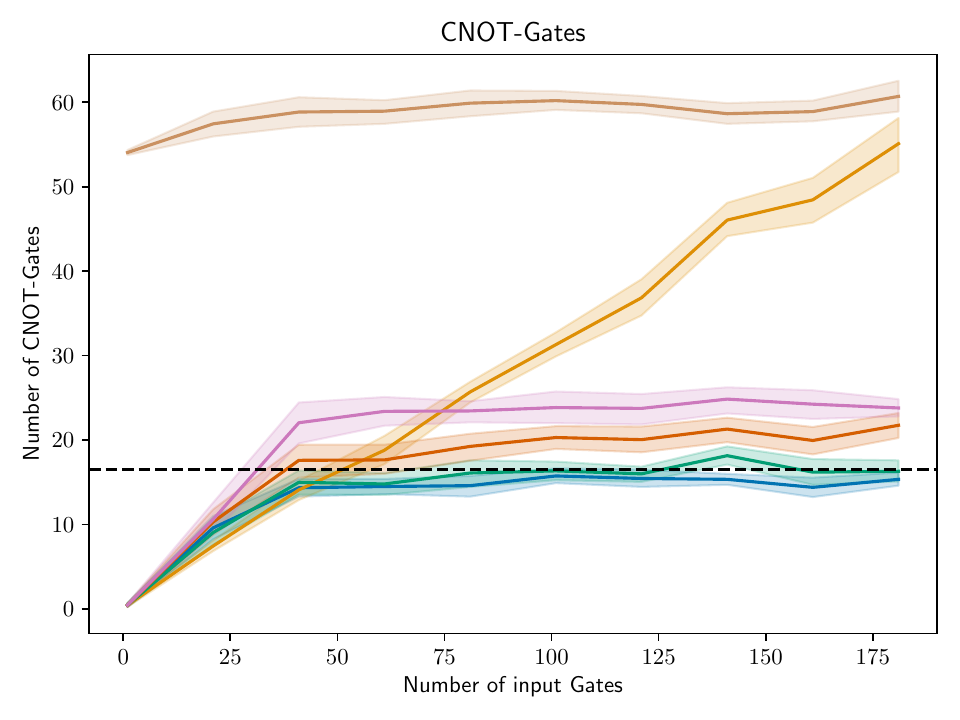}
    		\caption{Complete (5 Qubits)}
    	\end{subfigure}
    \hfill
     	\begin{subfigure}{0.47\textwidth}
    	    \centering
    		\includegraphics[width=\textwidth]{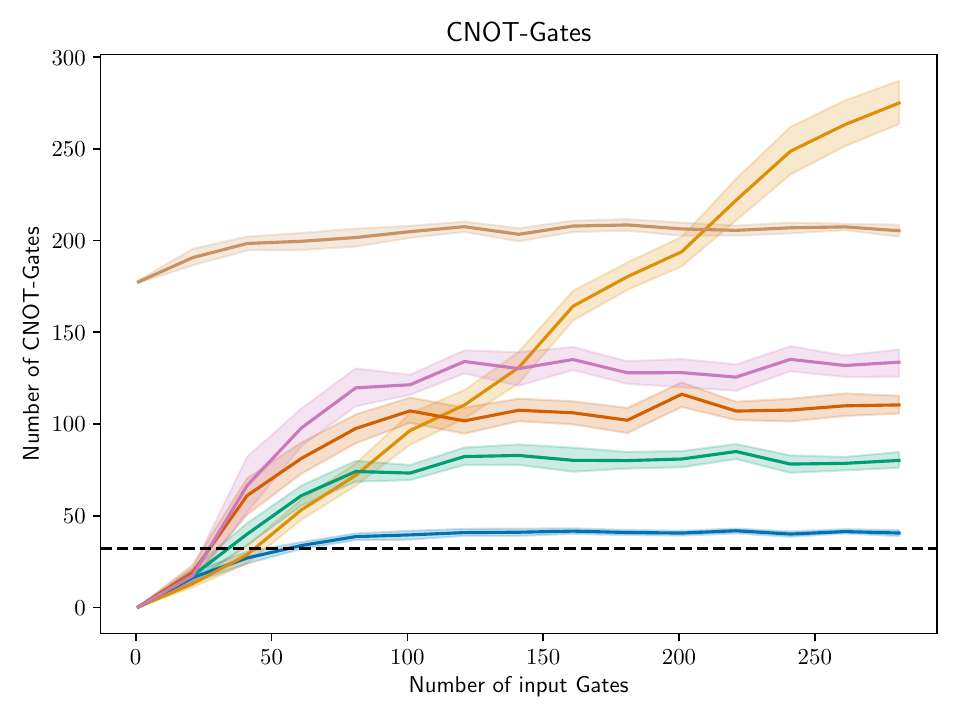}
    		\caption{Nairobi (7 Qubits)}
    	\end{subfigure}
     \hfill
     	\begin{subfigure}{0.47\textwidth}
    	    \centering
    		\includegraphics[width=\textwidth]{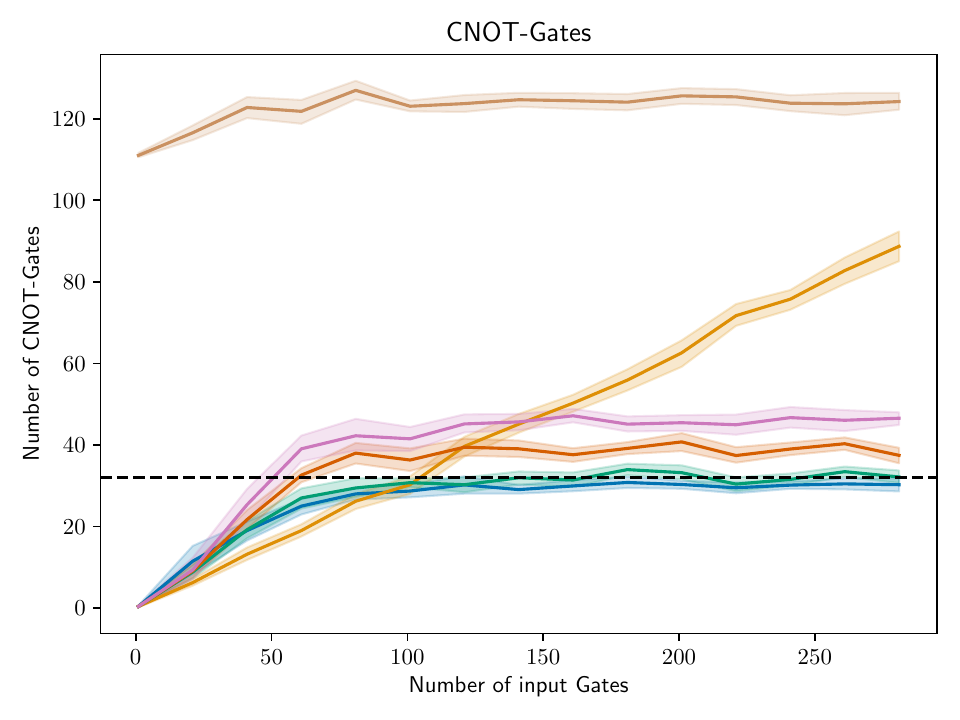}
    		\caption{Complete (7 Qubits)}
    	\end{subfigure}
     \hfill
    	\begin{subfigure}{0.47\textwidth}
    	    \centering
    		\includegraphics[width=\textwidth]{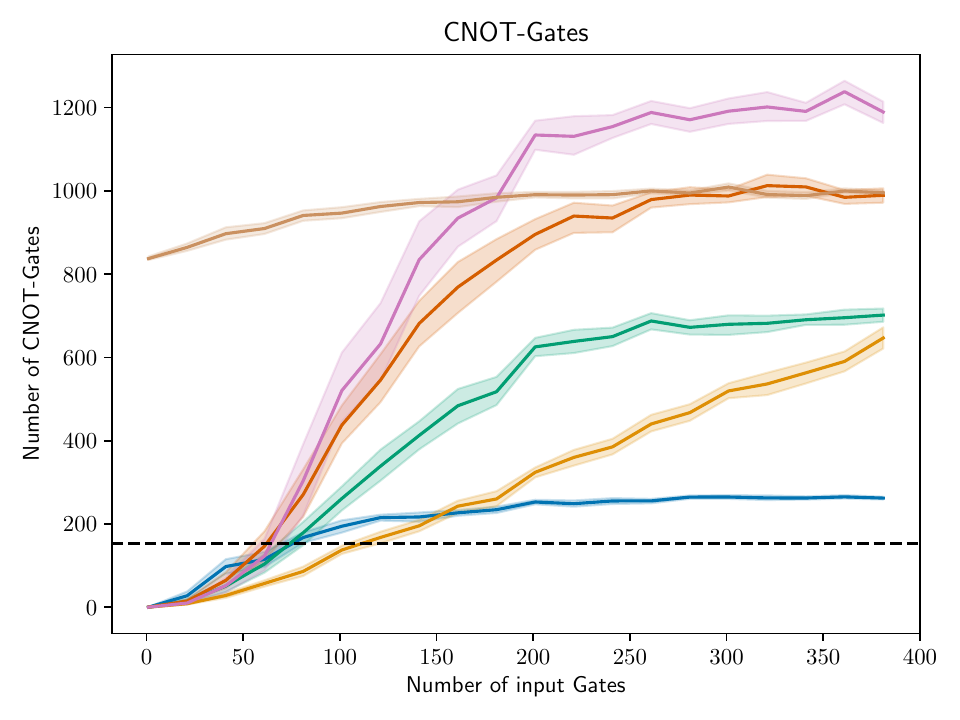}
    		\caption{Guadalupe (16 Qubits)}
    	\end{subfigure}
     \hfill
        \begin{subfigure}{0.47\textwidth}
    	    \centering
    		\includegraphics[width=\textwidth]{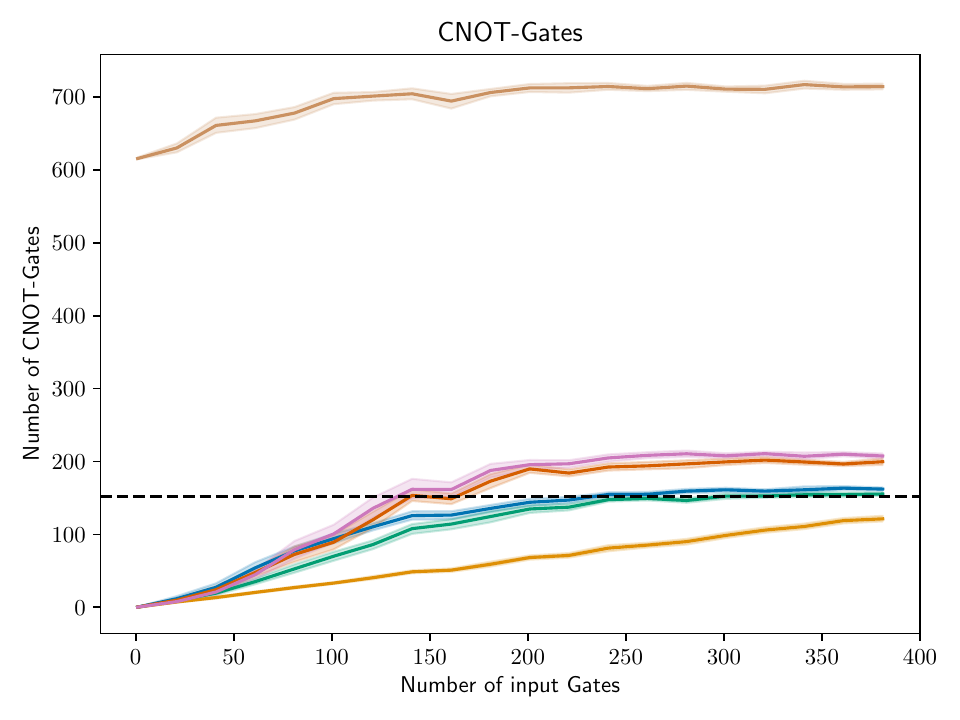}
    		\caption{Complete (16 Qubits)}
    	\end{subfigure}
    \end{subfigure}
	\caption{CNOT count for smaller architectures up to 16 qubits. Specifically, the backends Guadalupe, Nairobi, and Quito were used. We additionally reported the CNOT count of the complete architectures of the same size. The dashed black line describes the CNOT count upon the convergence of \citet{Bravyi_2021} in both pictures.}\label{fig:cx_comparison_smaller_arch}
\end{figure}

\begin{figure}[bt]
    \centering
    \vspace{-1em}
    \begin{subfigure}{\textwidth}
    \centering
    \includegraphics[width=0.8\textwidth]{img_new_methods/legend_experiments.pdf}
    \end{subfigure}
    \vspace{-1em}
	\begin{subfigure}{\textwidth}
        \begin{subfigure}{0.47\textwidth}
    	    \centering
    		\includegraphics[width=\textwidth]{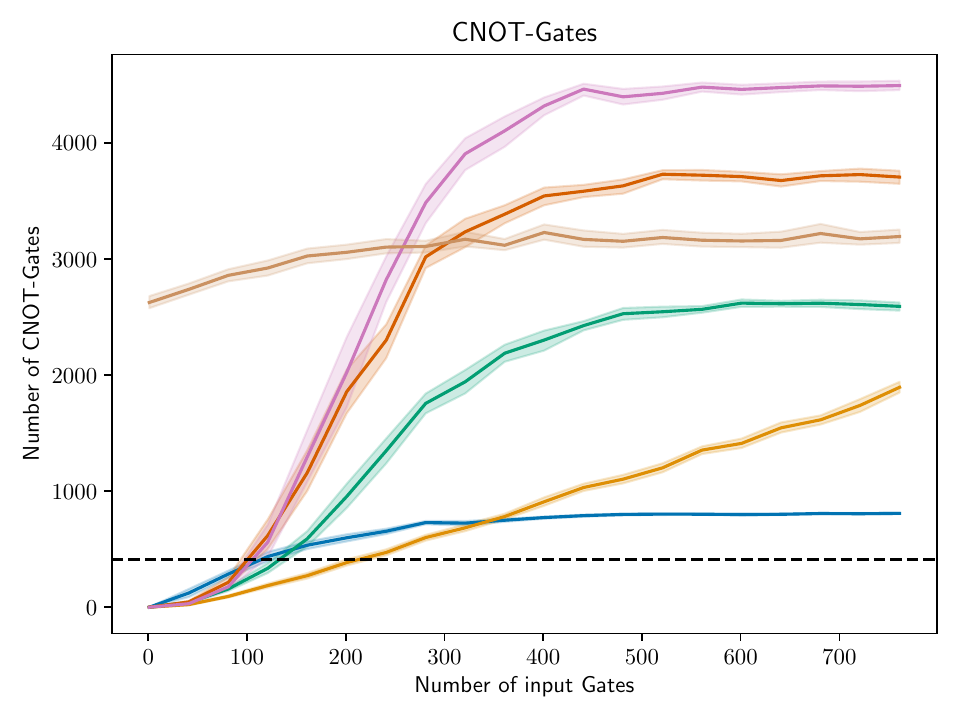}
    		\caption{Mumbai (27 Qubits)}
    	\end{subfigure}
     \hfill
        \begin{subfigure}{0.47\textwidth}
    	    \centering
            \includegraphics[width=\textwidth]{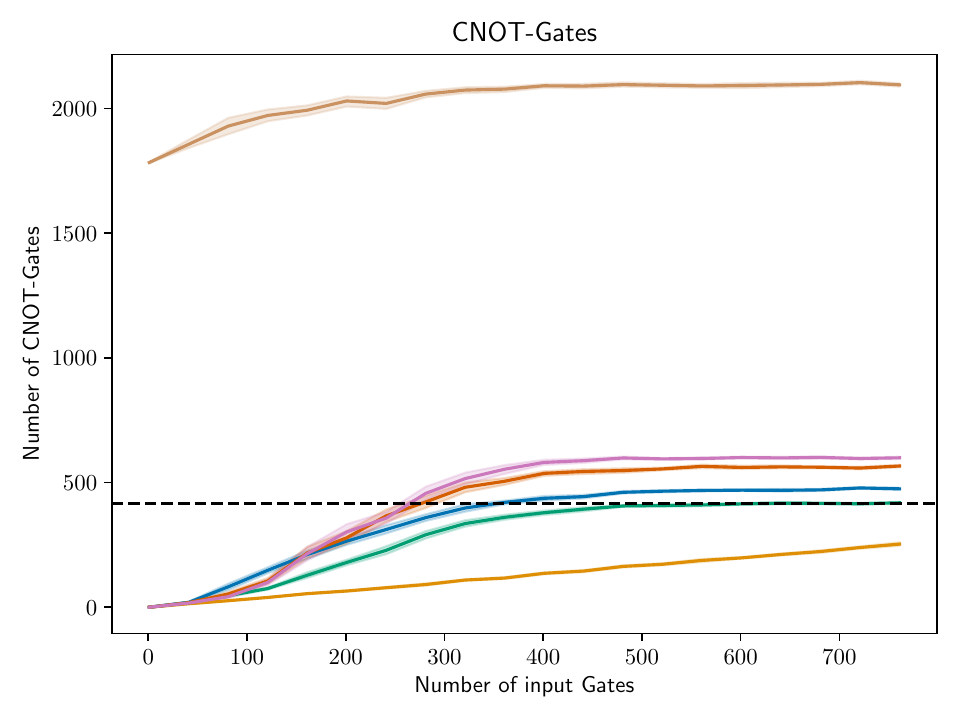}
    		\caption{Complete (27 Qubits)}
    	\end{subfigure}
        \begin{subfigure}{0.47\textwidth}
    	    \centering
    		\includegraphics[width=\textwidth]{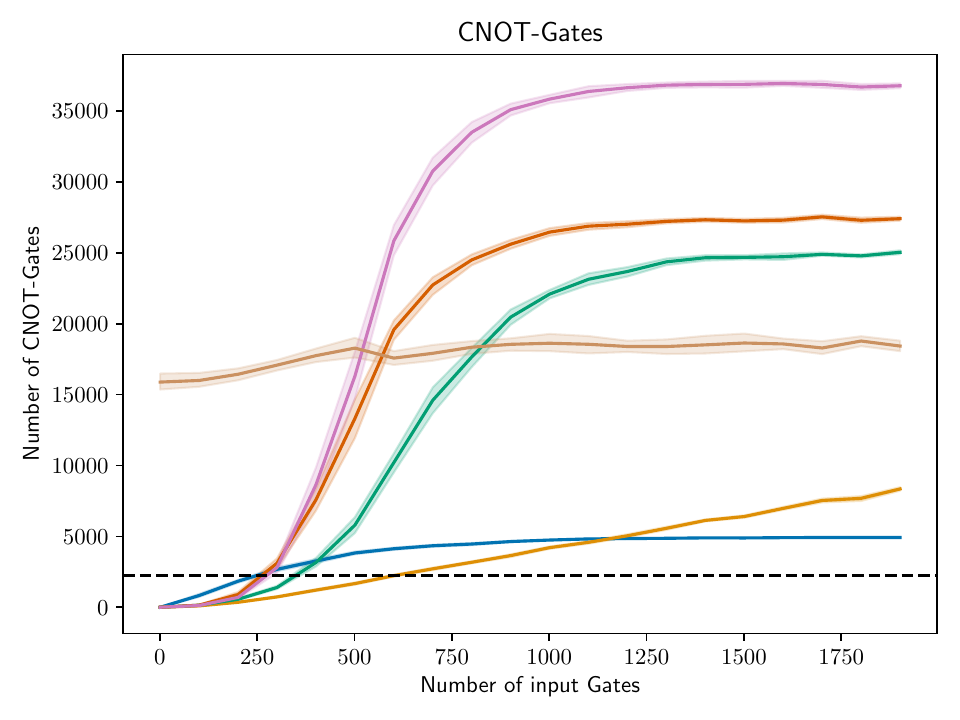}
    		\caption{Ithaca (65 Qubits)}
    	\end{subfigure}
     \hfill
        \begin{subfigure}{0.47\textwidth}
    	    \centering
            \includegraphics[width=\textwidth]{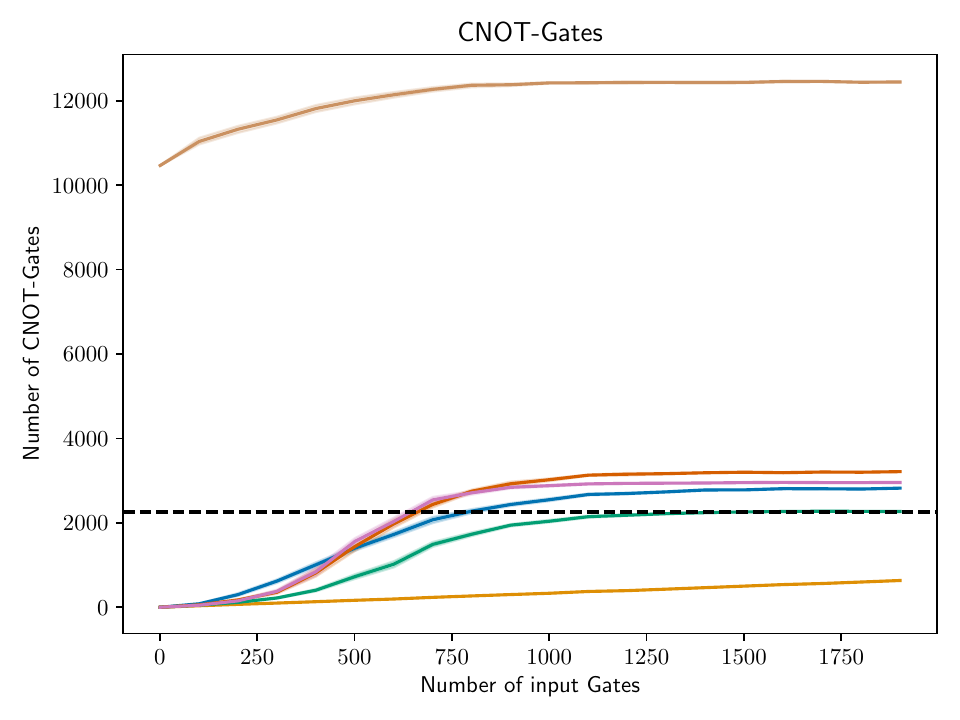}
    		\caption{Complete (65 Qubits)}
    	\end{subfigure}
        \begin{subfigure}{0.47\textwidth}
    	    \centering
    		\includegraphics[width=\textwidth]{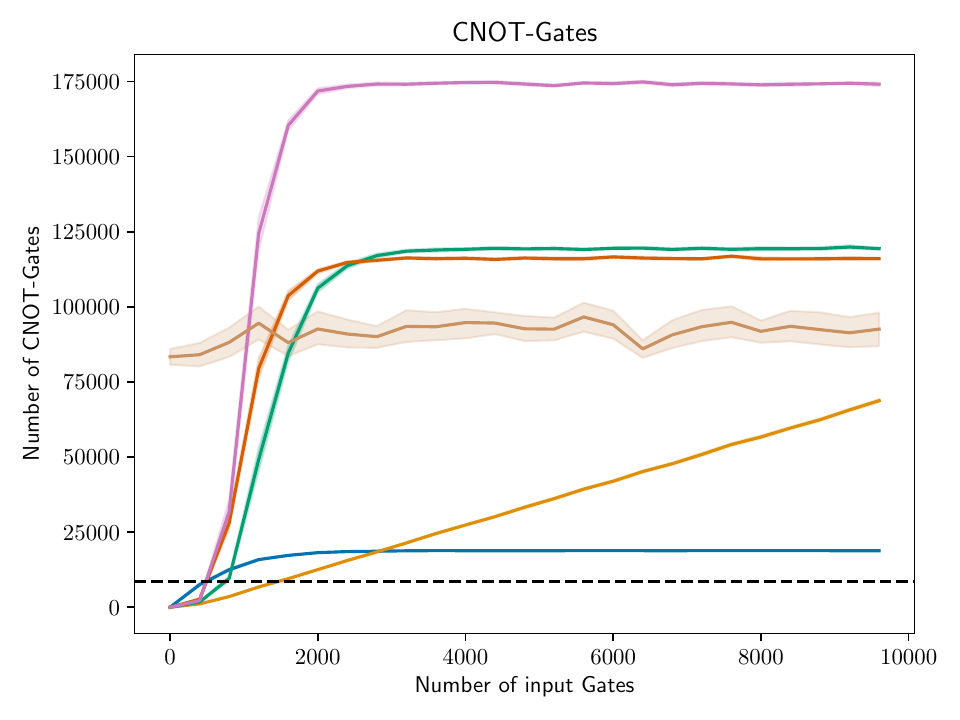}
    		\caption{Brisbane (127 Qubits)}
    	\end{subfigure}
        \hfill
        \begin{subfigure}{0.47\textwidth}
    	    \centering
            \includegraphics[width=\textwidth]{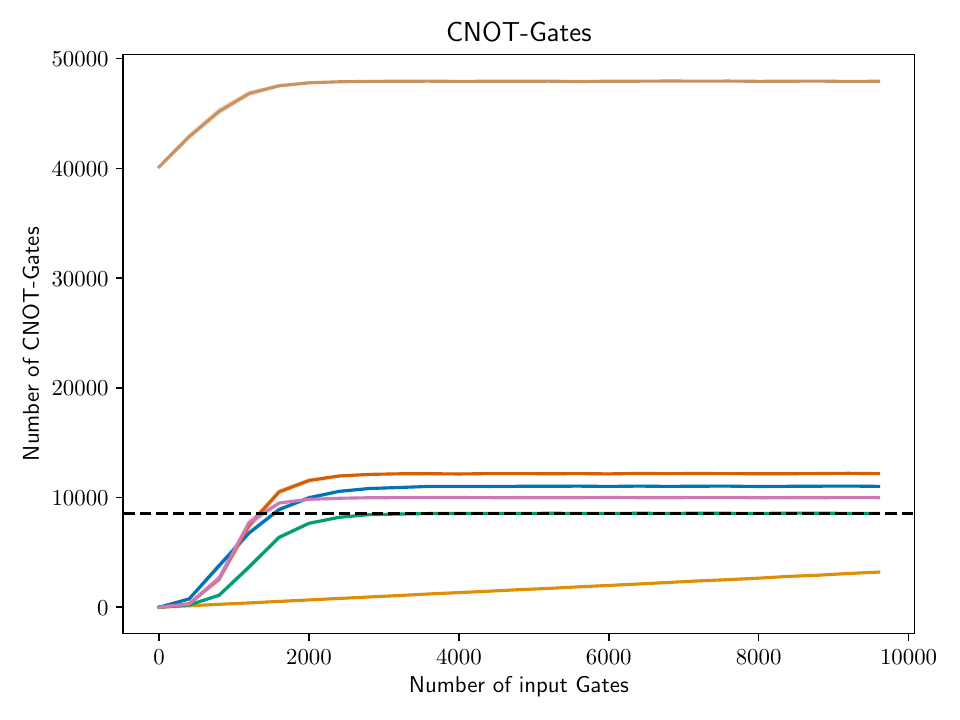}
    		\caption{Complete (127 Qubits)}
    	\end{subfigure}
    \end{subfigure}
    \caption{CNOT count for larger architectures up to 127 qubits. Specifically, the backends Brisbane, Ithaca, and Mumbai were used. We additionally reported the CNOT count of the complete architectures of the same size. The dashed black line describes the CNOT-Count upon the convergence of \citet{Bravyi_2021} in both pictures.}
    \label{fig:cx_comparison_larger_arch}
\end{figure}

As can be seen in \autoref{fig:cx_comparison_smaller_arch} and \autoref{fig:cx_comparison_larger_arch}, our method displays comparable results on a fully connected architecture with the benchmark algorithms; our method generally outperforms \citet{berg2021simple}, \citet{Duncan_2020}, and \citet{Maslov_2023} and is outperformed by \citet{Bravyi_2021}. We note that the comparative performance also tends to worsen with larger architectures. The horizontal dashed black line in each graph indicates the approximate asymptotic behavior of \citet{Bravyi_2021} on a complete architecture and allows a quick visual comparison of CNOT count before and after routing to a specific architecture. 

However, when we compare the CNOT count in the case of a specific target architecture, the extra CNOTs needed to adhere to the connectivity constraints are much fewer with our proposed method when compared to the baseline synthesis algorithms. 

We further observe in architectures with higher qubit count, such as Guadalupe and all architectures in \autoref{fig:cx_comparison_larger_arch}, that our algorithm is outperformed by the other methods for small circuits. We suspect that this is caused by the greedy heuristic used to choose our pivot.

\subsubsection{Comparison of CNOT count against theoretical bounds}
Note that the standard \textit{transpile}-method of Qiskit~\cite{qiskit} on a complete connectivity performs better than the synthesis methods. 
This shows that none of these methods synthesize an optimal circuit. This motivated us to compare the empirical convergence behavior of the synthesis methods, with respect to the theoretical bounds for CNOT synthesis.
Since the method by \citet{Bravyi_2021} synthesizes layers of CNOTs and the other synthesis methods use a RowCol-like structure, these methods should either generate $\mathcal{O}(n^2)$~\cite{Wu_2023} CNOTs or $\mathcal{O}(\nicefrac{n^2}{\log (n)})$~\cite{patel2008optimal}. 

\autoref{tab:asymp_bounds} shows the median CNOT count of the circuits in our experiments after convergence\footnote{all circuits with originally $\geq 75$ gates for Quito, Nairobi $\geq 110$, Guadalupe $\geq 250$, Mumbai $\geq 500$, Ithaca $\geq 1250$ and Brisbane $\geq 3000$.} and how that relates to the two theoretical CNOT synthesis bounds: the naive $\mathcal{O}(n^2)$ and the optimal $\mathcal{O}(\nicefrac{n^2}{\log (n)})$. The table shows that all methods seem to follow the naive $\mathcal{O}(n^2)$ bound, but with a much lower scalar than what would be expected in the worst case (i.e. $1$ CNOT per qubit per row = $2n^2$ per CNOT block). However, the increasing scalar with respect to the optimal bound shows that the theoretical bounds are not saturated in practice. Thus, empirical studies of the performance of quantum circuit synthesis are needed to get a realistic representation of the algorithm's performance.

\begin{table}[h!]
\centering
\adjustbox{width=\textwidth}{
\begin{tabular}{ccc|ccc|ccc|ccc} \hline
     &            &  & \multicolumn{3}{c|}{Ours}                                    & \multicolumn{3}{c|}{\citet{Bravyi_2021}}                                  & \multicolumn{3}{c}{\citet{berg2021simple}}                                    \\
Qubits & $\text{Bound}_l$ ($\frac{n^2}{\log_2n}$) & $\text{Bound}_u$ ($n^2$) & Empirical & $\nicefrac{\text{Empirical}}{\text{Bound}_l}$ & $\nicefrac{\text{Empirical}}{\text{Bound}_u}$ & Empirical & $\nicefrac{\text{Empirical}}{\text{Bound}_l}$ & $\nicefrac{\text{Empirical}}{\text{Bound}_u}$ & Empirical & $\nicefrac{\text{Empirical}}{\text{Bound}_l}$ & $\nicefrac{\text{Empirical}}{\text{Bound}_u}$ \\ \hline
5      & 11                    & 25    & 15        & 1.36        & 0.60      & 17        & 1.55     & 0.68   & 20        & 1.82     & 0.80  \\
7      & 17                    & 49    & 31        & 1.82        & 0.63      & 31        & 1.82     & 0.63   & 39        & 2.29     & 0.80  \\
16     & 64                    & 256   & 167       & 2.61        & 0.65      & 153       & 2.39     & 0.6    & 197       & 3.08     & 0.77  \\
27     & 153                   & 729   & 488       & 3.19        & 0.67      & 413       & 2.70     & 0.57   & 557       & 3.64     & 0.76  \\
65     & 702                   & 4225  & 2901      & 4.13        & 0.69      & 2252      & 3.21     & 0.53   & 3194      & 4.55     & 0.76  \\
127    & 2308                  & 16129 & 11396     & 4.94        & 0.71      & 8551      & 3.70     & 0.53   & 12173     & 5.27     & 0.75  \\
        \hline        
\end{tabular}
}
\vspace{0.2em}
\caption{The theoretic asymptotic bounds of CNOT counts for each architecture size using $\mathcal{O}(\nicefrac{n^2}{\log_2 (n)})$~\cite{patel2008optimal} or naive $\mathcal{O}(n^2)$~\cite{Wu_2023}, the observed bounds and their fractions with respect to each bound for each synthesis algorithm in the all-to-all case.}\label{tab:asymp_bounds}
\end{table}

\subsubsection{Investigation of circuit routing overhead}
We further evaluate the number of CNOTs introduced when transpiling to adhere to the connectivity constraints. For this, we define a metric called the \textit{routing portion} that describes the percentage of the CNOT count that is added when targeting a specific device rather than assuming all-to-all connectivity: 
\begin{equation}
\text{Routing portion} = \frac{\#CX_{r} - \#CX_{fc}}{\#CX_{r}},
\label{eq:routing_overhead}
\end{equation}
with $\#CX_{fc}$ the number of CNOTs on the fully connected architecture and $\#CX_{r}$ the number of CNOTs after transpiling to a device-specific architecture. The routing portion of the baseline synthesis methods upon convergence can be found in \autoref{tab:reduction_count}. We note that since \citet{Bravyi_2021}, \citet{berg2021simple}, \citet{Duncan_2020}, and \citet{Maslov_2023} do not synthesize to a target architecture, the routing portion for these methods represents the ability of the Qiskit transpiler to route the circuits obtained with these methods.

\begin{table}[h!]
\centering
\begin{tabular}{l | p{4em}p{4em}p{4em}p{4em}p{4em}p{4em}}
\hline
\backslashbox{Algorithm}{Architecture} & Quito 5~qubits & Nairobi 7~qubits & Guadalupe 16~qubits & Mumbai 27~qubits & Ithaca 65~qubits & Brisbane 127~qubits\\
\hline
\citet{Bravyi_2021}     & 46.03\% & 59.91\%   & 77.87\%     & 84.07\%  & 90.90\%  & 92.83\%    \\
\citet{berg2021simple}  & 46.77\% & 63.05\%   & 79.97\%     & 84.90\%  & 88.34\%  & 89.52\%    \\
\citet{Duncan_2020}  & 50.72\% & 64.35\%   & 82.41\%     & 86.53\%  & 91.96\%  & 94.27\%    \\
\citet{Maslov_2023}  & 32.23\% & 39.19\%   & 28.50\%     & 33.92\%  & 33.09\%  & 48.38\%    \\
proposed                     & 20.04\% & 25.57\%   & 39.02\%     & 41.82\%  & 42.93\%  & 41.49\%    \\
\hline
\end{tabular}
\vspace{0.2em}
\caption{Routing portion (\autoref{eq:routing_overhead}) for all circuits with originally $\geq 75$ gates for Quito, Nairobi $\geq 110$, Guadalupe $\geq 250$, Mumbai $\geq 500$, Ithaca $\geq 1250$ and Brisbane $\geq 3000$.}\label{tab:reduction_count}
\end{table}

We observe that the routing portion of our proposed algorithm is less than half the routing portion when synthesizing with a previous method and then routing it. Additionally, the routing portion of all methods tends to increase as the number of qubits in the architecture increases. We suspect that this is related to the distance between the qubits in the architecture which also increases as the number of qubits increases. 

\FloatBarrier
\subsection{Evaluation of Two-qubit Gate Depth}
We additionally evaluate the two-qubit gate depth, defined as the number of incident two-qubit gates along the longest path through the circuit when traversing only wires or two-qubit operations. 
This metric applies to near-term hardware devices where multi-qubit operations can be performed in parallel if they are applied to disjoint sets of qubits. For completeness, the depth of the full circuit, regardless of gate type, for the various algorithms used in our experiments, are shown in \autoref{apndx:full_depth}.

\begin{figure}[bt]
	\centering
    \begin{subfigure}{\textwidth}
    \centering
    \includegraphics[width=0.8\textwidth]{img_new_methods/legend_experiments.pdf}
    \end{subfigure}
    \vspace{-1em}
    \begin{subfigure}{\textwidth}
        \begin{subfigure}{0.47\textwidth}
    	    \centering
    		\includegraphics[width=\textwidth]{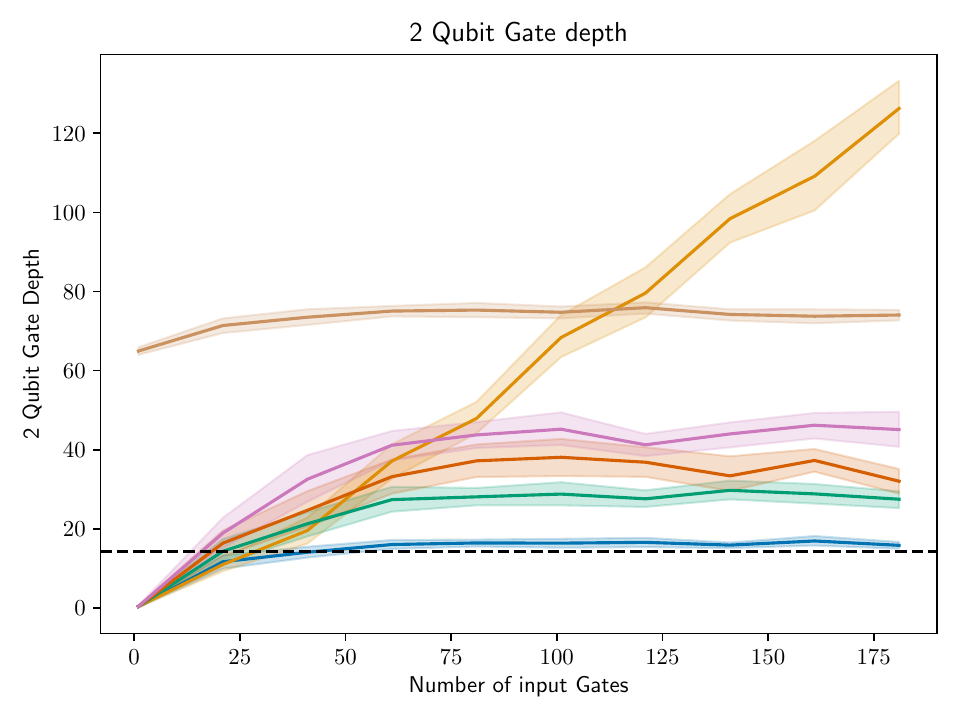}
    		\caption{Quito (5 Qubits)}
    	\end{subfigure}
    \hfill
        \begin{subfigure}{0.47\textwidth}
    	    \centering
            \includegraphics[width=\textwidth]{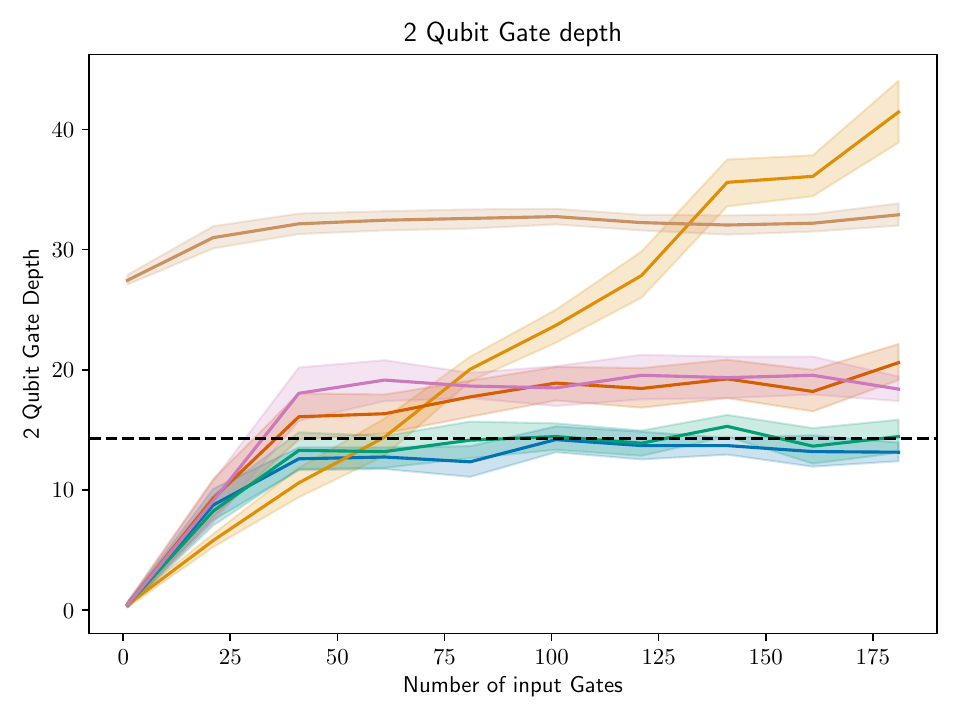}
    		\caption{Complete (5 Qubits)}
    	\end{subfigure}
    \hfill
     	\begin{subfigure}{0.47\textwidth}
    	    \centering
    		\includegraphics[width=\textwidth]{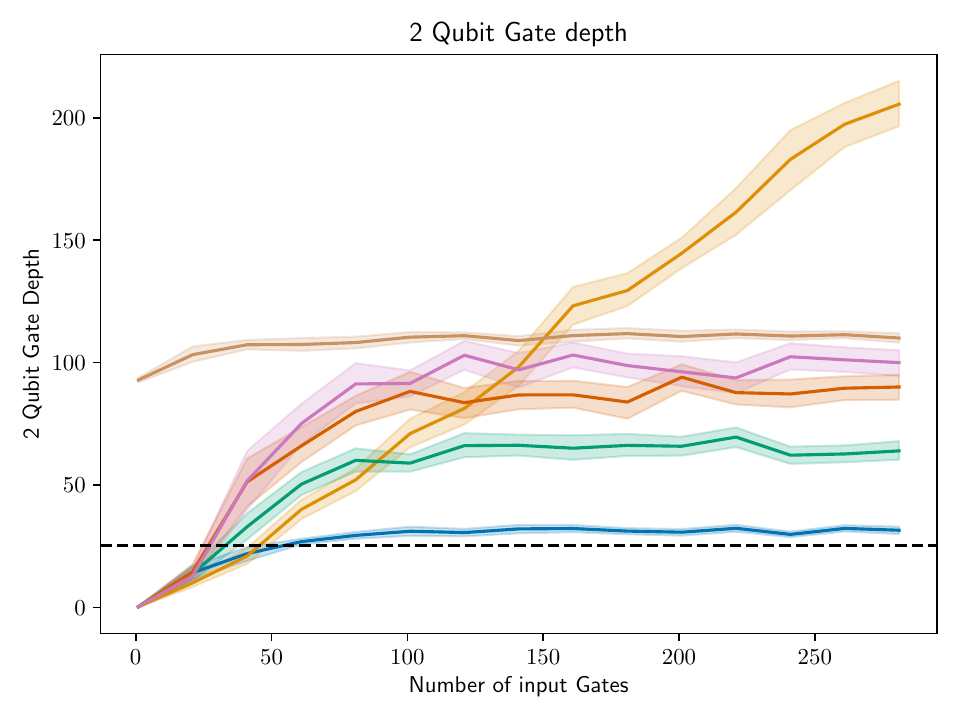}
    		\caption{Nairobi (7 Qubits)}
    	\end{subfigure}
     \hfill
     	\begin{subfigure}{0.47\textwidth}
    	    \centering
    		\includegraphics[width=\textwidth]{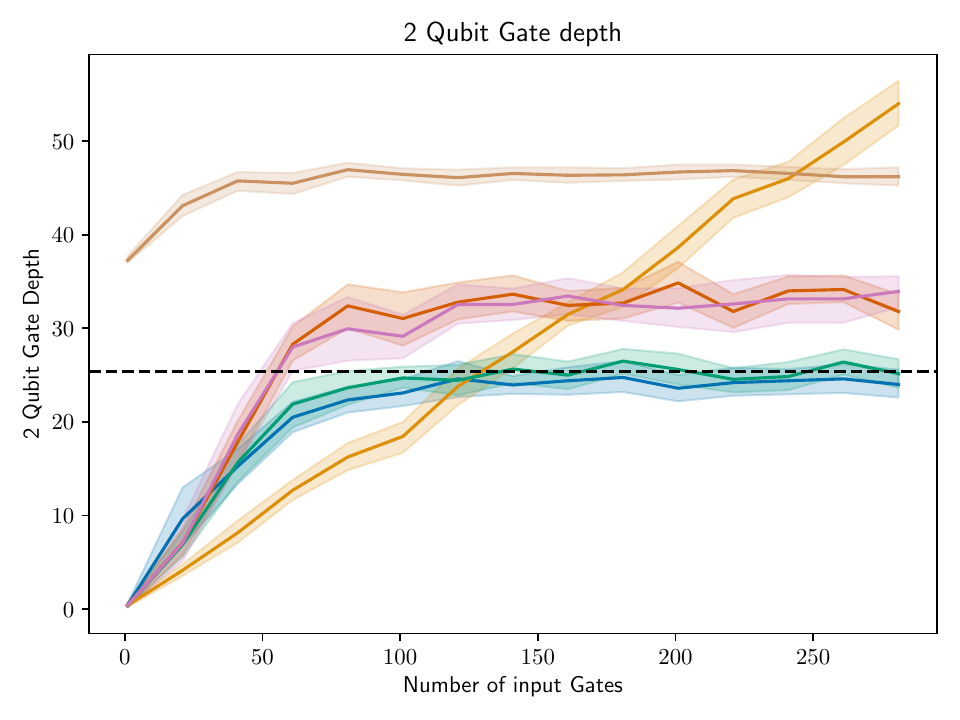}
    		\caption{Complete (7 Qubits)}
    	\end{subfigure}
     \hfill
    	\begin{subfigure}{0.47\textwidth}
    	    \centering
    		\includegraphics[width=\textwidth]{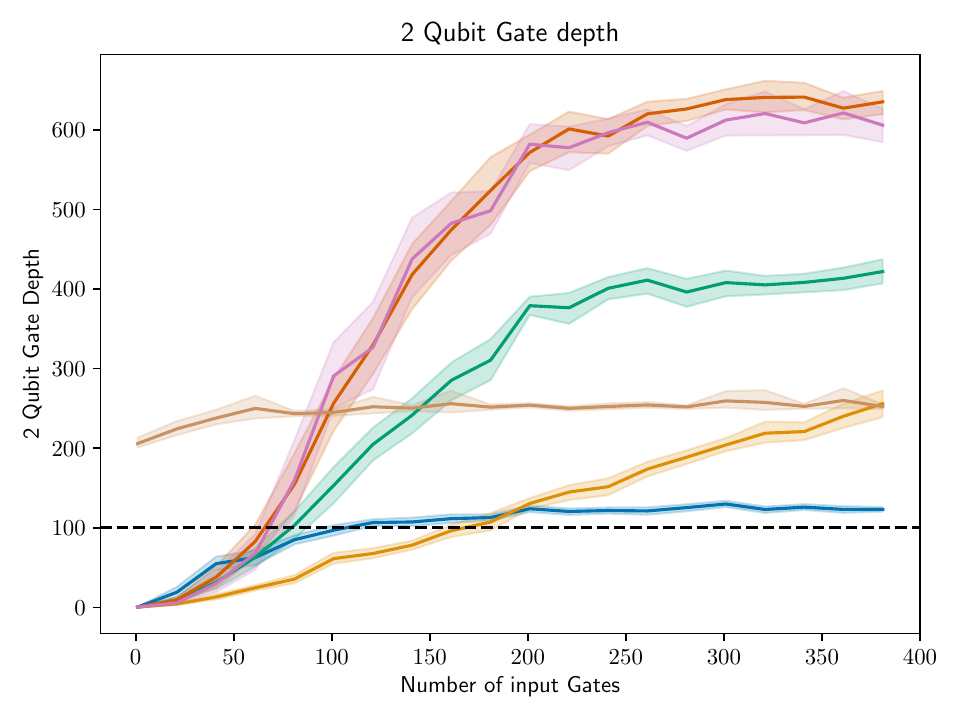}
    		\caption{Guadalupe (16 Qubits)}
    	\end{subfigure}
     \hfill
        \begin{subfigure}{0.47\textwidth}
    	    \centering
    		\includegraphics[width=\textwidth]{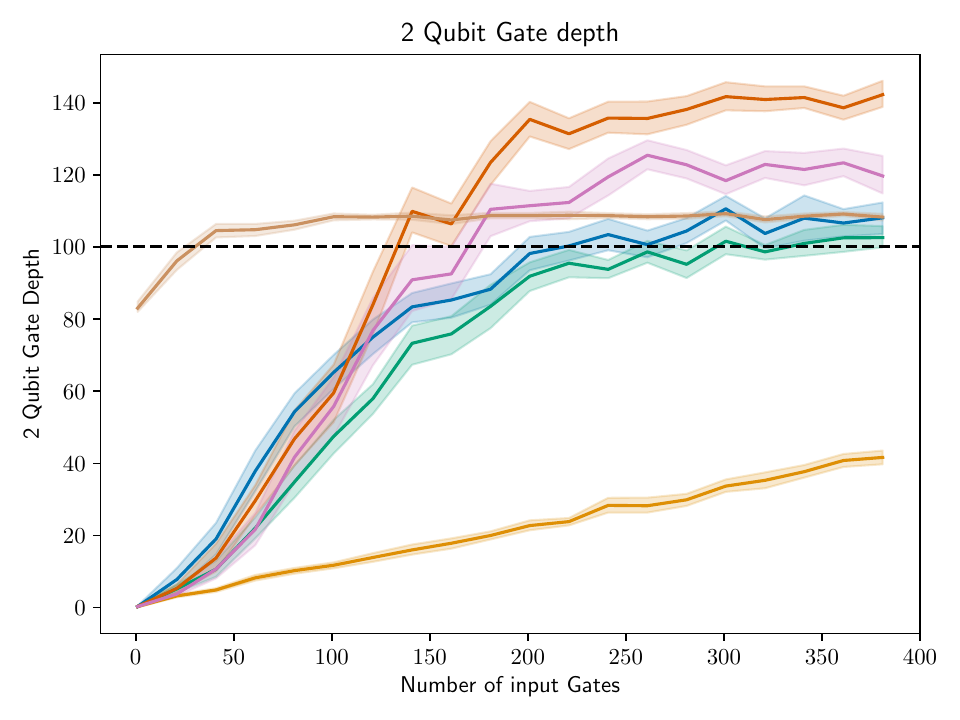}
    		\caption{Complete (16 Qubits)}
    	\end{subfigure}
    \end{subfigure}
	\caption{CNOT depth for smaller architectures up to 16 qubits. Specifically, the backends Guadalupe, Nairobi, and Quito were used. We additionally reported the CNOT count of the complete architectures of the same size.}\label{fig:2q_depth_comparison_smaller_arch}
    \vspace{3em}
\end{figure}

\begin{figure}[bt]
    \centering
    \begin{subfigure}{\textwidth}
    \centering
    \includegraphics[width=0.8\textwidth]{img_new_methods/legend_experiments.pdf}
    \end{subfigure}
    \vspace{-1em}
	\begin{subfigure}{\textwidth}
        \begin{subfigure}{0.47\textwidth}
    	    \centering
    		\includegraphics[width=\textwidth]{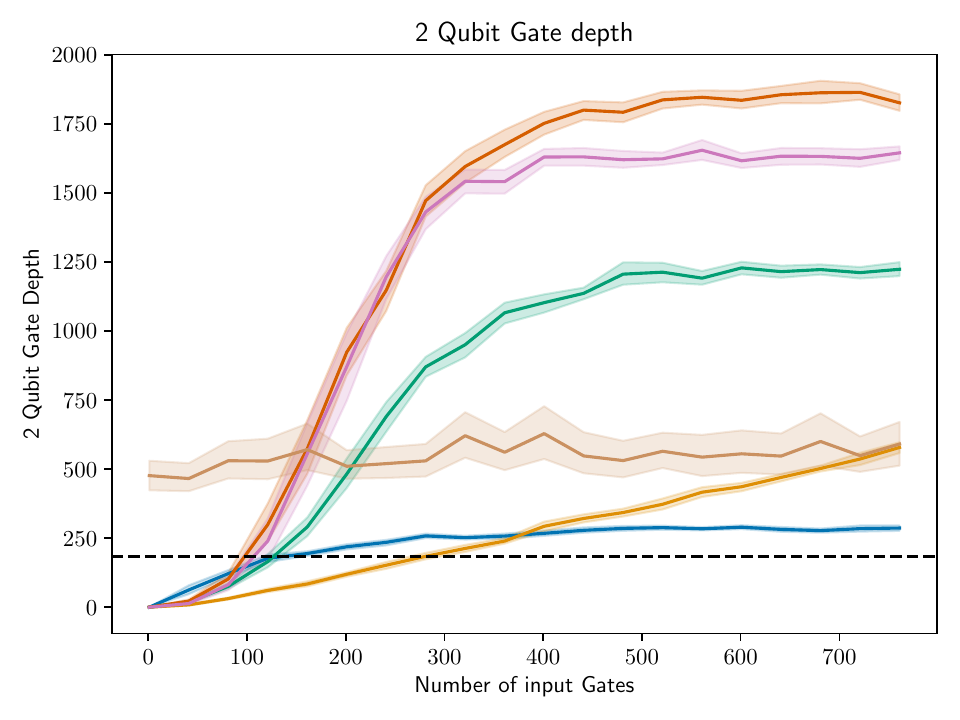}
    		\caption{Mumbai (27 Qubits)}
    	\end{subfigure}
     \hfill
        \begin{subfigure}{0.47\textwidth}
    	    \centering
            \includegraphics[width=\textwidth]{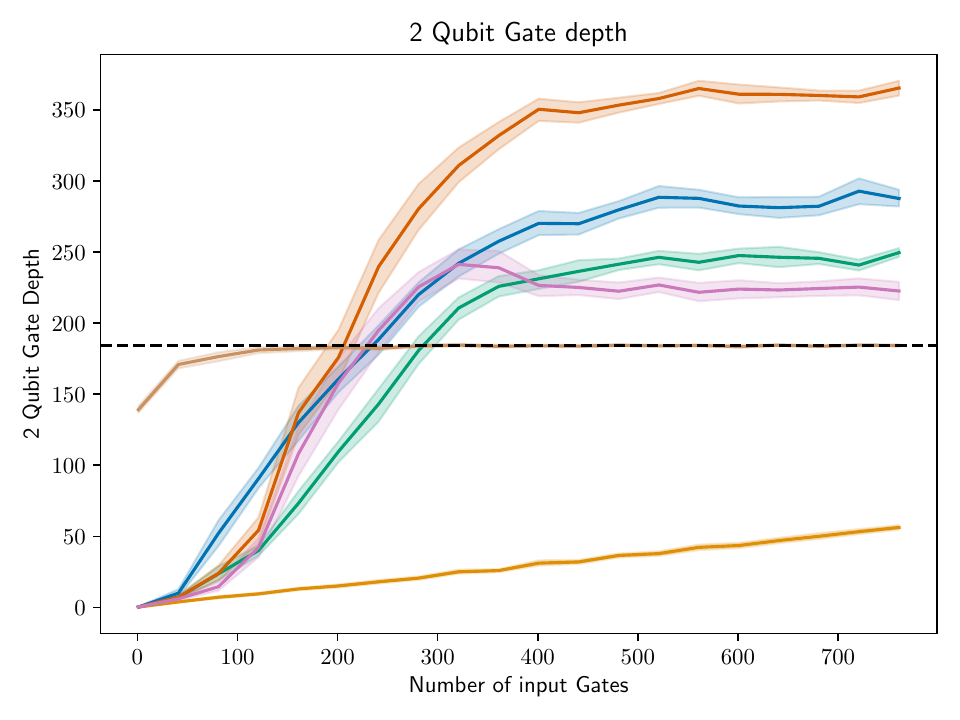}
    		\caption{Complete (27 Qubits)}
    	\end{subfigure}
        \begin{subfigure}{0.47\textwidth}
    	    \centering
    		\includegraphics[width=\textwidth]{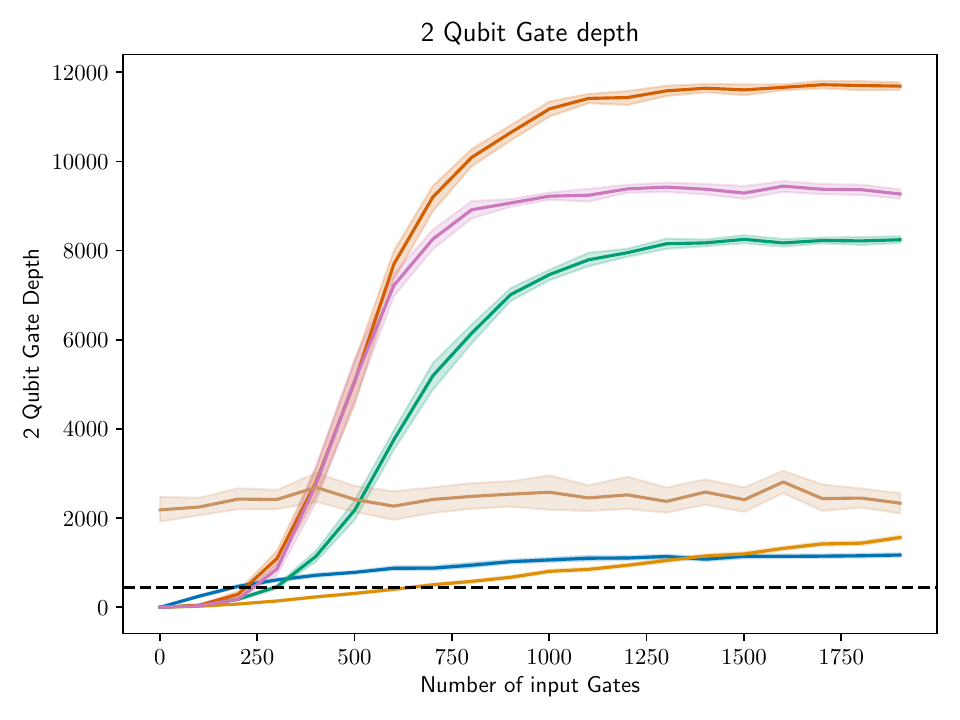}
    		\caption{Ithaca (65 Qubits)}
    	\end{subfigure}
     \hfill
        \begin{subfigure}{0.47\textwidth}
    	    \centering
            \includegraphics[width=\textwidth]{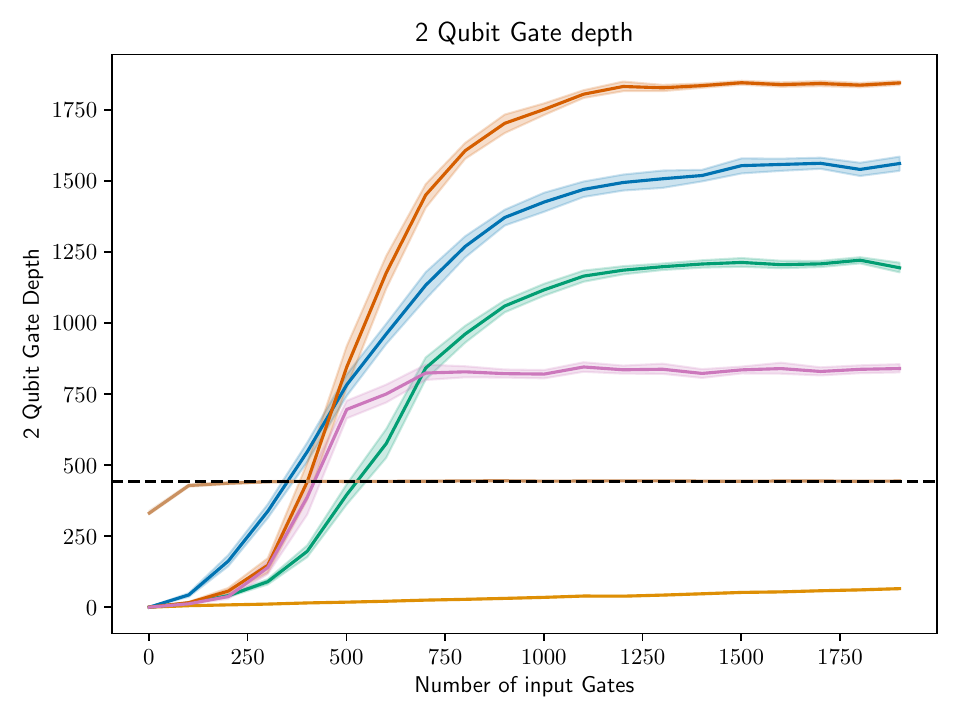}
    		\caption{Complete (65 Qubits)}
    	\end{subfigure}
        \begin{subfigure}{0.47\textwidth}
    	    \centering
    		\includegraphics[width=\textwidth]{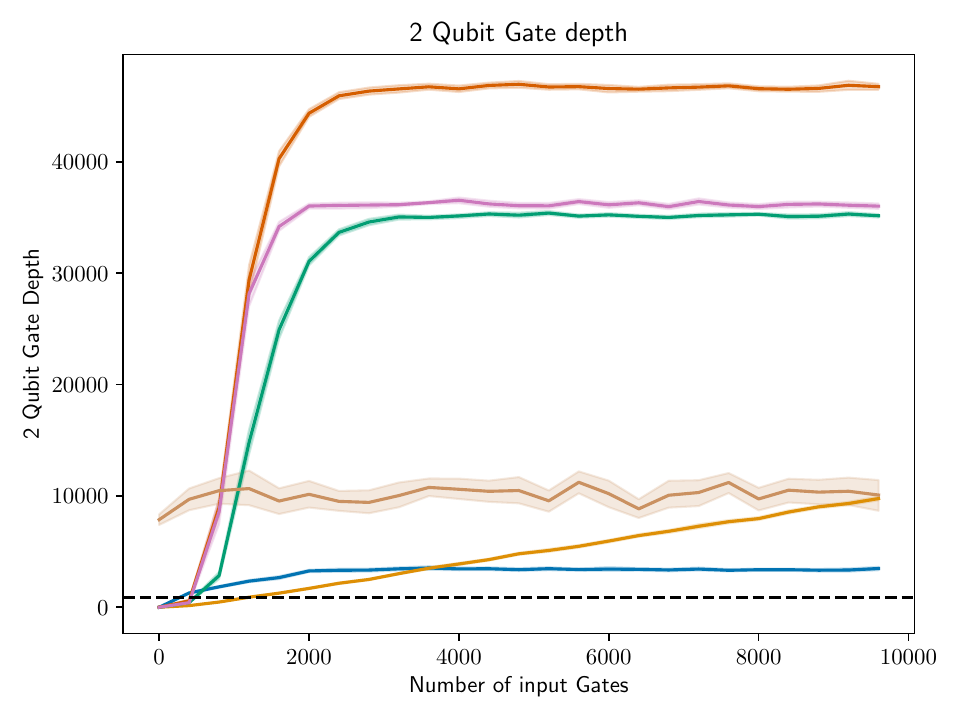}
    		\caption{Brisbane (127 Qubits)}
    	\end{subfigure}
        \hfill
        \begin{subfigure}{0.47\textwidth}
    	    \centering
            \includegraphics[width=\textwidth]{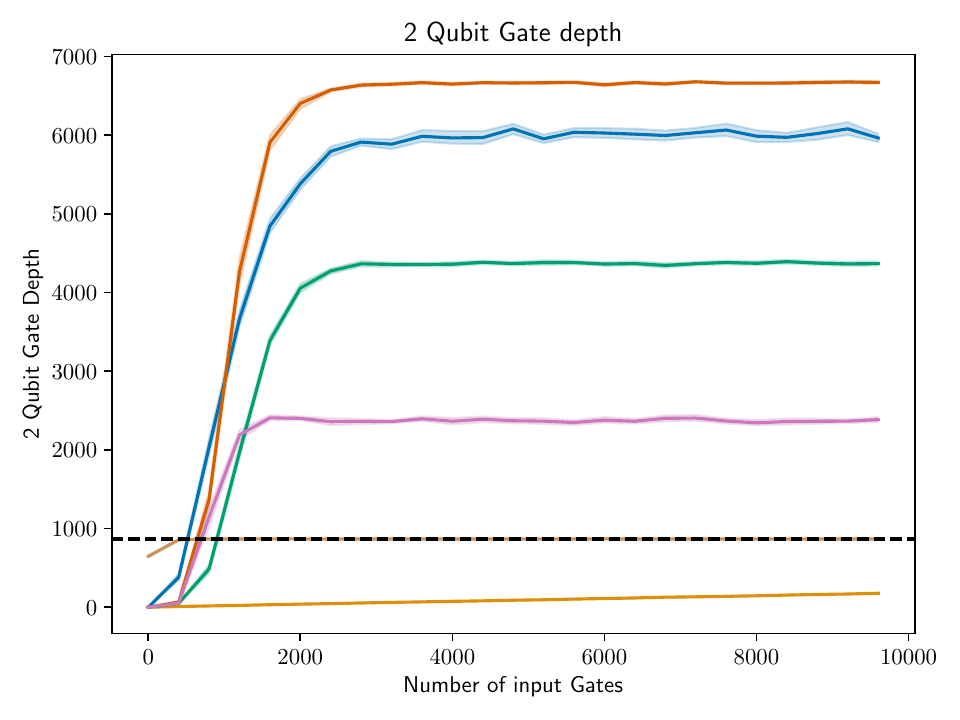}
    		\caption{Complete (127 Qubits)}
    	\end{subfigure}
    \end{subfigure}
    \caption{CNOT depth for larger architectures up to 127 qubits. Specifically, the backends Brisbane, Ithaca, and Mumbai were used. We additionally reported the CNOT count of the complete architectures of the same size.}
    \label{fig:2q_depth_comparison_larger_arch}
    \vspace{3em}
\end{figure}

As seen in \autoref{fig:2q_depth_comparison_smaller_arch} and \autoref{fig:2q_depth_comparison_larger_arch}, we once again observe comparable behavior between the various algorithms. The horizontal dashed black line in each graph indicates the asymptotic behavior of the best-performing algorithm on complete architectures, namely \citet{Bravyi_2021} for 16 qubits or less and \citet{Maslov_2023} for 27 qubits or more. 

In order to determine the effect of routing on circuit depth, we modify the previously defined routing portion to consider two-qubit gate depth instead of gate count:
\begin{equation}
\text{Routing depth portion} = \frac{d_{r} - d_{fc}}{d_{r}},
\label{eq:routing_depth_overhead}
\end{equation}
where $d_c$ is the two-qubit gate depth on a complete architecture and $d_r$ is the depth after routing. Surprisingly, we see that synthesizing to a restricted architecture actually improves the performance of the proposed algorithm with respect to two-qubit depth. We hypothesize that this is due to the greedy pivot selection of our proposed method that may repeatedly choose to run over the same qubit multiple times during synthesis. 

\begin{figure}
    \centering
    \begin{subfigure}{\linewidth}
    \centering
        \includegraphics[width=0.45\textwidth]{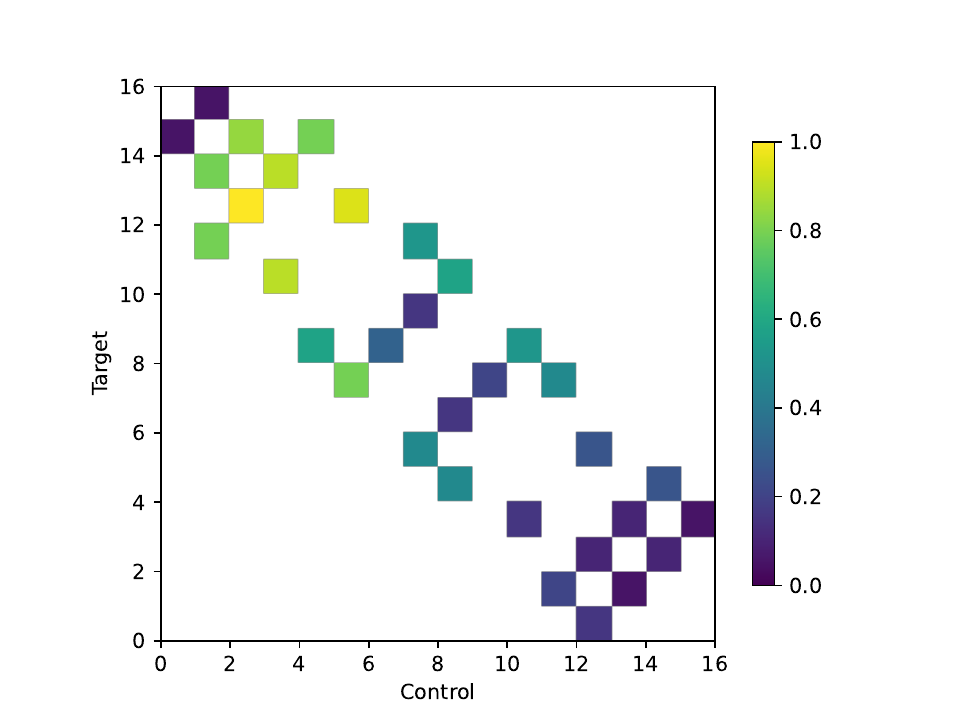}
        \includegraphics[width=0.45\textwidth]{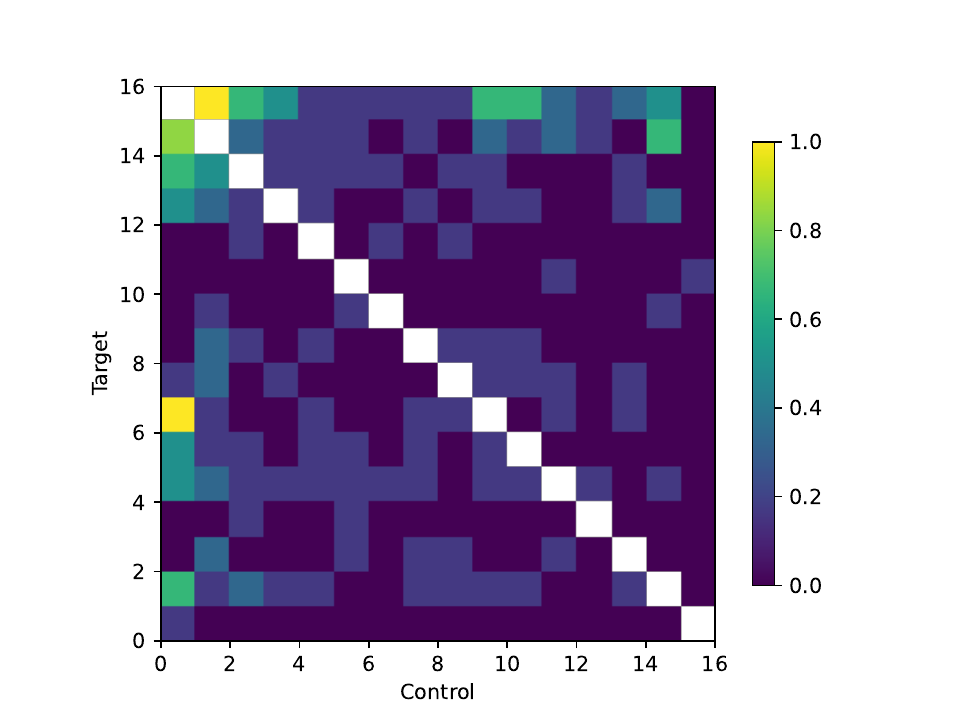}
        \caption{Guadalupe and complete 16}
    \end{subfigure}

    \begin{subfigure}{\linewidth}
    \centering
        \includegraphics[width=0.45\textwidth]{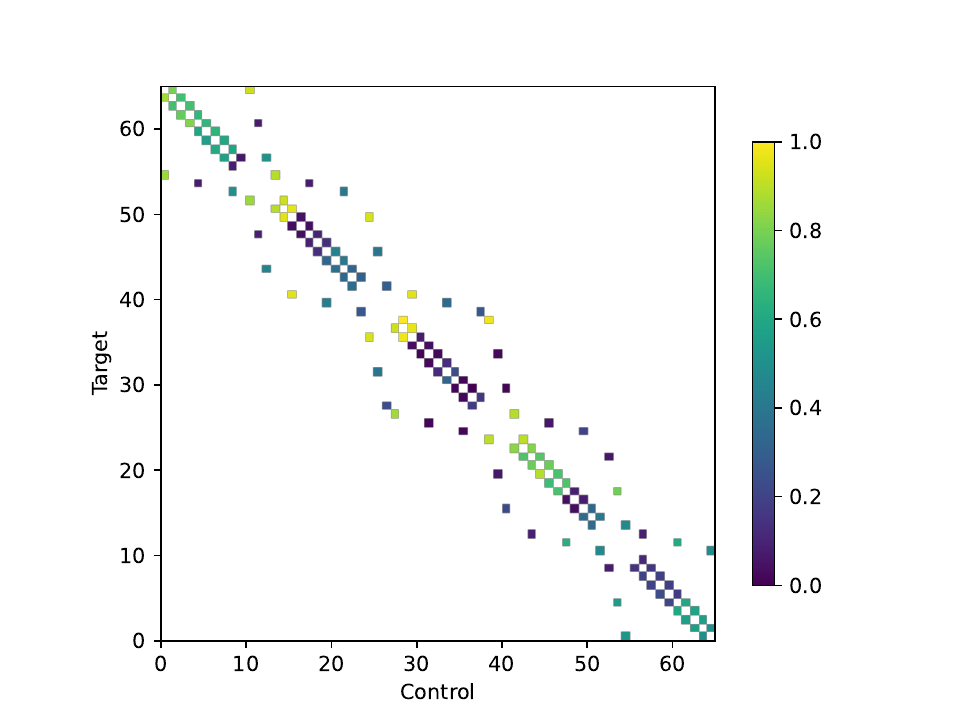}
        \includegraphics[width=0.45\textwidth]{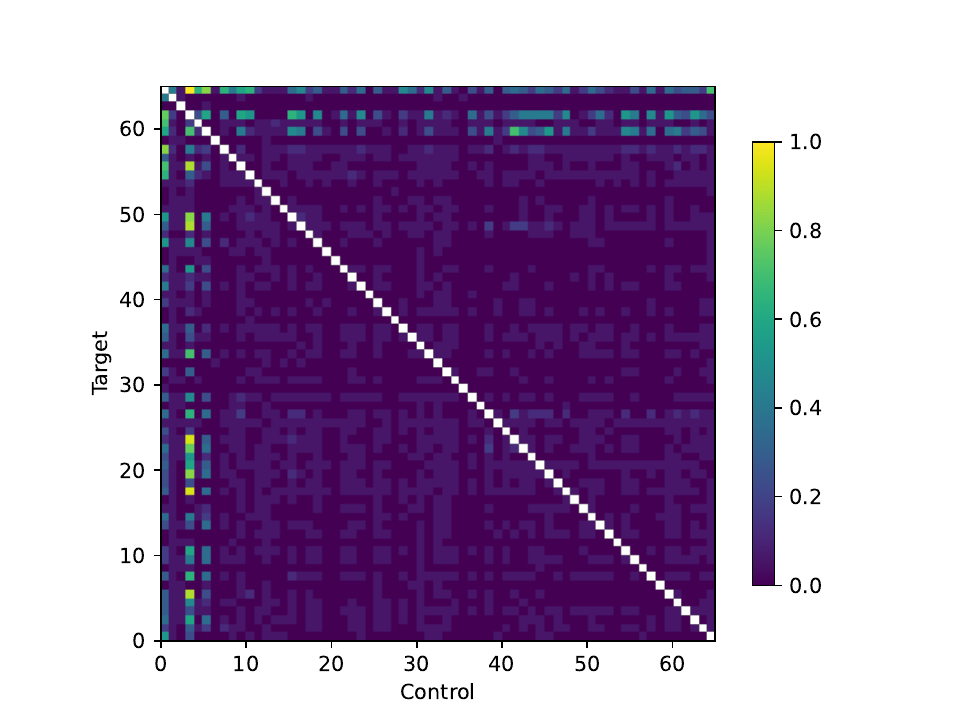}
        \caption{Ithaca and complete 65}
    \end{subfigure}
    
    \caption{A heatmap of the two-qubit gate placement for Guadalupe and Ithaca combined with their complete counterpart next to it using the proposed algorithm.}
    \label{fig:heatmaps}
\end{figure}

We investigated this hypothesis by counting the density of two-qubit gates on edges in the connectivity graph. A heatmap of the two-qubit gate placement for Guadalupe and Ithaca can be seen in \autoref{fig:heatmaps}. The figure shows that while for the constrained case, the CNOTs are placed more uniformly, computing the Steiner tree for the complete case will provide the tendency of entangling the first qubits with all others, which outlines a star-like shape that increases the depth of the overall circuit. Hence, the depth for larger architectures, given our method, will be increased on a fully-connected architecture which explains why the routing portion for the depth is negative.
\begin{table}[h!]
\centering
\begin{tabular}{l | p{4em}p{4em}p{4em}p{4em}p{4em}p{4em}}
\hline
\backslashbox{Algorithm}{Architecture} & Quito 5~qubits & Nairobi 7~qubits & Guadalupe 16~qubits & Mumbai 27~qubits & Ithaca 65~qubits & Brisbane 127~qubits\\
\hline
\citet{Bravyi_2021}     & 48.85\% & 60.59\%   & 75.65\%     & 79.85\%  & 85.35\%  & 87.58\%    \\
\citet{berg2021simple}  & 46.01\% & 61.12\%   & 77.63\%     & 80.45\%  & 84.30\%  & 85.73\%    \\
\citet{Duncan_2020}  & 55.44\% & 66.55\%   & 79.93\%     & 86.16\%  & 91.05\%  & 93.45\%    \\
\citet{Maslov_2023}  & 56.03\% & 57.73\%   & 57.36\%     & 66.39\%  & 82.40\%  & 91.56\%    \\
proposed                   & 20.50\% & 22.25\%   & 15.30\%     & 1.43\%  & -37.73\%  & -75.11\%    \\
\hline
\end{tabular}
\vspace{0.2em}
\caption{Routing depth portion (\autoref{eq:routing_depth_overhead}) for all circuits with originally $\geq 75$ gates for Quito, Nairobi $\geq 110$, Guadalupe $\geq 250$, Mumbai $\geq 500$, Ithaca $\geq 1250$ and Brisbane $\geq 3000$.}\label{tab:depth_reduction_count}
\end{table}

\FloatBarrier
\subsection{Evaluation against a depth optimal compiler}
\citet{peham2023depthoptimal} have used an SAT-Solver to provide a depth optimal compilation algorithm for specifically Clifford tableaus. In this work, we utilized this algorithm implemented in the Munich quantum toolkit to provide further insights into the limitations of our approach. We compare the methods of \citet{Bravyi_2021_simp}, \citet{Maslov_2023}, and our proposed method against randomly sampled Clifford tableaus. This is because \citet{Maslov_2023} provides the best-known asymptotic upper bound at the time of writing in terms of circuit depth for linear nearest neighbour (LNN) architectures. However, none of the IBM devices used in our experiments have a native LNN structure. As we have seen in \autoref{fig:2q_depth_comparison_smaller_arch} and \autoref{fig:2q_depth_comparison_larger_arch} for small architectures the LNN assumption still creates a large routing overhead and the depth-optimal algorithm from \citet{Maslov_2023} is outperformed by the unconstrained algorithms for small architectures. On the flip-side, the optimal solver from \citet{peham2023depthoptimal} is unable to find the depth-optimal circuit within reasonable time for architectures that are large enough for \citet{Maslov_2023} to be the best baseline algorithm. Thus, we also include \citet{Bravyi_2021} in our comparisons, since their method performed best in the CNOT-depth experiments.

Our results can be found in \autoref{tab:sec4:depth_optimal_reduction_count}. We can see that for smaller qubits, our method improves depth-optimal circuits with transpilation when targeting constrained connectivities. This is because the additional SWAP gates needed to route the circuits will worsen the results of optimally compiling a Clifford tableau. This can be observed among all the various architectures provided in this evaluation. 

Note that we have rebased the output circuit the gate-set to $\{H, S, CX\}$-Gates for comparability, which is why the depth is higher compared to the average depth proposed in \citet{peham2023depthoptimal}, e.g., $5.70$ for three qubits. On a test trial with $20$ samples for a complete architecture with three qubits, we obtained an average depth of $5.5 \pm 0.5$ without rebasing the gates. 

\begin{table}[h!]
\centering
\begin{tabular}{l | p{6em}p{6em}p{6em}|p{8em}}
\multicolumn{5}{c}{Depth} \\
\hline
Architecture                            & \multicolumn{3}{ c |}{Line / Complete }             & Quito / Complete \\
\hline
\backslashbox{Algorithm}{No. Qubits}    & 3             & 4               & 5                 & 5 \\
\hline
\citet{peham2023depthoptimal}           & 16.00 / \textbf{9.70}& 27.00 / \textbf{12.25}& 43.05 / \textbf{15.05}& \textbf{29.00} / \textbf{15.05} \\
\citet{Bravyi_2021}                     & 21.85 / 17.20& 43.50 / 20.45& 65.50 / 25.80& 46.25 / 25.80 \\
\citet{Maslov_2023}                     & 33.95 / 34.95& 46.20 / 48.30& 58.05 / 58.55& 103.75 / 58.55 \\
proposed                                & \textbf{15.55} / 14.55& \textbf{23.00} / 22.90& \textbf{29.50} / 28.40& 30.35 / 28.40 \\
\hline
\multicolumn{5}{c}{} \\
\multicolumn{5}{c}{CNOT Count} \\
\hline
Architecture & \multicolumn{3}{ c |}{Line / Complete } & Quito / Complete \\
\hline
\backslashbox{Algorithm}{No. Qubits} & 3 & 4 & 5 & 5 \\
\hline
\citet{peham2023depthoptimal}   & 5.90 / 4.15& 15.15 / \textbf{7.55}& 29.00 / \textbf{11.65}& 24.95 / \textbf{11.65} \\
\citet{Bravyi_2021}             & 5.50 / \textbf{3.40}& 20.20 / 10.95& 34.60 / 16.20& 31.65 / 16.20 \\
\citet{Maslov_2023}             & 18.40 / 18.60& 35.95 / 36.70& 59.75 / 60.25& 87.85 / 60.25 \\
proposed                        & \textbf{4.95} / 4.85& \textbf{10.75} / 9.70& \textbf{18.35} / 15.40& \textbf{18.40} / 15.40 \\
\hline
\end{tabular}
\vspace{0.2em}
\caption{Average depth and CNOT count for randomly sampled Clifford tableaus over a Line 3, 4, and 5 architecture in combination with its complete counterpart. As well as for the Quito device}\label{tab:sec4:depth_optimal_reduction_count}
\end{table}

\FloatBarrier
\subsection{Evaluation on actual hardware}
\begin{figure}[h! bt]
\vspace{-0.5em}
	\centering
	\includegraphics[width=0.45\textwidth]{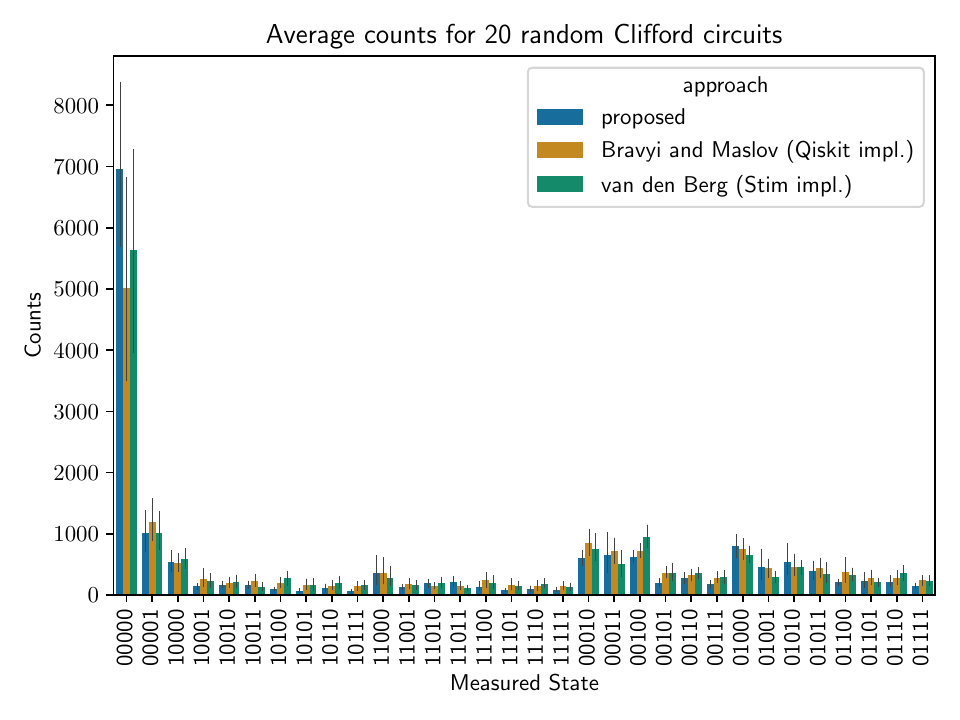}
	\caption{Average Counts of our simulation of $CC^\dagger$ for 20 random Clifford circuits on Quito (5 qubits) and Nairobi (7 qubits).}\label{fig:average_count_clifford}
 \vspace{-1em}
\end{figure}

To the impact of reducing the CNOT count on the final fidelity of our circuit, we executed the random Clifford circuits on real hardware. Because each random Clifford circuit creates a different quantum state, it is difficult to aggregate their fidelity into a single picture. Therefore, we executed each Clifford circuit $C$ followed by their inverse $C^\dagger$ on IBM hardware, which should precisely return input state: $\ket{0\dots0}$, which we can measure the fidelity on. 

We generated 20 random circuits $C$ and executed $CC^\dagger$ on the quantum device with $16.000$ shots. We used the devices \textit{quito} and \textit{nairobi} to verify our results since those were the only devices available to us at that time. For this experiment, we generated $25$ random gates per circuit for \textit{quito} and $30$ random gates for \textit{nairobi}, based on the maximum circuit size that each device can execute without decohering or accruing too much error. 

We re-synthesized each circuit with the method of \citet{Bravyi_2021}, \citet{berg2021simple}, and our approach. Then, we transpiled the circuit to target architecture for the two baseline methods before finally executing the resulting circuit $C$ on the quantum device as $CC^\dagger$.

We report the averaged shots in \autoref{fig:average_count_clifford} to outline the results. On average, we can observe that our method performs best regarding the final distribution of shots. 
This is also visible when computing the average Hellinger fidelity of our experiments in \autoref{tab:quito}. Particularly, we show that our method obtains the highest Hellinger fidelity most of the time ($13$ circuit for Quito and $14$ circuits for Nairobi). However, we note that due to the decoherence times and the gate fidelities of the Nairobi device, the observed Hellinger fidelities are close to zero for most circuits in all methods.
Specifically experiment $12$ on Nairobi is interesting. Here both methods \citet{berg2021simple} and \citet{Bravyi_2021} showed a much better fidelity compared to our method. Upon evaluating the resulting circuits, we found that both \citet{berg2021simple} and \citet{Bravyi_2021} provided an overhead of $8$ CNOTs, and our method an overhead of $30$ CNOTs\footnote{We counted the CNOTs of the complete circuit describing $CC^\dagger$.}. We suspect that this is a special case where our heuristics choose the wrong pivot qubits or qubit mapping resulting in a lower fidelity caused by the extra CNOTs.

Lastly, we compared the on-device execution times our circuit experiments required as reported by IBM for each device. Noting that the overall coherence times are measured in the hundreds of $\mu s$, a difference in the order of seconds is notable between experiments. Overall, we observed that our algorithm slightly improves the execution time compared to the methods of \citet{Bravyi_2021} and \citet{berg2021simple}. This is more noticeable for the Nairobi device than the Quito device. We suspect that the observed difference is due to the smaller number of CNOTs in circuits compiled using our methodology.

\begin{table}[h!bt]
    \centering
\adjustbox{max width=\textwidth}{
\begin{tabular}{l|lll|lll}
\hline
& \multicolumn{3}{c}{$F(Q, P)$} & \multicolumn{3}{c}{execution-time (seconds)} \\
Quito &  proposed  & \begin{tabular}{l} \citet{Bravyi_2021} \\ (Qiskit impl.) \end{tabular} & \begin{tabular}{l} \citet{berg2021simple} \\ (Stim impl.) \end{tabular}  & proposed  & \begin{tabular}{l} \citet{Bravyi_2021} \\ (Qiskit impl.) \end{tabular} & \begin{tabular}{l} \citet{berg2021simple} \\ (Stim impl.) \end{tabular}\\
\hline
0          & \textbf{0.1976} & 0.1813          & 0.1508          & 5.4142          & \textbf{5.3069} & 5.4108          \\
1          & \textbf{0.3421} & 0.0870          & 0.0901          & \textbf{5.2946} & 5.5140          & 5.4343          \\
2          & \textbf{0.2848} & 0.1199          & 0.1070          & \textbf{5.3327} & 5.3977          & 5.5118          \\
3          & \textbf{0.4168} & 0.0758          & 0.1129          & \textbf{5.3183} & 5.4965          & 5.3873          \\
4          & \textbf{0.5678} & 0.4650          & 0.4349          & 5.3805          & \textbf{5.3481} & 5.3808          \\
5          & 0.2486          & \textbf{0.3476} & 0.3259          & 5.3758          & 5.4728          & \textbf{5.2968} \\
6          & 0.3704          & 0.1124          & \textbf{0.4283} & 5.3785          & \textbf{5.3341} & 5.3498          \\
7          & 0.5652          & 0.4539          & \textbf{0.7320} & 5.2031          & 5.4315          & \textbf{5.1910} \\
8          & \textbf{0.8101} & 0.7481          & 0.7466          & \textbf{5.1780} & 5.2014          & 5.3413          \\
9          & \textbf{0.8038} & 0.7333          & 0.7303          & \textbf{5.1142} & 5.1569          & 5.3420          \\
10         & \textbf{0.4197} & 0.3029          & 0.2579          & \textbf{5.2879} & 5.4189          & 5.5622          \\
11         & 0.4072          & \textbf{0.4546} & 0.4318          & 5.3605          & \textbf{5.3541} & 5.4032          \\
12         & \textbf{0.5393} & 0.1974          & 0.3840          & 5.3731          & 5.3787          & \textbf{5.3239} \\
13         & \textbf{0.3275} & 0.0683          & 0.0853          & \textbf{5.4525} & 5.5729          & 5.4827          \\
14         & \textbf{0.7494} & 0.3492          & 0.3321          & 5.2307          & \textbf{5.2301} & 5.4411          \\
15         & 0.1579          & 0.1183          & \textbf{0.2268} & 5.4612          & \textbf{5.3384} & 5.4455          \\
16         & \textbf{0.2581} & 0.1401          & 0.0579          & \textbf{5.3082} & 5.4267          & 5.5682          \\
17         & 0.4746          & 0.7039          & \textbf{0.7208} & 5.2876          & \textbf{5.2581} & 5.3087          \\
18         & 0.5395          & 0.5657          & \textbf{0.6166} & 5.3176          & \textbf{5.2079} & 5.2365          \\
19         & \textbf{0.2246} & 0.0534          & 0.0675          & \textbf{5.4311} & 5.4546          & 5.5747          \\
            \hline \hline
           & 0.4353          & 0.3139          & 0.3520          & 5.3250          & 5.3650          & 5.3996           \\
           \hline
           \multicolumn{7}{ c }{}\\
\hline
& \multicolumn{3}{c}{$F(Q, P)$} & \multicolumn{3}{c}{execution-time (seconds)} \\
Nairobi &  proposed  & \begin{tabular}{l} \citet{Bravyi_2021} \\ (Qiskit impl.) \end{tabular} & \begin{tabular}{l} \citet{berg2021simple} \\ (Stim impl.) \end{tabular}  & proposed  & \begin{tabular}{l} \citet{Bravyi_2021} \\ (Qiskit impl.) \end{tabular} & \begin{tabular}{l} \citet{berg2021simple} \\ (Stim impl.) \end{tabular}\\
\hline
0       & \textbf{0.1040}  & 0.0354             & 0.0298           & \textbf{6.9084}      & 7.2458                 & 7.0425               \\
1       & 0.0864           & 0.1010             & \textbf{0.1164}  & \textbf{6.7632}      & 7.0183                 & 8.9713               \\
2       & \textbf{0.1666}  & 0.0766             & 0.0159           & 6.7101               & \textbf{6.6981}        & 8.7126               \\
3       & 0.0409           & \textbf{0.0626}    & 0.0118           & \textbf{7.1935}      & 7.2377                 & 9.1640               \\
4       & \textbf{0.0816}  & 0.0162             & 0.0119           & 6.9697               & \textbf{6.9378}        & 7.0088               \\
5       & \textbf{0.4677}  & 0.1309             & 0.4353           & \textbf{6.7681}      & 6.9456                 & 8.5064               \\
6       & \textbf{0.0667}  & 0.0311             & 0.0232           & \textbf{6.9118}      & 7.1538                 & 8.3971               \\
7       & \textbf{0.2140}  & 0.0184             & 0.0124           & \textbf{6.6833}      & 6.9575                 & 8.9666               \\
8       & \textbf{0.1384}  & 0.0195             & 0.0409           & \textbf{6.8453}      & 6.9347                 & 8.3863               \\
9       & \textbf{0.0265}  & 0.0206             & 0.0260           & 7.2114               & \textbf{6.9228}        & 8.6713               \\
10      & 0.0681           & \textbf{0.1176}    & 0.0496           & \textbf{6.8542}      & 6.9490                 & 8.9581               \\
11      & \textbf{0.0551}  & 0.0192             & 0.0117           & \textbf{6.8284}      & 10.4572                & 8.5234               \\
12      & 0.1163           & 0.7146             & \textbf{0.7220}  & \textbf{6.7571}      & 6.9508                 & 8.2320               \\
13      & \textbf{0.1134}  & 0.0408             & 0.0779           & \textbf{6.9502}      & 7.1364                 & 12.8277              \\
14      & \textbf{0.0775}  & 0.0172             & 0.0078           & \textbf{6.5106}      & 6.9187                 & 8.6433               \\
15      & \textbf{0.0389}  & 0.0337             & 0.0273           & 7.4033               & \textbf{7.0621}        & 7.1139               \\
16      & 0.0381           & \textbf{0.0566}    & 0.0408           & \textbf{6.7696}      & 7.0401                 & 8.3049               \\
17      & \textbf{0.0281}  & 0.0256             & 0.0116           & 7.0195               & \textbf{6.9862}        & 7.0484               \\
18      & \textbf{0.0909}  & 0.0129             & 0.0143           & \textbf{6.9081}      & 6.9627                 & 8.5572               \\
19      & 0.0376           & 0.0522             & \textbf{0.0886}  & \textbf{6.8788}      & 7.0341                 & 8.2870               \\
\hline \hline
        & 0.1028           & 0.0801             & 0.0888           & 6.8922               & 7.1775                 & 8.5161                \\
        \hline
\end{tabular}
}
\vspace{1em}
\caption{Fidelities, denoted as $F(Q, P) = (\sum_{i} \sqrt{p_iq_i})^2$), and experiment run-times for the \textbf{IBM Quito} and \textbf{IBM Nairobi device}. Each row marks the execution of a single circuit generated from $25$ or $30$ Clifford gates for Quito and Nairobi, respectively. The last row in each table shows the average fidelity and run-time. The highest fidelity and shortest run-time in each row are marked in bold.}
\label{tab:quito}
\vspace{6.65049em}
\end{table}

%% file: struct/conclusion.tex
In our work, we created a novel, architecture-aware algorithm for synthesizing Clifford tableaus under arbitrary connectivity constraints. We observed that our algorithm reduces the number of CNOT gates by a significant factor when targeting architecturally constrained devices. However, in the case of all-to-all connectivity, our proposed algorithm performed worse than the state-of-the-art method by \citet{Bravyi_2021}. Since we expect many quantum devices to remain architecturally constrained in the future, architecture-aware synthesis is needed to address the qubit routing problem. Our evaluations on real hardware have shown a similar picture, specifically that reducing the number of multi-qubit gates executed on a quantum circuit reduces execution time and increases the final fidelity.

Our results show that using a depth-optimal strategy for a connectivity graph that does not match the intended connectivity is premature optimization. When using the proposed algorithm in a constrained case, it outperformed the depth-optimal method from \citet{peham2023depthoptimal} because of the required routing overhead. Similarly, we outperformed the LNN depth-optimized method from \citet{Maslov_2023} when routing is added. 

Nevertheless, we can see that our heuristic still has room for improvement, specifically for the larger architectures and for Clifford circuits with few gates. Since none of the synthesis algorithms were able to recover a similar CNOT count from the complete architecture, there seems to be some inherent inefficiency within all Clifford synthesis algorithms. This could potentially show that these algorithms are not yet fully leveraging the structure of the stabilizer and destabilizer relationships in the Clifford tableaus. Therefore, we suggest integrating the block-wise strategy of \citet{patel2008optimal} towards the synthesis of Clifford tableaus as much as possible. Combining this with an architecture-aware synthesis method could yield both optimal asymptotic behaviour and further reduction of CNOTs when compiling the circuit towards a specific architecture.

For completeness, we want to emphasize that stabilizer circuits are classically simulable and therefore not universal for quantum computation. As such, our algorithm needs to be combined with other algorithms to be used for the compilation of quantum circuits. One way to do this is to recognize the similarities between Clifford tableaus and parity maps and realize that parity maps can be generalized to phase polynomials~\cite{Amy_2018,degriend2020architectureaware,Nash_2020,debrugière2023shallower,vandaele2022phase,kutin2007computation,amy2013meetinthemiddle,Nam_2018}. This would generalize Clifford tableaus to Pauli string operators, in a way similar to the method proposed by \citet{Martiel_2022}. In fact, it might be possible to use our algorithm directly to optimize their final, complicated Clifford operator for cases where the expected value is not calculated with respect to a specific Hamiltonian. Alternatively, we can adapt existing Pauli string compilation methods~\cite{Cowtan_2020, li2022paulihedral} to be made architecture-aware, or extend existing methods that use the mixed ZX-phase polynomial form~\cite{gogioso2022annealing,winderl2023recursively,vandegriend2023towards} to so-called Pauli gadgets~\cite{Cowtan_2020,yeung2020diagrammatic}.

Similarly, it is possible to adapt our proposed algorithm in the compilation of quantum error correction codes because quantum error correction codes consist of stabilizers and the algorithm synthesizes stabilizers. Additionally, our algorithm might be useful for the architecture-aware extraction of quantum circuits from ZX-diagrams in ZX-calculus~\cite{Duncan_2020,backens2021there}.

A final interesting point of comparison could be to provide an optimal baseline by adapting methods like \citet{peham2023depthoptimal} to be architecture-aware. Furthermore, extending layered synthesis approaches like \citet{Duncan_2020,Aaronson_2004,Bravyi_2021} can be adapted by applying architecture-aware synthesis methods already present to those approaches. This could bring further insights to the performance of our method and architecture-aware synthesis in general.

In summary, the synthesis of Clifford tableaus is an important primitive for the efficient compilation of quantum programs. Using our proposed algorithm, we synthesize to a target architecture directly without the need for added SWAP gates. We have shown that this holistic approach reduces routing overhead tremendously, a key factor in improving the performance of quantum hardware in the near term.

%% file: main.bbl
\providecommand{\noopsort}[1]{}

%% file: struct/appendix.tex
\section{Tableau Example}\label{apenx:tableau_example}
\FloatBarrier
Here, we present an example of steps (2)-(5) of the introduced algorithm. We start with an arbitrary 3-qubit Clifford tableau corresponding to $C^\dagger$, where the chosen pivot is the first qubit $q_1$. We limit the underlying architecture to a line architecture where $(q_1,q_2)$ and $(q_2, q_3)$ are adjacent pairs of qubits.
\[C^\dagger = 
\begin{bNiceArray}{lcc|ccc}[first-row,first-col]
   &  &  & &  &  &\\
   & 1 & 0 & 1 & 1 & 0 & 0\\
   & 0 & 0 & 0 & 0 & 1 &  0\\
   & 1 & 0 & 1 & 0 & 0  & 1\\
\hline
    & 1 & 0  & 0  & 0 & 0 & 0\\
    & 0 & 1  & 0  & 0 & 1 & 0\\
    & 1 & 0 & 0   & 0 & 0 & 1
\end{bNiceArray}
\]
\textbf{Step (2)}: We sanitize the first destabilizer in the first row by applying $S_1$ 
\[
\begin{bNiceArray}{lcc|ccc}[first-row,first-col]
    \CodeBefore
    \cellcolor{gray!80}{-1, -4}
    \Body
   &  &  & &  &  &\\
   & \tikzmarknode{a}{1} & 0 & 1 & \tikzmarknode{b}{1} & 0 & 0\\
   & 0 & 0 & 0 & 0  & 1 &  0\\
   & 1 & 0 & 1 & 0 & 0  & 1\\
\hline
    & 1 & 0  & 0  & 0 & 0 & 0\\
    & 0 & 1  & 0  & 0 & 1 & 0\\
    & 1 & 0 & 0   & 0 & 0 & 1
\end{bNiceArray}
\xrightarrow{S_1} 
\begin{bNiceArray}{lcc|ccc}[first-row,first-col]
\CodeBefore
    \cellcolor{yellow!60}{-4}
    \Body
   &  &  & &  &  &\\
   & 1 & 0 & 1 & 0 & 0 & 0\\
   & 0 & 0 & 0 & 0  & 1 &  0\\
   & 1 & 0 & 1 & 1 & 0  & 1\\
\hline
    & 1 & 0  & 0  & 1 & 0 & 0\\
    & 0 & 1  & 0  & 0 & 1 & 0\\
    & 1 & 0 & 0   & 1 & 0 & 1
\end{bNiceArray}
\]
\begin{tikzpicture}[remember picture,overlay]
\draw[-latex]([yshift=.5ex]a.north) to[bend left]node[above]{$\oplus$} ([yshift=.5ex]b.north);
\end{tikzpicture}
\textbf{Step (3)}: 
We then remove the interaction of the first qubit $q_1$ with the third qubit $q_3$. Due to the restricted architecture, we traverse the Steiner tree from leaf to root, applying $CNOT_{3,2}$ to fill in the intermediate Steiner tree nodes.
$$
\begin{bNiceArray}{lcc|ccc}[first-row,first-col]
\CodeBefore
    \cellcolor{gray!80}{-2, -3}
    \cellcolor{gray!40}{-5, -6}
    \Body
   &  &  & &  &  &\\
   & 1 & \tikzmarknode{a}{0} & \tikzmarknode{b}{1} & 0 &\tikzmarknode{c}{0} & \tikzmarknode{d}{0} \\
   & 0 & 0 & 0 & 0  & 1 &  0\\
   & 1 & 0 & 1 & 1 & 0  & 1\\
\hline
    & 1 & 0  & 0  & 1 & 0 & 0\\
    & 0 & 1  & 0  & 0 & 1 & 0\\
    & 1 & 0 & 0   & 1 & 0 & 1
\end{bNiceArray}
\xrightarrow{CNOT_{3,2}} 
\begin{bNiceArray}{lcc|ccc}[first-row,first-col]
\CodeBefore
    \cellcolor{yellow!60}{-2}
    \cellcolor{orange!35}{-6}
    \Body
   &  &  & &  &  &\\
   & 1 & 1 & 1 & 0 & 0 & 0\\
   & 0 & 0 & 0 & 0 & 1 &  1\\
   & 1 & 1 & 1 & 1 & 0  & 1\\
\hline
    & 1 & 0  & 0  & 1 & 0 & 0\\
    & 0 & 1  & 0  & 0 & 1 & 1\\
    & 1 & 0 & 0   & 1 & 0 & 1
\end{bNiceArray}
$$
\begin{tikzpicture}[remember picture,overlay]
\draw[-latex]([yshift=.5ex]b.north) to[bend right]node[above]{$\oplus$} ([yshift=.5ex]a.north);
\draw[-latex]([yshift=.5ex]c.north) to[bend left]node[above]{$\oplus$} ([yshift=.5ex]d.north);
\end{tikzpicture}

Then, we apply $CNOT_{2,3}$ and $CNOT_{1,2}$ to remove interactions entirely and set the first destabilizer to an identity row.
\[
\begin{bNiceArray}{lcc|ccc}[first-row,first-col]
\CodeBefore
    \cellcolor{gray!80}{-2, -3}
    \cellcolor{gray!40}{-5,-6}
    \Body
   &  &  & &  &  &\\
   & 1 & \tikzmarknode{a}{1} & \tikzmarknode{b}{1} & 0 & \tikzmarknode{c}{0} & \tikzmarknode{d}{0} \\
   & 0 & 0 & 0 & 0  & 1 &  1\\
   & 1 & 1 & 1 & 1 & 0  & 1\\
\hline
    & 1 & 0  & 0  & 1 & 0 & 0\\
    & 0 & 1  & 0  & 0 & 1 & 1\\
    & 1 & 0 & 0   & 1 & 0 & 1
\end{bNiceArray}
\xrightarrow{CNOT_{2,3}} 
\begin{bNiceArray}{lcc|ccc}[first-row,first-col]
\CodeBefore
    \cellcolor{yellow!60}{-3}
    \cellcolor{orange!35}{-5}
    \cellcolor{gray!80}{-1, -2}
    \cellcolor{gray!40}{-4}
    \Body
   &  &  & &  &  &\\
   & \tikzmarknode{e}{1} & \tikzmarknode{f}{1} & 0  & \tikzmarknode{g}{0} & \tikzmarknode{h}{0}& 0\\
   & 0 & 0 & 0 & 0  & 0 &  1\\
   & 1 & 1 & 0 & 1 & 1  & 1\\
\hline
    & 1 & 0  & 0  & 1 & 0 & 0\\
    & 0 & 1  & 1  & 0 & 0 & 1\\
    & 1 & 0 & 0   & 1 & 1 & 1
\end{bNiceArray}
\xrightarrow{CNOT_{1,2}} 
\begin{bNiceArray}{lcc|ccc}[first-row,first-col]
\CodeBefore
    \cellcolor{yellow!60}{-2}
    \cellcolor{orange!35}{-4}
    \rowcolor{green!60}{1}
    \Body
   &  &  & &  &  &\\
   & 1 & 0 & 0 & 0 & 0 & 0\\
   & 0 & 0 & 0 & 0 & 0 &  1\\
   & 1 & 0 & 0 & 0 & 1  & 1\\
\hline
    & 1 & 1  & 0  & 1 & 0 & 0\\
    & 0 & 1  & 1  & 0 & 0 & 1\\
    & 1 & 1  & 0  & 0 & 1 & 1
\end{bNiceArray}
\]
\begin{tikzpicture}[remember picture,overlay]
\draw[-latex]([yshift=.5ex]a.north) to[bend left]node[above]{$\oplus$} ([yshift=.5ex]b.north);
\draw[-latex]([yshift=.5ex]d.north) to[bend right]node[above]{$\oplus$} ([yshift=.5ex]c.north);
\draw[-latex]([yshift=.5ex]e.north) to[bend left]node[above]{$\oplus$} ([yshift=.5ex]f.north);
\draw[-latex]([yshift=.5ex]h.north) to[bend right]node[above]{$\oplus$} ([yshift=.5ex]g.north);
\end{tikzpicture}

\newpage
\textbf{Step (4)}: 
We then move to sanitize the corresponding stabilizer. We do this, sanitizing qubits from left to right; first, we apply $H_1S_1H_1$ to ensure that no fill-in occurs on the corresponding destabilizer.
\[
\begin{bNiceArray}{lcc|ccc}[first-row,first-col]
\CodeBefore
    \cellcolor{gray!80}{-1,-4}
    \Body
   &  &  & &  &  &\\
   & \tikzmarknode{a}{1} & 0 & 0 & \tikzmarknode{b}{0} & 0 & 0\\
   & 0 & 0 & 0 & 0 & 0 &  1\\
   & 1 & 0 & 0 & 0 & 1  & 1\\
\hline
    & 1 & 1  & 0  & 1 & 0 & 0\\
    & 0 & 1  & 1  & 0 & 0 & 1\\
    & 1 & 1  & 0  & 0 & 1 & 1\\
\end{bNiceArray}
\xrightarrow{H_1} 
\begin{bNiceArray}{lcc|ccc}[first-row,first-col]
\CodeBefore
    \cellcolor{yellow!60}{-1,-4}
    \Body
   &  &  & &  &  &\\
   & \tikzmarknode{c}{0} & 0 & 0 & \tikzmarknode{d}{1} & 0 & 0\\
   & 0 & 0 & 0 & 0 & 0 &  1\\
   & 0 & 0 & 0 & 1 & 1  & 1\\
\hline
    & 1 & 1  & 0  & 1 & 0 & 0\\
    & 0 & 1  & 1  & 0 & 0 & 1\\
    & 0 & 1  & 0  & 1 & 1 & 1\\
\end{bNiceArray}
\xrightarrow{S_1} 
\begin{bNiceArray}{lcc|ccc}[first-row,first-col]
\CodeBefore
    \cellcolor{yellow!60}{-1,-4}
    \Body
   &  &  & &  &  &\\
   & \tikzmarknode{e}{0} & 0 & 0 & \tikzmarknode{f}{1} & 0 & 0\\
   & 0 & 0 & 0 & 0 & 0 &  1\\
   & 0 & 0 & 0 & 1 & 1  & 1\\
\hline
    & 1 & 1  & 0  & 0 & 0 & 0\\
    & 0 & 1  & 1  & 0 & 0 & 1\\
    & 0 & 1  & 0  & 1 & 1 & 1\\
\end{bNiceArray}
\xrightarrow{H_1} 
\begin{bNiceArray}{lcc|ccc}[first-row,first-col]
\CodeBefore
    \cellcolor{yellow!60}{-1,-4}
    \Body
   &  &  & &  &  &\\
   & 1 & 0 & 0 & 0 & 0 & 0\\
   & 0 & 0 & 0 & 0 & 0 &  1\\
   & 1 & 0 & 0 & 0 & 1  & 1\\
\hline
    & 0 & 1  & 0  & 1 & 0 & 0\\
    & 0 & 1  & 1  & 0 & 0 & 1\\
    & 1 & 1  & 0  & 0 & 1 & 1\\
\end{bNiceArray}
\]
\begin{tikzpicture}[remember picture,overlay]
\draw[latex-latex]([yshift=.5ex]a.north) to[bend left]node[above]{} ([yshift=.5ex]b.north);
\draw[-latex]([yshift=.5ex]c.north) to[bend left]node[above]{$\oplus$} ([yshift=.5ex]d.north);
\draw[latex-latex]([yshift=.5ex]e.north) to[bend left]node[above]{} ([yshift=.5ex]f.north);
\end{tikzpicture}

 Then, $H_2$ is applied to sanitize the second qubit.
\[
\begin{bNiceArray}{lcc|ccc}[first-row,first-col]
\CodeBefore
    \cellcolor{gray!80}{-2,-5}
    \Body
   &  &  & &  &  &\\
   & 1 & \tikzmarknode{a}{0} & 0 & 0 & \tikzmarknode{b}{0} & 0\\
   & 0 & 0 & 0 & 0 & 0 &  1\\
   & 1 & 0 & 0 & 0 & 1  & 1\\
\hline
    & 0 & 1  & 0  & 1 & 0 & 0\\
    & 0 & 1  & 1  & 0 & 0 & 1\\
    & 1 & 1  & 0  & 0 & 1 & 1\\
\end{bNiceArray}
\xrightarrow{H_2} 
\begin{bNiceArray}{lcc|ccc}[first-row,first-col]
\CodeBefore
    \cellcolor{yellow!60}{-2,-5}
    \Body
   &  &  & &  &  &\\
   & 1 & 0 & 0 & 0 & 0 & 0\\
   & 0 & 0 & 0 & 0 & 0 &  1\\
   & 1 & 1 & 0 & 0 & 0  & 1\\
\hline
    & 0 & 0  & 0  & 1 & 1 & 0\\
    & 0 & 0  & 1  & 0 & 1 & 1\\
    & 1 & 1  & 0  & 0 & 1 & 1\\
\end{bNiceArray}
\]
\begin{tikzpicture}[remember picture,overlay]
\draw[latex-latex]([yshift=.5ex]a.north) to[bend left]node[above]{} ([yshift=.5ex]b.north);
\end{tikzpicture}

\textbf{Step (5)}: 
Finally, as $q_1$ and $q_2$ are adjacent, we apply $CNOT_{2,1}$ to remove the last interacting term.
\[
\begin{bNiceArray}{lcc|ccc}[first-row,first-col]
\CodeBefore
    \cellcolor{gray!60}{-1,-2}
    \cellcolor{gray!35}{-4,-5}
    \Body
   &  &  & &  &  &\\
   & \tikzmarknode{a}{1} & \tikzmarknode{b}{0} & 0 & \tikzmarknode{c}{0} & \tikzmarknode{d}{0} & 0\\
   & 0 & 0 & 0 & 0 & 0 &  1\\
   & 1 & 1 & 0 & 0 & 0  & 1\\
\hline
    & 0 & 0  & 0  & 1 & 1 & 0\\
    & 0 & 0  & 1  & 0 & 1 & 1\\
    & 1 & 1  & 0  & 0 & 1 & 1\\
\end{bNiceArray}
\xrightarrow{CNOT_{2,1}} 
\begin{bNiceArray}{lcc|ccc}[first-row,first-col]
\CodeBefore
    \cellcolor{green!60}{-1,-4}
    \rowcolor{green!60}{1,4}
    \Body
   &  &  & &  &  &\\
   & 1 & 0 & 0 & 0 & 0 & 0\\
   & 0 & 0 & 0 & 0 & 0 &  1\\
   & 0 & 1 & 0 & 0 & 0  & 1\\
\hline
   & 0 & 0  & 0  & 1 & 0 & 0\\
   & 0 & 0  & 1  & 0 & 1 & 1\\
   & 0 & 1  & 0  & 0 & 1 & 1\\
\end{bNiceArray}
\]
\begin{tikzpicture}[remember picture,overlay]
\draw[-latex]([yshift=.5ex]b.north) to[bend right]node[above]{$\oplus$} ([yshift=.5ex]a.north);
\draw[-latex]([yshift=.5ex]c.north) to[bend left]node[above]{$\oplus$} ([yshift=.5ex]d.north);
\end{tikzpicture}

We additionally observe that the rows and columns corresponding to $q_1$, highlighted in green, are now all identity rows and columns. Hence, we can remove $q_1$ from the tableau and repeat the steps (2)-(5) with the next chosen pivot.

Tracking the appended gates, we obtain an architecture-conforming Clifford circuit that disentangles the first qubit:
\begin{figure}[h]
    \centering
    \includegraphics[width=0.6\textwidth]{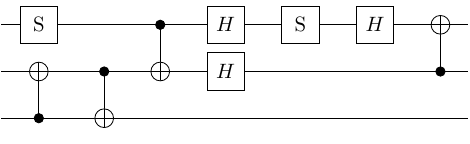}
    \caption{Synthesized Clifford circuit disentangling $q_1$ from the remainder of the qubits}
    \label{fig:synthesis_example}
\end{figure}
\FloatBarrier
\newpage

\section{Evaluation of the Single Qubit gate count}\label{apenx:eval_single_qubit}
\begin{figure}[hbt!]
    \centering
        \begin{subfigure}{\textwidth}
    \centering
    \includegraphics[width=0.8\textwidth]{img_new_methods/legend_experiments.pdf}
    \end{subfigure}
    \begin{subfigure}{\textwidth}
        \centering
        \includegraphics[width=0.3\textwidth]{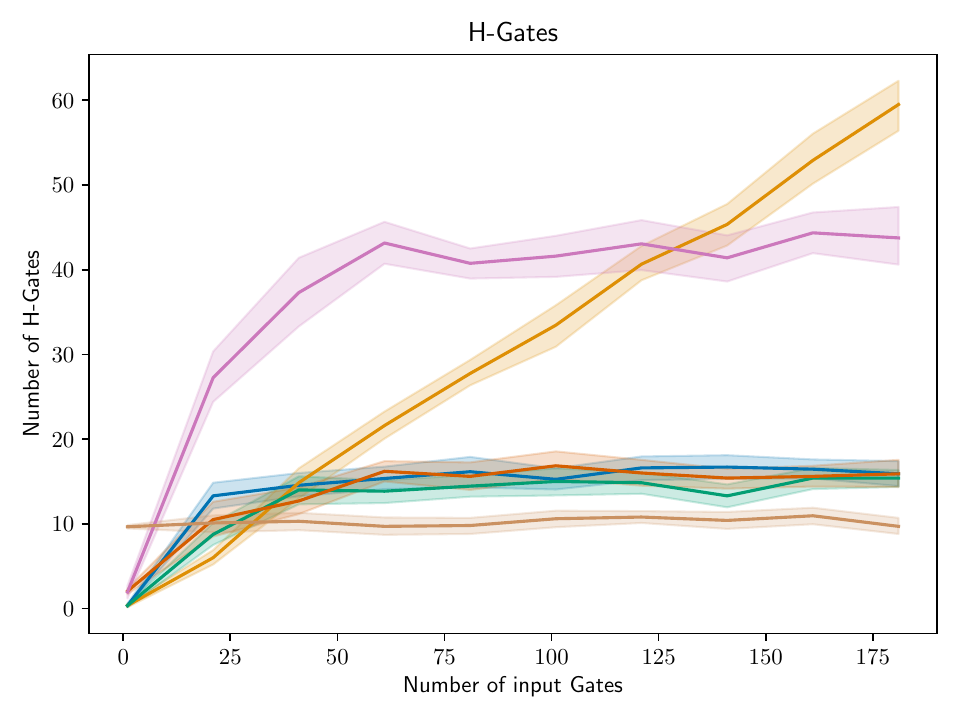}
        \includegraphics[width=0.3\textwidth]{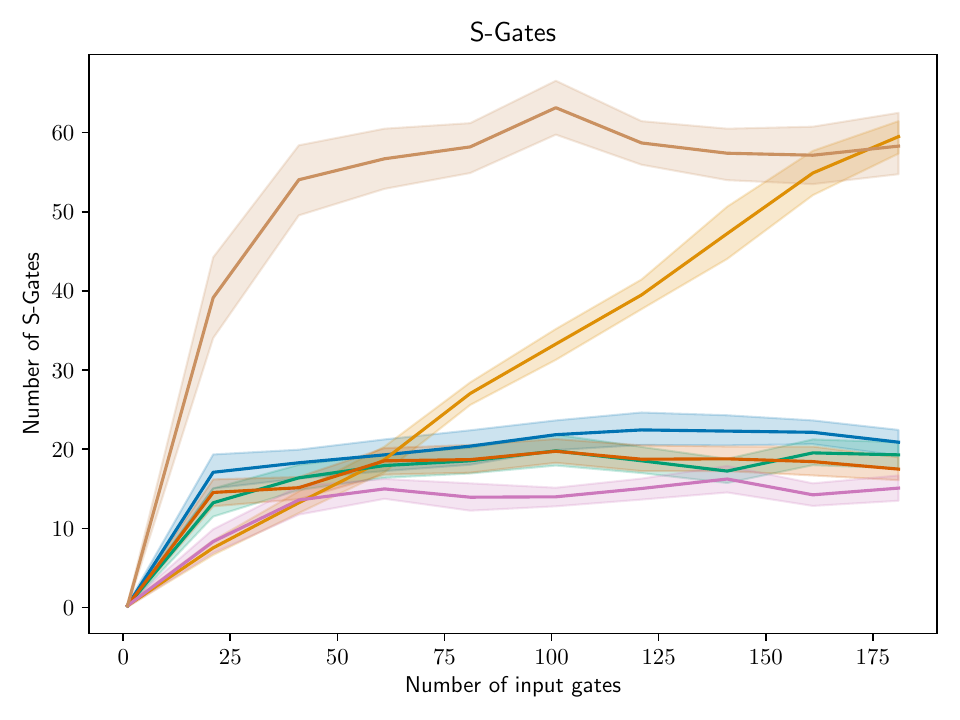}
        \includegraphics[width=0.3\textwidth]{img_new_methods/random_quito_cx.pdf}
        \caption{Quito (5 Qubits)}
    \end{subfigure}
    \begin{subfigure}{\textwidth}
        \centering
        \includegraphics[width=0.3\textwidth]{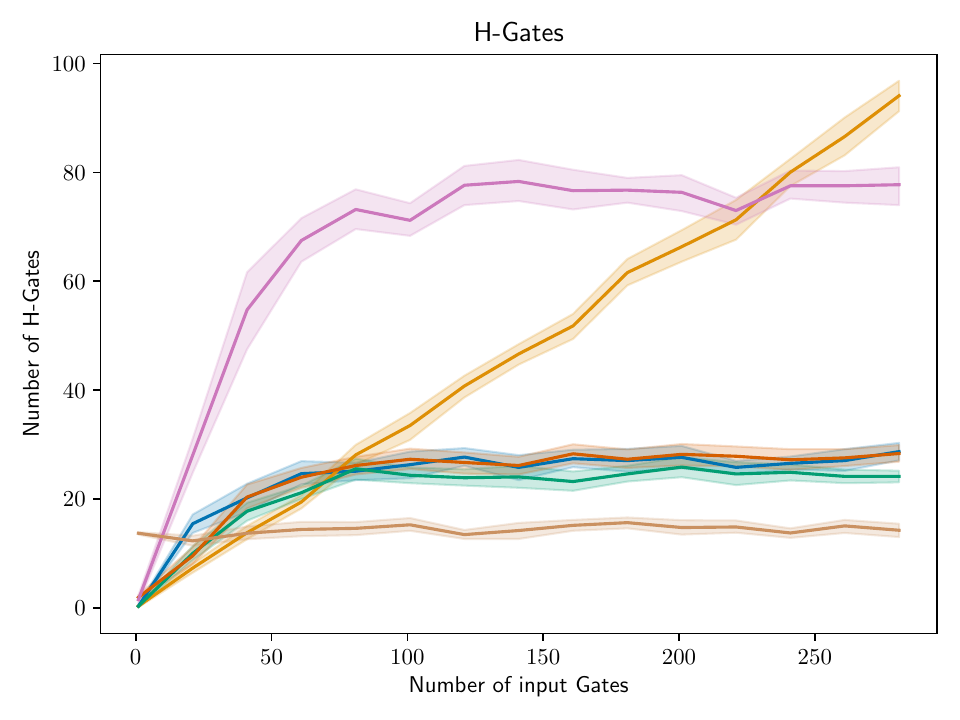}
        \includegraphics[width=0.3\textwidth]{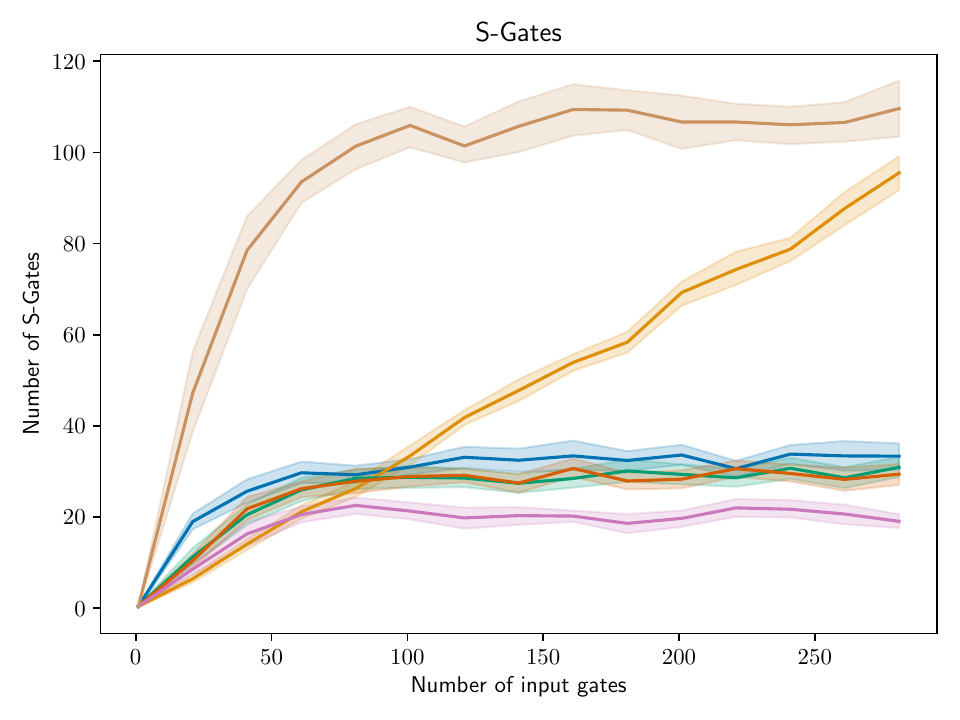}
        \includegraphics[width=0.3\textwidth]{img_new_methods/random_nairobi_cx.pdf}
        \caption{Nairobi (7 Qubits)}
    \end{subfigure}
    \begin{subfigure}{\textwidth}
        \centering
        \includegraphics[width=0.3\textwidth]{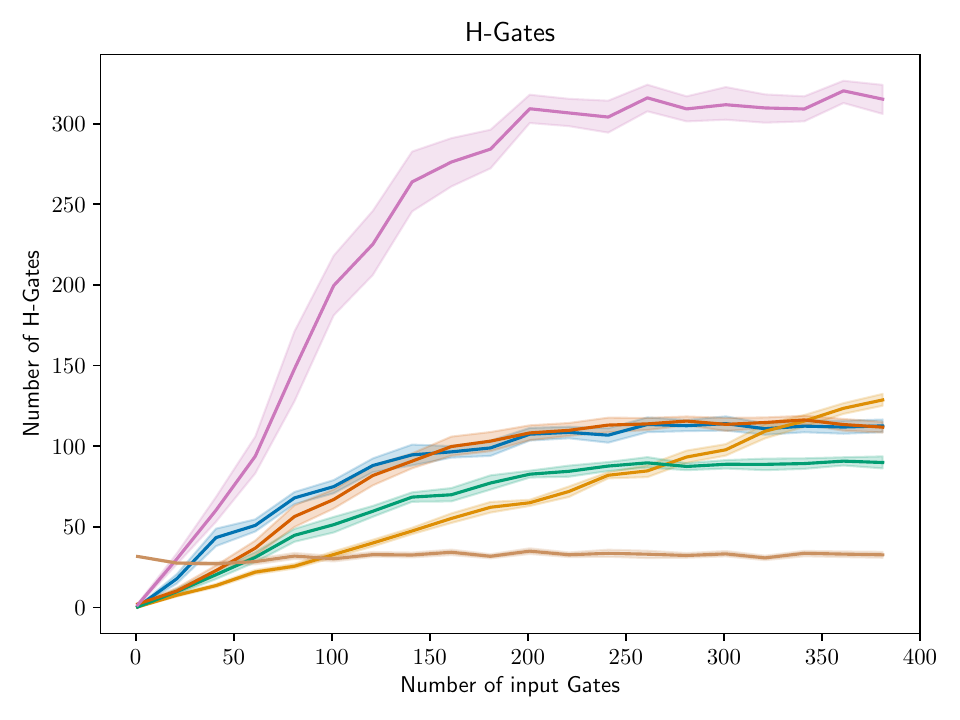}
        \includegraphics[width=0.3\textwidth]{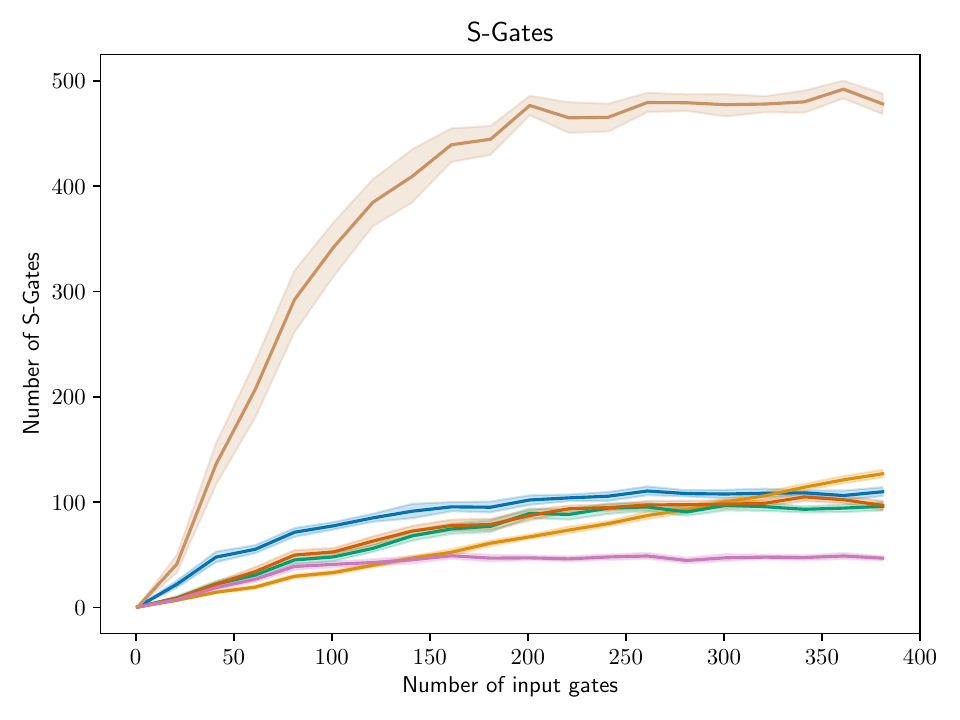}
        \includegraphics[width=0.3\textwidth]{img_new_methods/random_guadalupe_cx.pdf}
        \caption{Guadalupe (16 Qubits)}
    \end{subfigure}
    \caption{H-, S- and CNOT-Count for smaller architectures up to 16 qubits. Specifically, the backends Guadalupe, Nairobi, and Quito were used.}
    \label{fig:all_gates_small_device}
\end{figure}

\begin{figure}[hbt!]
    \centering
        \begin{subfigure}{\textwidth}
    \centering
    \includegraphics[width=0.8\textwidth]{img_new_methods/legend_experiments.pdf}
    \end{subfigure}
    \begin{subfigure}{\textwidth}
        \centering
        \includegraphics[width=0.3\textwidth]{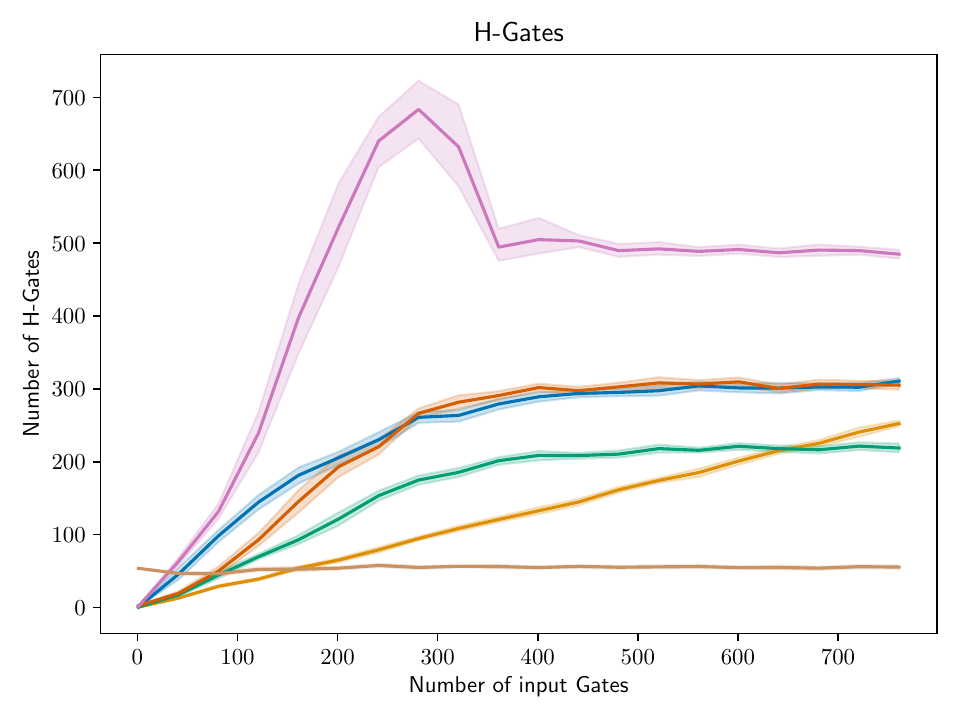}
        \includegraphics[width=0.3\textwidth]{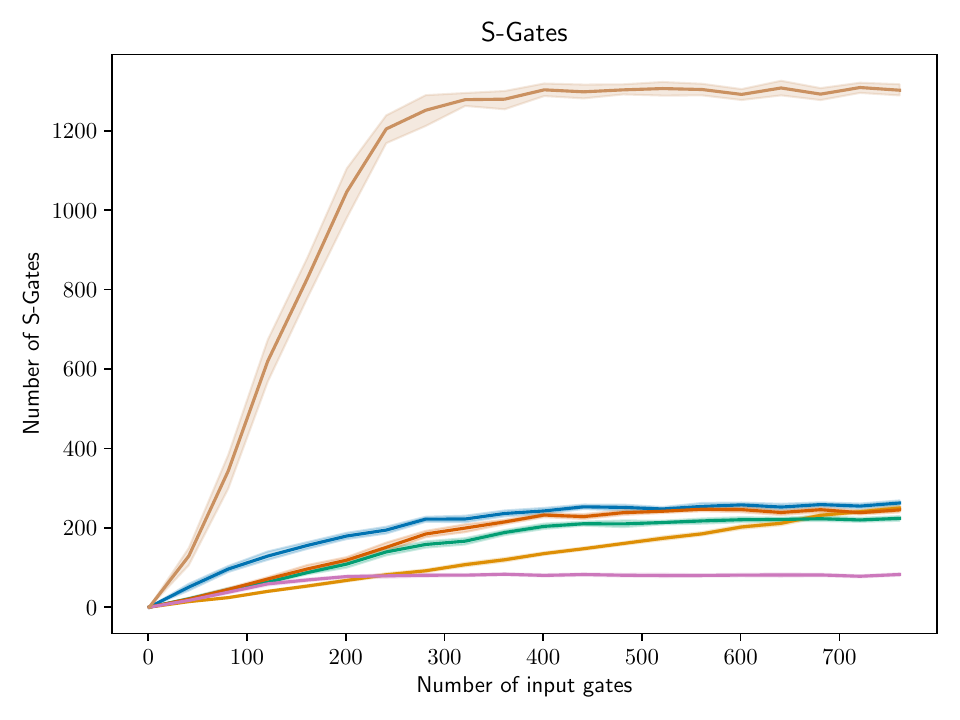}
        \includegraphics[width=0.3\textwidth]{img_new_methods/random_mumbai_cx.pdf}
        \caption{Mumbai (27 Qubits)}
    \end{subfigure}
    \begin{subfigure}{\textwidth}
        \centering
        \includegraphics[width=0.3\textwidth]{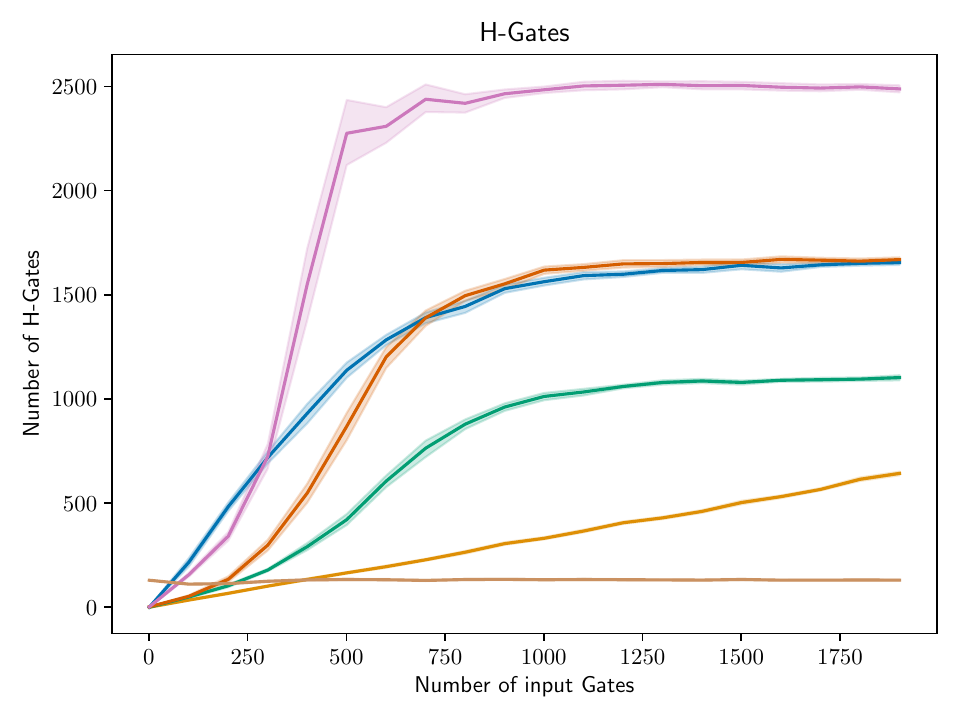}
        \includegraphics[width=0.3\textwidth]{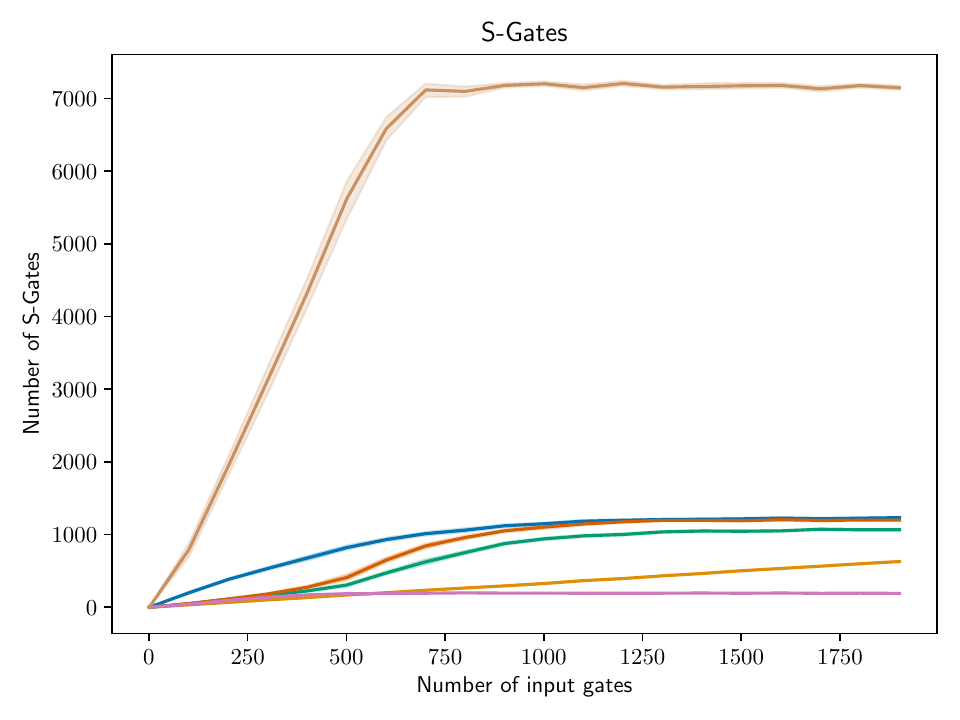}
        \includegraphics[width=0.3\textwidth]{img_new_methods/random_ithaca_cx.pdf}
        \caption{Ithaca (65 Qubits)}
    \end{subfigure}
    \begin{subfigure}{\textwidth}
        \centering
        \includegraphics[width=0.3\textwidth]{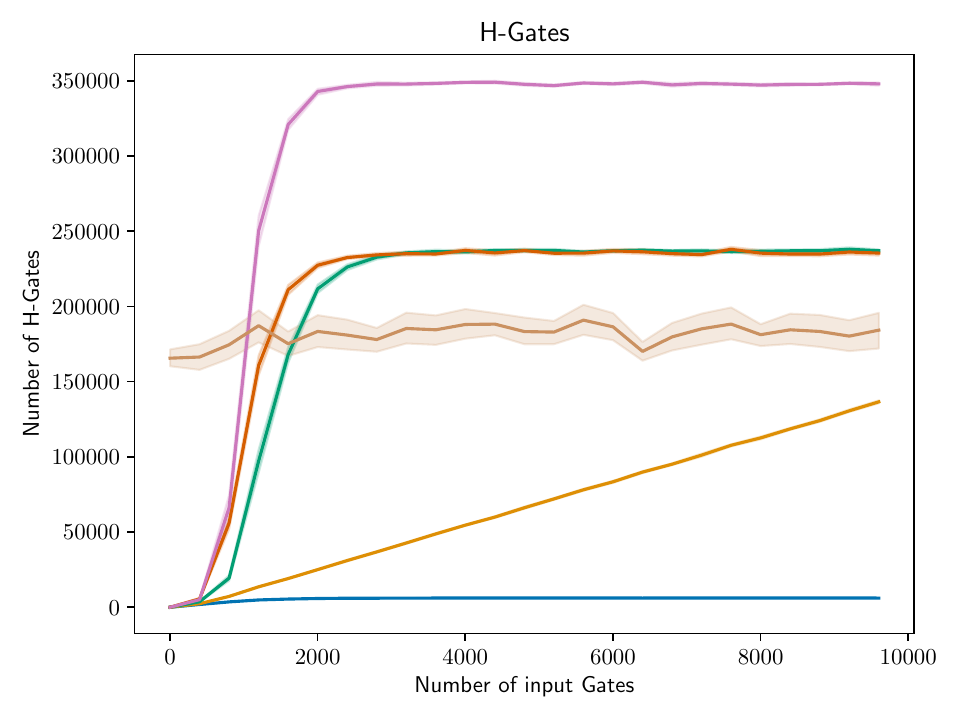}
        \includegraphics[width=0.3\textwidth]{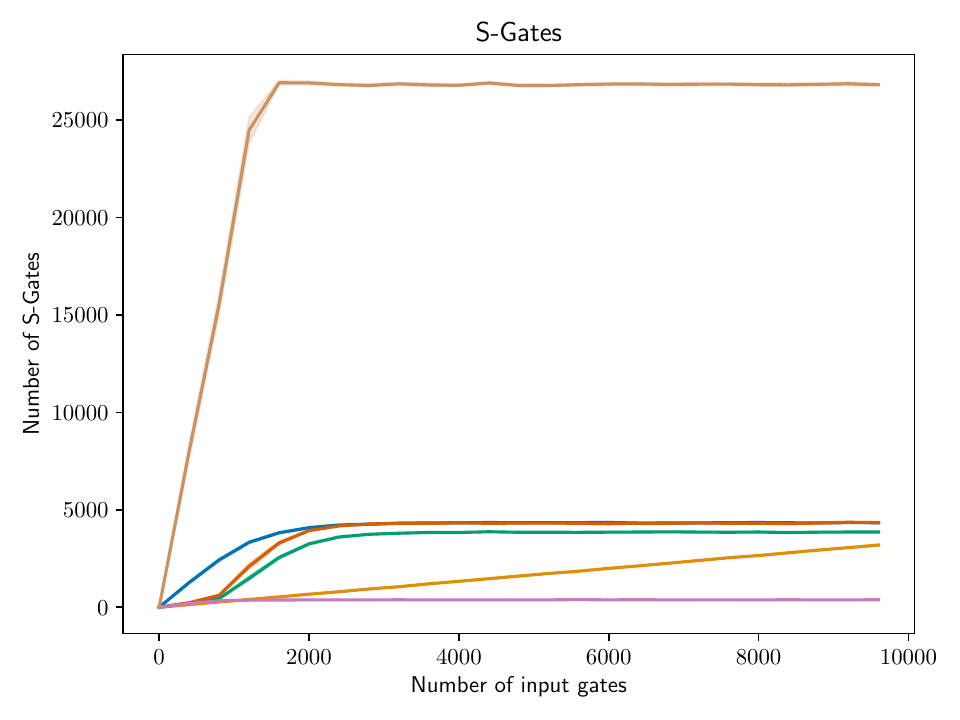}
        \includegraphics[width=0.3\textwidth]{img_new_methods/random_brisbane_cx.pdf}
        \caption{Brisbane (127 Qubits)}
    \end{subfigure}
    \caption{H-, S- and CNOT-Count for larger architectures up to 127 qubits. Specifically, the backends Brisbane, Ithaca, and Mumbai were used. }
    \label{fig:all_gates_large_device}
\end{figure}

\FloatBarrier
While we have focused on checking the CNOT count in our evaluation, we want to report our S- and H-Gates results. To allow a complete view of the performance of our algorithm, observe \autoref{fig:all_gates_small_device} and \autoref{fig:all_gates_large_device}. Overall, we could see that while our algorithm improves among the CNOT-Count, it shows the tendency to increase both H- and S-Gates. Specifically, this phenomenon tends to be visible for larger architectures like Mumbai and Ithaca. while the S- and H-Count for smaller architectures are equivalent. This is due to the sanitization process and because the sign column of the Clifford tableau is adjusted after synthesis, which is quite expensive in terms of H and S-Gates. 
\FloatBarrier
\clearpage

\section{Evaluation of total circuit depth}\label{apndx:full_depth}
\begin{figure}[bt]
	\centering
    \begin{subfigure}{\textwidth}
    \centering
    \includegraphics[width=0.8\textwidth]{img_new_methods/legend_experiments.pdf}
    \end{subfigure}
    \vspace{-1em}
    \begin{subfigure}{\textwidth}
        \begin{subfigure}{0.47\textwidth}
    	    \centering
    		\includegraphics[width=\textwidth]{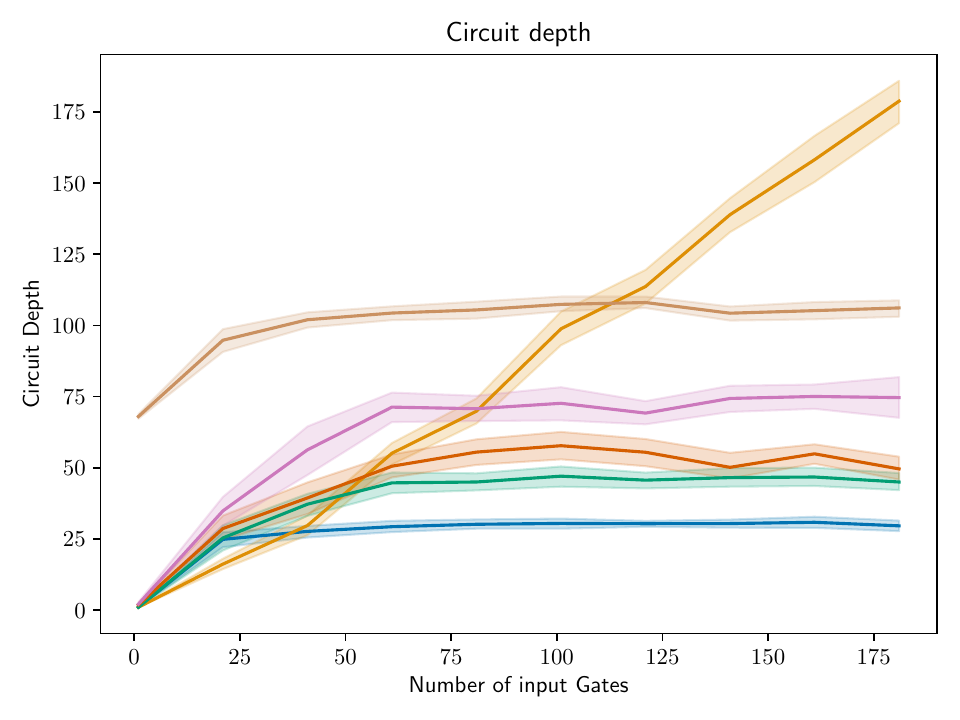}
    		\caption{Quito (5 Qubits)}
    	\end{subfigure}
    \hfill
        \begin{subfigure}{0.47\textwidth}
    	    \centering
            \includegraphics[width=\textwidth]{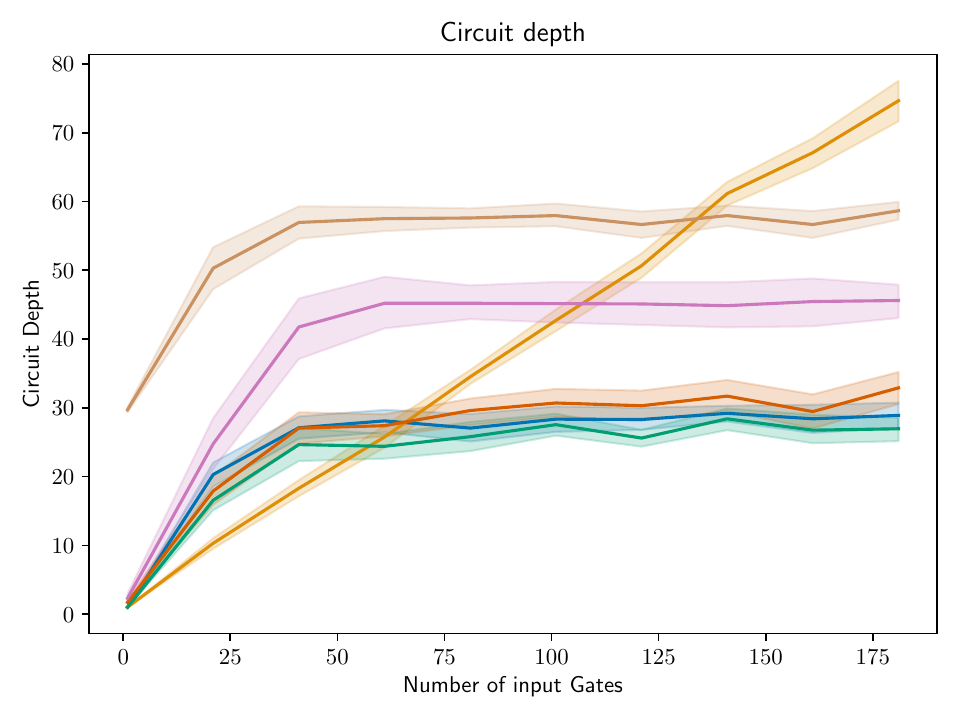}
    		\caption{Complete (5 Qubits)}
    	\end{subfigure}
    \hfill
     	\begin{subfigure}{0.47\textwidth}
    	    \centering
    		\includegraphics[width=\textwidth]{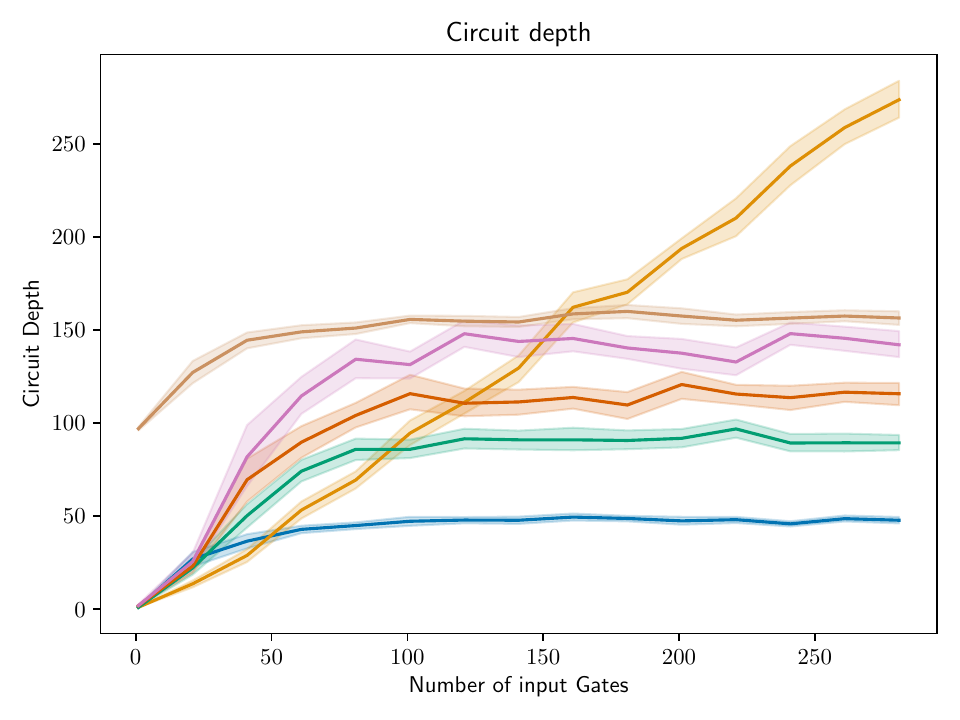}
    		\caption{Nairobi (7 Qubits)}
    	\end{subfigure}
     \hfill
     	\begin{subfigure}{0.47\textwidth}
    	    \centering
    		\includegraphics[width=\textwidth]{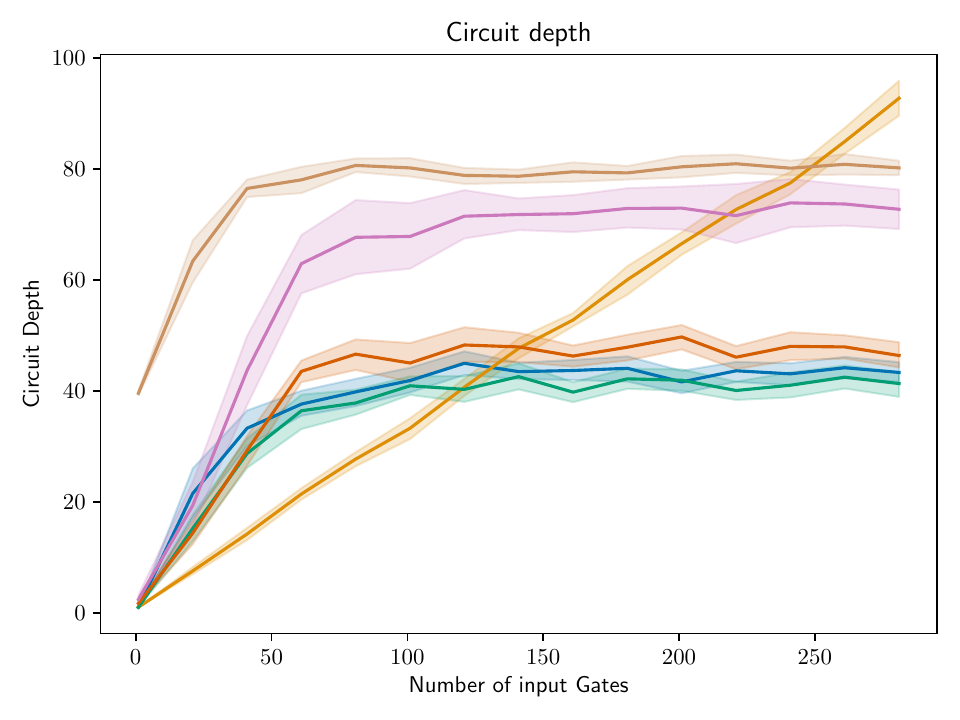}
    		\caption{Complete (7 Qubits)}
    	\end{subfigure}
     \hfill
    	\begin{subfigure}{0.47\textwidth}
    	    \centering
    		\includegraphics[width=\textwidth]{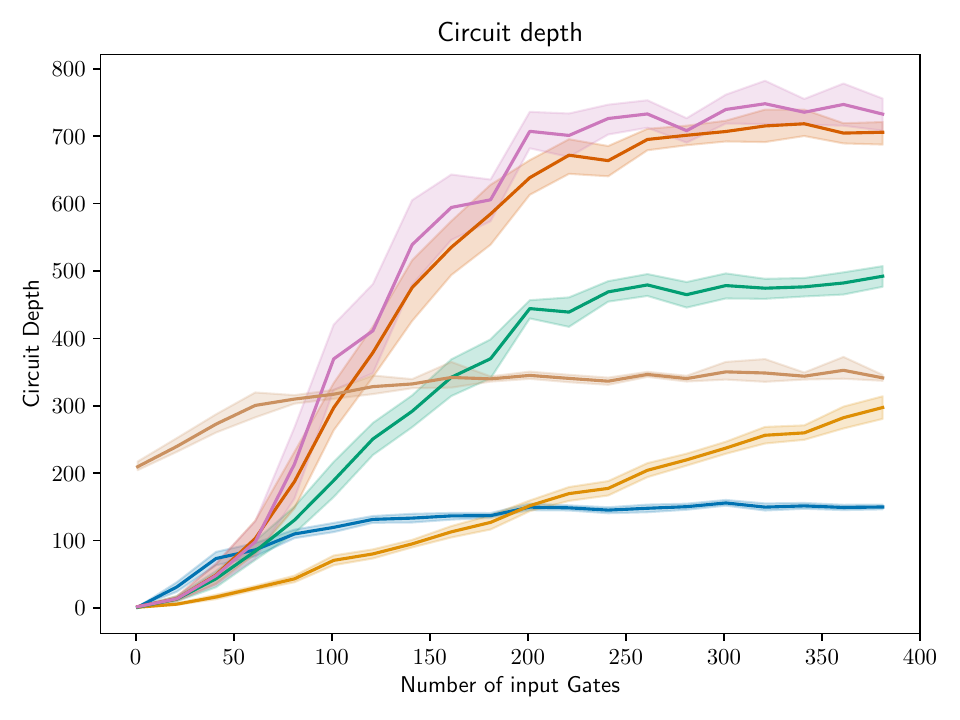}
    		\caption{Guadalupe (16 Qubits)}
    	\end{subfigure}
     \hfill
        \begin{subfigure}{0.47\textwidth}
    	    \centering
    		\includegraphics[width=\textwidth]{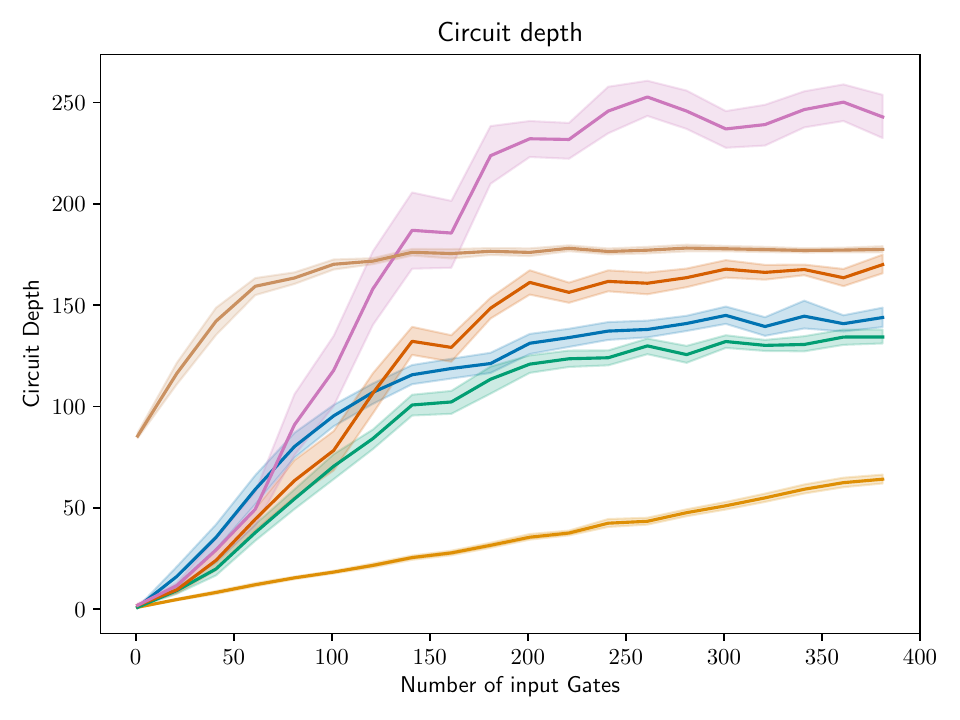}
    		\caption{Complete (16 Qubits)}
    	\end{subfigure}
    \end{subfigure}
	\caption{Circuit depth for smaller architectures up to 16 qubits. Specifically, the backends Guadalupe, Nairobi, and Quito were used. We additionally reported the CNOT count of the complete architectures of the same size.}\label{fig:depth_comparison_smaller_arch}
    \vspace{3em}
\end{figure}

\begin{figure}[bt]
    \centering
    \begin{subfigure}{\textwidth}
    \centering
    \includegraphics[width=0.8\textwidth]{img_new_methods/legend_experiments.pdf}
    \end{subfigure}
    \vspace{-1em}
	\begin{subfigure}{\textwidth}
        \begin{subfigure}{0.47\textwidth}
    	    \centering
    		\includegraphics[width=\textwidth]{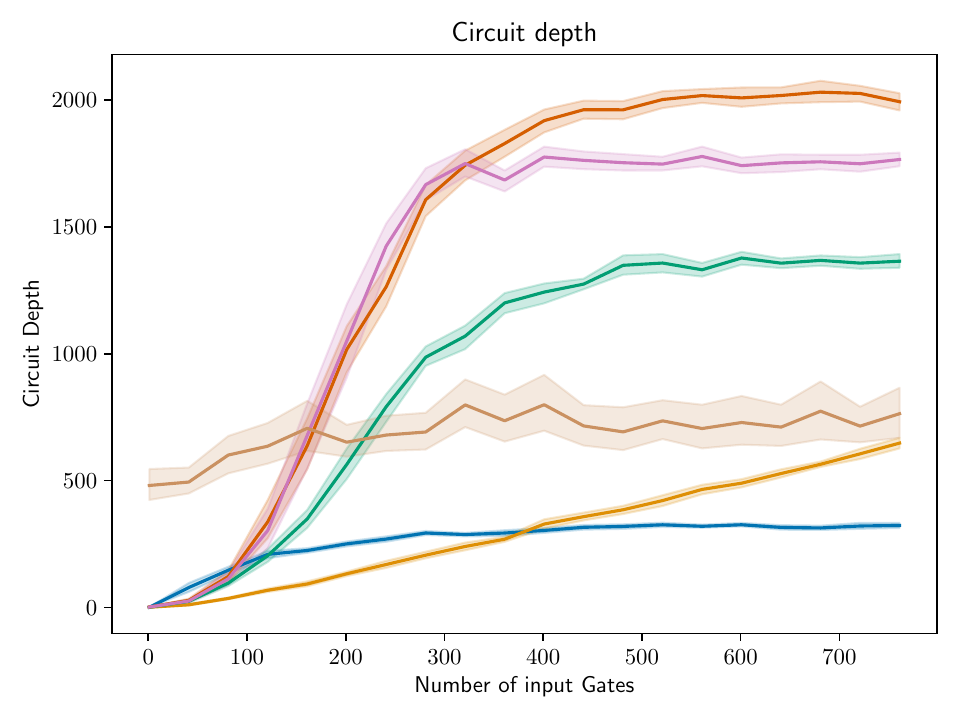}
    		\caption{Mumbai (27 Qubits)}
    	\end{subfigure}
     \hfill
        \begin{subfigure}{0.47\textwidth}
    	    \centering
            \includegraphics[width=\textwidth]{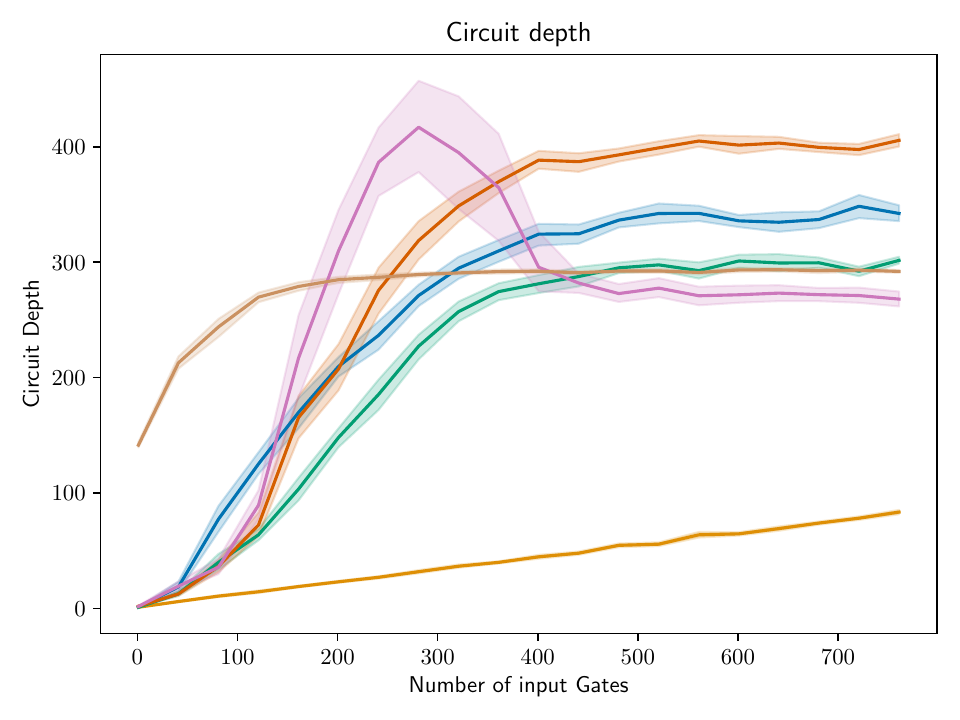}
    		\caption{Complete (27 Qubits)}
    	\end{subfigure}
        \begin{subfigure}{0.47\textwidth}
    	    \centering
    		\includegraphics[width=\textwidth]{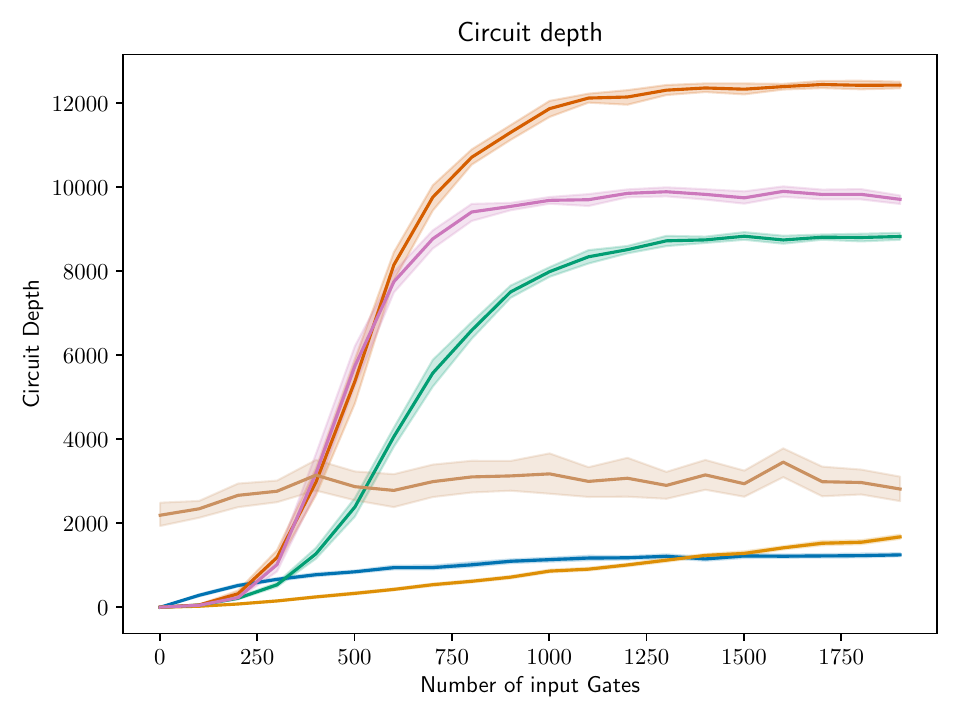}
    		\caption{Ithaca (65 Qubits)}
    	\end{subfigure}
     \hfill
        \begin{subfigure}{0.47\textwidth}
    	    \centering
            \includegraphics[width=\textwidth]{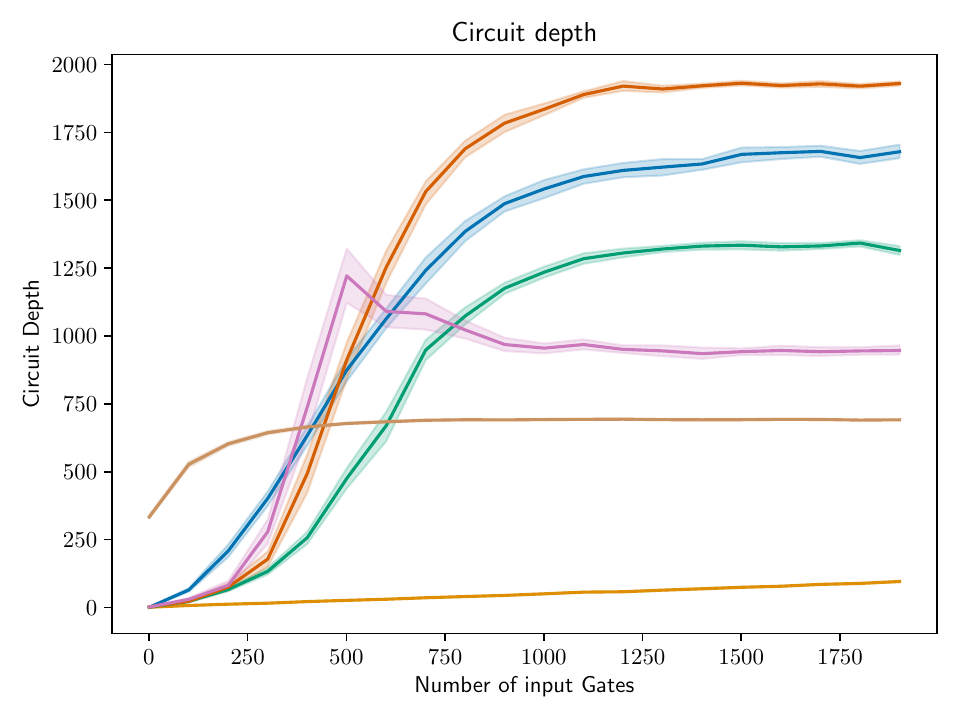}
    		\caption{Complete (65 Qubits)}
    	\end{subfigure}
        \begin{subfigure}{0.47\textwidth}
    	    \centering
    		\includegraphics[width=\textwidth]{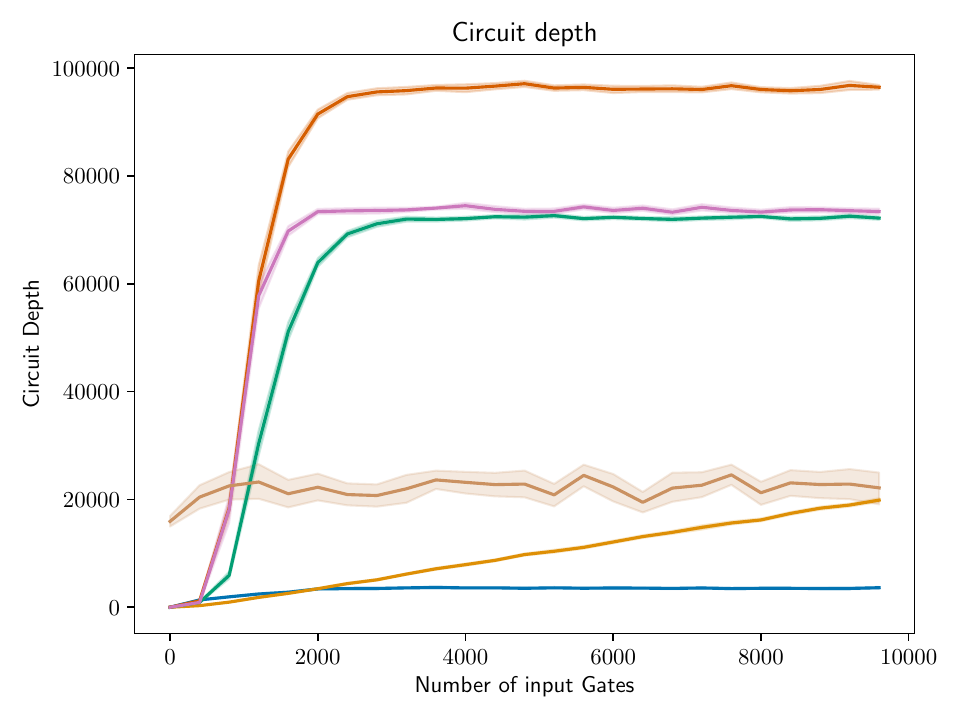}
    		\caption{Brisbane (127 Qubits)}
    	\end{subfigure}
        \hfill
        \begin{subfigure}{0.47\textwidth}
    	    \centering
            \includegraphics[width=\textwidth]{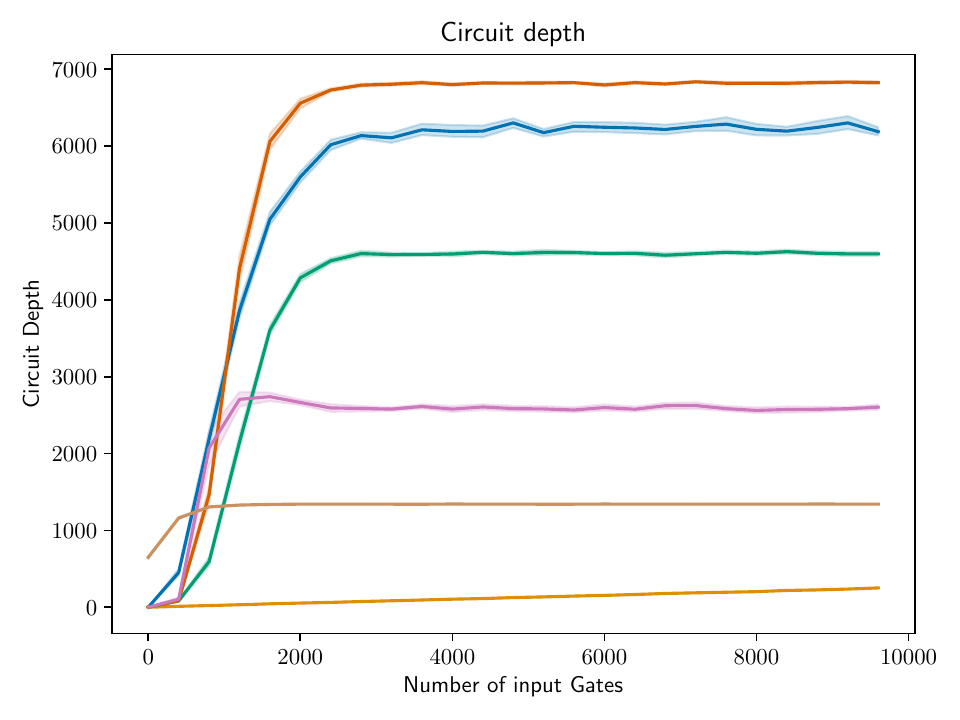}
    		\caption{Complete (127 Qubits)}
    	\end{subfigure}
    \end{subfigure}
    \caption{Circuit depth for larger architectures up to 127 qubits. Specifically, the backends Brisbane, Ithaca, and Mumbai were used. We additionally reported the CNOT count of the complete architectures of the same size.}
    \label{fig:depth_comparison_larger_arch}
    \vspace{3em}
\end{figure}

For completeness, we plot the total circuit depth in \autoref{fig:depth_comparison_smaller_arch} and \autoref{fig:depth_comparison_larger_arch} for the proposed algorithm and the baseline algorithms.

\FloatBarrier
\clearpage

\section{Architectures used in this study}\label{apndx:arch_study}
\begin{figure}[h]
	\centering
    \begin{subfigure}{0.3\textwidth}
    \centering
        \includegraphics[width=0.9\textwidth]{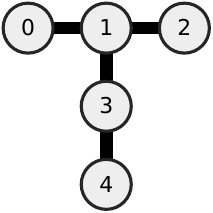}
        \caption{Quito (5 Qubits)}
    \end{subfigure}
    \hspace{0.135\textwidth}
    \begin{subfigure}{0.3\textwidth}
    \centering
        \includegraphics[width=0.9\textwidth]{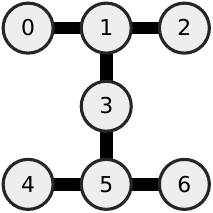}
        \caption{Nairobi (7 Qubits)}
    \end{subfigure}
    \begin{subfigure}{0.8\textwidth}
    \centering
        \includegraphics[width=0.9\textwidth]{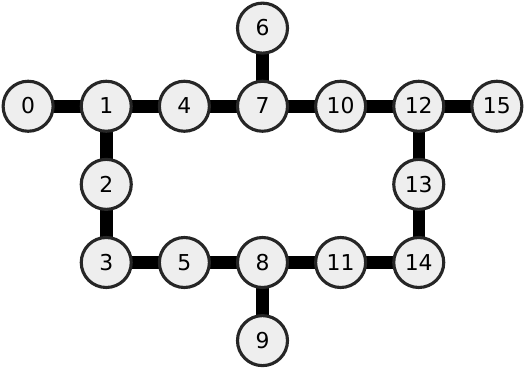}
        \caption{Guadalupe (16 Qubits)}
    \end{subfigure}
	
	\caption{Smaller architectures used in our evaluation.}\label{fig:architectures_clifford_small}
 
\end{figure}

\FloatBarrier
In this work, we focused on IBM hardware and consequently executed our algorithms solely on IBM architectures. To show adequate scaling differences, we evaluated our algorithm on many architectures of different sizes. We varied the size from 5 to 127 while providing real-world architectures instead of model graphs, like a square, a line, or a circle. In \autoref{fig:architectures_clifford_small} and \autoref{fig:all_gates_large_device}, one can find the graphs provided by qiskit. 

\begin{figure}
    \centering
    \begin{subfigure}{0.6\textwidth}
    \centering
        \includegraphics[width=0.9\textwidth]{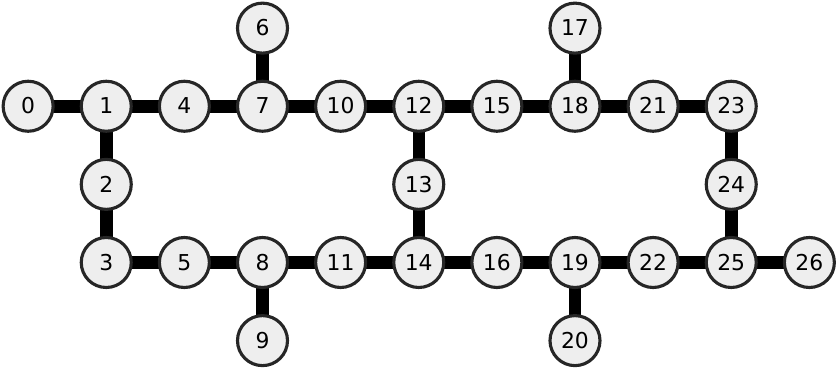}
        \caption{Mumbai (27 Qubits)}
    \end{subfigure}
    \vspace{1em}
    \begin{subfigure}{0.6\textwidth}
    \centering
        \includegraphics[width=0.9\textwidth]{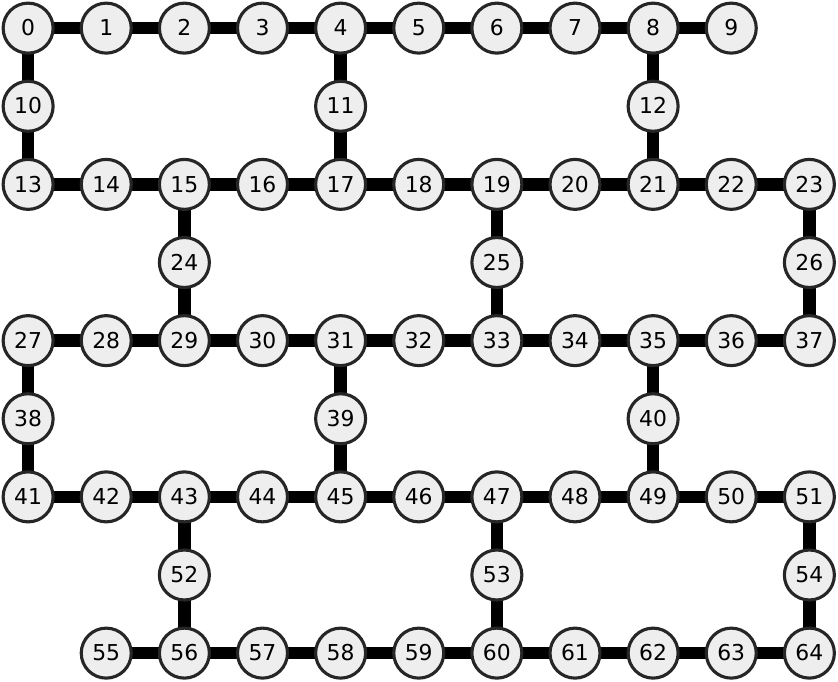}
        \caption{Ithaca (65 Qubits)}
    \end{subfigure}
    \vspace{1em}
    \begin{subfigure}{0.6\textwidth}
    \centering
        \includegraphics[width=0.9\textwidth]{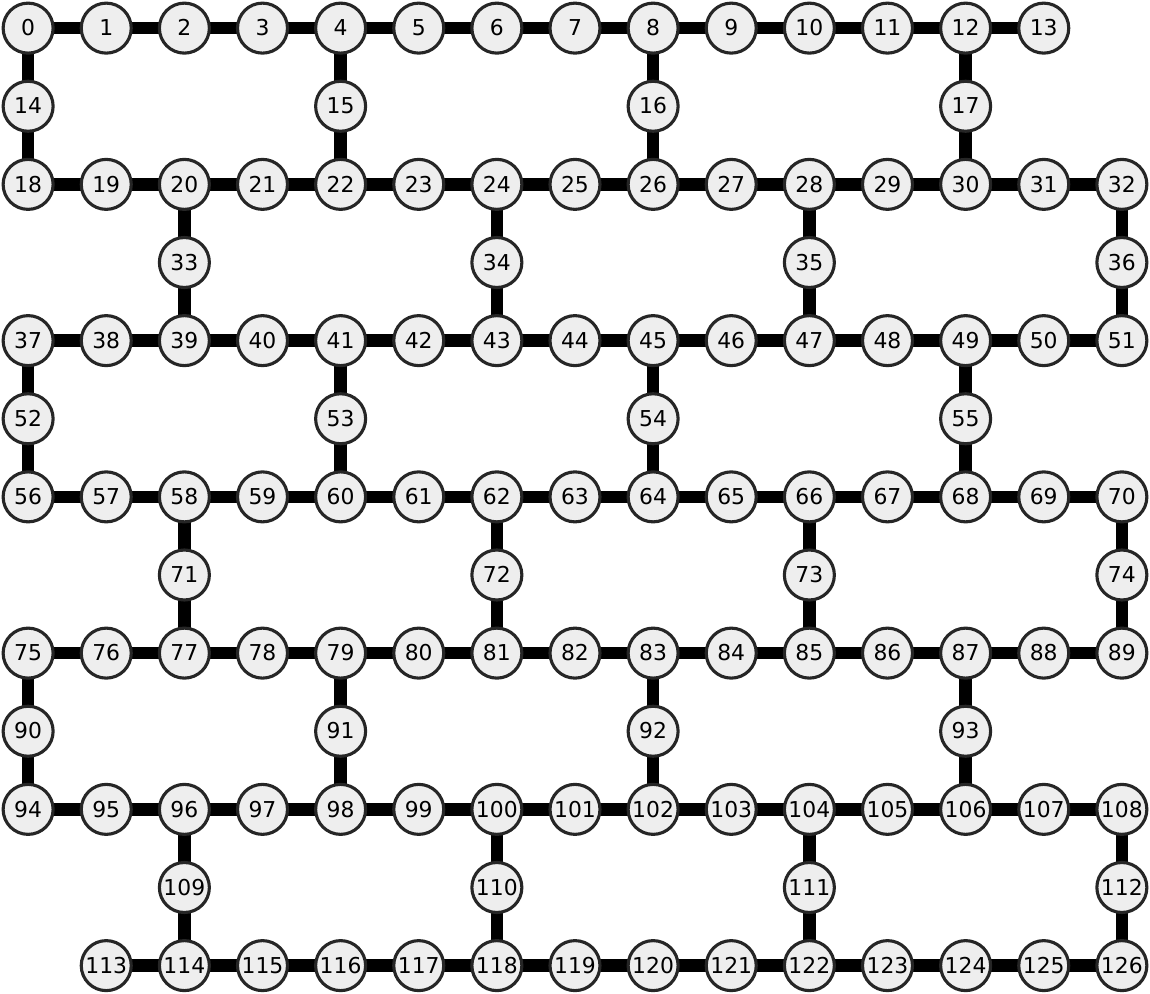}
        \caption{Brisbane (127 Qubits)}
    \end{subfigure}
    \caption{Larger architectures used in our evaluation.}\label{fig:architectures_clifford_large}
\end{figure}

%% file: main.bbl
\begin{thebibliography}{43}
\providecommand{\natexlab}[1]{#1}
\providecommand{\url}[1]{\texttt{#1}}
\expandafter\ifx\csname urlstyle\endcsname\relax
  \providecommand{\doi}[1]{doi: #1}\else
  \providecommand{\doi}{doi: \begingroup \urlstyle{rm}\Url}\fi

\bibitem[Aaronson and Gottesman(2004)]{Aaronson_2004}
Scott Aaronson and Daniel Gottesman.
\newblock Improved simulation of stabilizer circuits.
\newblock \emph{Physical Review A}, 70\penalty0 (5), nov 2004.
\newblock \doi{10.1103/physreva.70.052328}.

\bibitem[Amy et~al.(2013{\natexlab{a}})Amy, Maslov, Mosca, and Roetteler]{Amy2013}
M.~Amy, D.~Maslov, M.~Mosca, and M.~Roetteler.
\newblock A meet-in-the-middle algorithm for fast synthesis of depth-optimal quantum circuits.
\newblock \emph{IEEE Transactions on Computer-Aided Design of Integrated Circuits and Systems}, 32\penalty0 (6):\penalty0 818–830, June 2013{\natexlab{a}}.
\newblock ISSN 1937-4151.
\newblock \doi{10.1109/tcad.2013.2244643}.
\newblock URL \url{http://dx.doi.org/10.1109/TCAD.2013.2244643}.

\bibitem[Amy et~al.(2013{\natexlab{b}})Amy, Maslov, Mosca, and Roetteler]{amy2013meetinthemiddle}
Matthew Amy, Dmitri Maslov, Michele Mosca, and Martin Roetteler.
\newblock A meet-in-the-middle algorithm for fast synthesis of depth-optimal quantum circuits, 2013{\natexlab{b}}.

\bibitem[Amy et~al.(2018)Amy, Azimzadeh, and Mosca]{Amy_2018}
Matthew Amy, Parsiad Azimzadeh, and Michele Mosca.
\newblock On the controlled-{NOT} complexity of controlled-{NOT}{\textendash}phase circuits.
\newblock \emph{Quantum Science and Technology}, 4\penalty0 (1):\penalty0 015002, sep 2018.
\newblock \doi{10.1088/2058-9565/aad8ca}.
\newblock URL \url{https://doi.org/10.1088\%2F2058-9565\%2Faad8ca}.

\bibitem[Backens et~al.(2021)Backens, Miller-Bakewell, de~Felice, Lobski, and van~de Wetering]{backens2021there}
Miriam Backens, Hector Miller-Bakewell, Giovanni de~Felice, Leo Lobski, and John van~de Wetering.
\newblock There and back again: A circuit extraction tale.
\newblock \emph{Quantum}, 5:\penalty0 421, 2021.

\bibitem[{\noopsort{Berg}}{van den Berg}(2021)]{berg2021simple}
Ewout {\noopsort{Berg}}{van den Berg}.
\newblock A simple method for sampling random {Clifford} operators, 2021.

\bibitem[Bravyi and Maslov(2021)]{Bravyi_2021}
Sergey Bravyi and Dmitri Maslov.
\newblock Hadamard-free circuits expose the structure of the {Clifford} group.
\newblock \emph{{IEEE} Transactions on Information Theory}, 67\penalty0 (7):\penalty0 4546--4563, jul 2021.
\newblock \doi{10.1109/tit.2021.3081415}.

\bibitem[Bravyi et~al.(2021)Bravyi, Shaydulin, Hu, and Maslov]{Bravyi_2021_simp}
Sergey Bravyi, Ruslan Shaydulin, Shaohan Hu, and Dmitri Maslov.
\newblock {Clifford} circuit optimization with templates and symbolic {Pauli} gates.
\newblock \emph{Quantum}, 5:\penalty0 580, nov 2021.
\newblock \doi{10.22331/q-2021-11-16-580}.
\newblock URL \url{https://doi.org/10.22331\%2Fq-2021-11-16-580}.

\bibitem[Cowtan et~al.(2020{\natexlab{a}})Cowtan, Dilkes, Duncan, Simmons, and Sivarajah]{Cowtan_2020}
Alexander Cowtan, Silas Dilkes, Ross Duncan, Will Simmons, and Seyon Sivarajah.
\newblock Phase gadget synthesis for shallow circuits.
\newblock \emph{Electronic Proceedings in Theoretical Computer Science}, 318:\penalty0 213--228, may 2020{\natexlab{a}}.
\newblock \doi{10.4204/eptcs.318.13}.

\bibitem[Cowtan et~al.(2020{\natexlab{b}})Cowtan, Simmons, and Duncan]{cowtan2020generic}
Alexander Cowtan, Will Simmons, and Ross Duncan.
\newblock A generic compilation strategy for the unitary coupled cluster ansatz, 2020{\natexlab{b}}.

\bibitem[de~Brugière and Martiel(2023)]{debrugière2023shallower}
Timothée~Goubault de~Brugière and Simon Martiel.
\newblock Shallower cnot circuits on realistic quantum hardware, 2023.

\bibitem[Dehaene and De~Moor(2003)]{Dehaene_2003}
Jeroen Dehaene and Bart De~Moor.
\newblock {Clifford} group, stabilizer states, and linear and quadratic operations over {GF}(2).
\newblock \emph{Phys. Rev. A}, 68:\penalty0 042318, Oct 2003.
\newblock \doi{10.1103/PhysRevA.68.042318}.
\newblock URL \url{https://link.aps.org/doi/10.1103/PhysRevA.68.042318}.

\bibitem[Duncan et~al.(2020)Duncan, Kissinger, Perdrix, and {\noopsort{wetering}}{van de Wetering}]{Duncan_2020}
Ross Duncan, Aleks Kissinger, Simon Perdrix, and John {\noopsort{wetering}}{van de Wetering}.
\newblock Graph-theoretic simplification of quantum circuits with the {ZX}-calculus.
\newblock \emph{Quantum}, 4:\penalty0 279, jun 2020.
\newblock \doi{10.22331/q-2020-06-04-279}.
\newblock URL \url{https://doi.org/10.223312Fq-2020-06-04-279}.

\bibitem[Floyd(1962)]{floyd1962algorithm}
Robert~W Floyd.
\newblock Algorithm 97: shortest path.
\newblock \emph{Communications of the ACM}, 5\penalty0 (6):\penalty0 345, 1962.

\bibitem[Gheorghiu et~al.(2022)Gheorghiu, Huang, Li, Mosca, and Mukhopadhyay]{gheorghiu2022reducing}
Vlad Gheorghiu, Jiaxin Huang, Sarah~Meng Li, Michele Mosca, and Priyanka Mukhopadhyay.
\newblock Reducing the {CNOT} count for {Clifford+T} circuits on {NISQ} architectures.
\newblock \emph{IEEE Transactions on Computer-Aided Design of Integrated Circuits and Systems}, 2022.

\bibitem[Gidney(2021)]{Gidney_2021}
Craig Gidney.
\newblock Stim: a fast stabilizer circuit simulator.
\newblock \emph{Quantum}, 5:\penalty0 497, jul 2021.
\newblock \doi{10.22331/q-2021-07-06-497}.

\bibitem[Gogioso and Yeung(2022)]{gogioso2022annealing}
Stefano Gogioso and Richie Yeung.
\newblock Annealing optimisation of mixed {ZX} phase circuits, 2022.

\bibitem[Gottesman(1997)]{gottesman1997stabilizer}
Daniel Gottesman.
\newblock Stabilizer codes and quantum error correction, 1997.

\bibitem[Grier and Schaeffer(2022)]{Grier_2022}
Daniel Grier and Luke Schaeffer.
\newblock The classification of {Clifford} gates over qubits.
\newblock \emph{Quantum}, 6:\penalty0 734, jun 2022.
\newblock \doi{10.22331/q-2022-06-13-734}.
\newblock URL \url{https://doi.org/10.223312Fq-2022-06-13-734}.

\bibitem[Karp(2010)]{karp2010reducibility}
Richard~M Karp.
\newblock \emph{Reducibility among combinatorial problems}.
\newblock Springer, 2010.

\bibitem[Kissinger and van~de Griend(2019)]{kissinger2019cnot}
Aleks Kissinger and Arianne~Meijer van~de Griend.
\newblock {CNOT} circuit extraction for topologically-constrained quantum memories, 2019.

\bibitem[Kissinger and {\noopsort{Wetering}}{van de Wetering}(2020)]{kissinger2020Pyzx}
Aleks Kissinger and John {\noopsort{Wetering}}{van de Wetering}.
\newblock {{PyZX}: Large Scale Automated Diagrammatic Reasoning}.
\newblock In Bob Coecke and Matthew Leifer, editors, \emph{{Proceedings 16th International Conference on} Quantum Physics and Logic, {Chapman University, Orange, CA, USA., 10-14 June 2019}}, volume 318 of \emph{Electronic Proceedings in Theoretical Computer Science}, pages 229--241. Open Publishing Association, 2020.
\newblock \doi{10.4204/EPTCS.318.14}.

\bibitem[Kou et~al.(1981)Kou, Markowsky, and Berman]{kou_1981}
L.~Kou, G.~Markowsky, and L.~Berman.
\newblock A fast algorithm for {Steiner} trees.
\newblock \emph{Acta Informatica}, 15\penalty0 (2):\penalty0 141--145, 1981.
\newblock \doi{10.1007/BF00288961}.
\newblock URL \url{https://doi.org/10.1007/BF00288961}.

\bibitem[Kutin et~al.(2007)Kutin, Moulton, and Smithline]{kutin2007computation}
Samuel~A. Kutin, David~Petrie Moulton, and Lawren~M. Smithline.
\newblock Computation at a distance, 2007.

\bibitem[Li et~al.(2019)Li, Ding, and Xie]{li2019tackling}
Gushu Li, Yufei Ding, and Yuan Xie.
\newblock Tackling the qubit mapping problem for nisq-era quantum devices, 2019.

\bibitem[Li et~al.(2022)Li, Wu, Shi, Javadi-Abhari, Ding, and Xie]{li2022paulihedral}
Gushu Li, Anbang Wu, Yunong Shi, Ali Javadi-Abhari, Yufei Ding, and Yuan Xie.
\newblock Paulihedral: a generalized block-wise compiler optimization framework for quantum simulation kernels.
\newblock In \emph{Proceedings of the 27th ACM International Conference on Architectural Support for Programming Languages and Operating Systems}, pages 554--569, 2022.

\bibitem[Martiel and de~Brugi{\`{e}}re(2022)]{Martiel_2022}
Simon Martiel and Timoth{\'{e} }e~Goubault de~Brugi{\`{e}}re.
\newblock Architecture aware compilation of quantum circuits via lazy synthesis.
\newblock \emph{Quantum}, 6:\penalty0 729, jun 2022.
\newblock \doi{10.22331/q-2022-06-07-729}.
\newblock URL \url{https://doi.org/10.22331\%2Fq-2022-06-07-729}.

\bibitem[Maslov and Roetteler(2018)]{maslov_2018}
Dmitri Maslov and Martin Roetteler.
\newblock Shorter stabilizer circuits via {Bruhat} decomposition and quantum circuit transformations.
\newblock \emph{IEEE Transactions on Information Theory}, 64\penalty0 (7):\penalty0 4729--4738, 2018.
\newblock \doi{10.1109/TIT.2018.2825602}.

\bibitem[Maslov and Yang(2023)]{Maslov_2023}
Dmitri Maslov and Willers Yang.
\newblock Cnot circuits need little help to implement arbitrary hadamard-free clifford transformations they generate.
\newblock \emph{npj Quantum Information}, 9\penalty0 (1), September 2023.
\newblock ISSN 2056-6387.
\newblock \doi{10.1038/s41534-023-00760-2}.
\newblock URL \url{http://dx.doi.org/10.1038/s41534-023-00760-2}.

\bibitem[Meijer van~de Griend(2023)]{vandegriend2023towards}
Arianne Meijer van~de Griend.
\newblock Towards a generic compilation approach for quantum circuits through resynthesis.
\newblock \emph{arXiv preprint arXiv:2304.08814}, 2023.

\bibitem[{\noopsort{Meijer-Griend}}{Meijer-van de Griend} and Duncan(2020)]{degriend2020architectureaware}
Arianne {\noopsort{Meijer-Griend}}{Meijer-van de Griend} and Ross Duncan.
\newblock Architecture-aware synthesis of phase polynomials for {NISQ} devices, 2020.

\bibitem[Nam et~al.(2018)Nam, Ross, Su, Childs, and Maslov]{Nam_2018}
Yunseong Nam, Neil~J. Ross, Yuan Su, Andrew~M. Childs, and Dmitri Maslov.
\newblock Automated optimization of large quantum circuits with continuous parameters.
\newblock \emph{npj Quantum Information}, 4\penalty0 (1), may 2018.
\newblock \doi{10.1038/s41534-018-0072-4}.
\newblock URL \url{https://doi.org/10.1038\%2Fs41534-018-0072-4}.

\bibitem[Nash et~al.(2020)Nash, Gheorghiu, and Mosca]{Nash_2020}
Beatrice Nash, Vlad Gheorghiu, and Michele Mosca.
\newblock Quantum circuit optimizations for {NISQ} architectures.
\newblock \emph{Quantum Science and Technology}, 5\penalty0 (2):\penalty0 025010, mar 2020.
\newblock \doi{10.1088/2058-9565/ab79b1}.

\bibitem[{\noopsort{Nest}}{van den Nest}(2009)]{nest2009classical}
Maarten {\noopsort{Nest}}{van den Nest}.
\newblock Classical simulation of quantum computation, the {Gottesman-Knill} theorem, and slightly beyond, 2009.

\bibitem[Patel et~al.(2008)Patel, Markov, and Hayes]{patel2008optimal}
Ketan~N Patel, Igor~L Markov, and John~P Hayes.
\newblock Optimal synthesis of linear reversible circuits.
\newblock \emph{Quantum Inf. Comput.}, 8\penalty0 (3):\penalty0 282--294, 2008.

\bibitem[Peham et~al.(2023)Peham, Brandl, Kueng, Wille, and Burgholzer]{peham2023depthoptimal}
Tom Peham, Nina Brandl, Richard Kueng, Robert Wille, and Lukas Burgholzer.
\newblock Depth-optimal synthesis of clifford circuits with sat solvers, 2023.

\bibitem[{Qiskit contributors}(2023)]{qiskit}
{Qiskit contributors}.
\newblock Qiskit: An open-source framework for quantum computing, 2023.

\bibitem[van~de Griend and Li(2023)]{vandegriend2023dynamic}
Arianne~Meijer van~de Griend and Sarah~Meng Li.
\newblock Dynamic qubit routing with {CNOT} circuit synthesis for quantum compilation, 2023.

\bibitem[Vandaele et~al.(2022)Vandaele, Martiel, and de~Brugi{\`e}re]{vandaele2022phase}
Vivien Vandaele, Simon Martiel, and Timoth{\'e}e~Goubault de~Brugi{\`e}re.
\newblock Phase polynomials synthesis algorithms for {NISQ} architectures and beyond.
\newblock \emph{Quantum Science and Technology}, 7\penalty0 (4):\penalty0 045027, 2022.

\bibitem[Warshall(1962)]{warshall1962theorem}
Stephen Warshall.
\newblock A theorem on boolean matrices.
\newblock \emph{Journal of the ACM (JACM)}, 9\penalty0 (1):\penalty0 11--12, 1962.

\bibitem[Winderl et~al.(2023)Winderl, Huang, and Mendl]{winderl2023recursively}
David Winderl, Qunsheng Huang, and Christian~B. Mendl.
\newblock A recursively partitioned approach to architecture-aware {ZX} polynomial synthesis and optimization, 2023.

\bibitem[Wu et~al.(2023)Wu, He, Yang, Shou, Tian, Zhang, and Sun]{Wu_2023}
Bujiao Wu, Xiaoyu He, Shuai Yang, Lifu Shou, Guojing Tian, Jialin Zhang, and Xiaoming Sun.
\newblock Optimization of {CNOT} circuits on limited-connectivity architecture.
\newblock \emph{Physical Review Research}, 5\penalty0 (1), jan 2023.
\newblock \doi{10.1103/physrevresearch.5.013065}.
\newblock URL \url{https://doi.org/10.1103\%2Fphysrevresearch.5.013065}.

\bibitem[Yeung(2020)]{yeung2020diagrammatic}
Richie Yeung.
\newblock Diagrammatic design and study of ans\"{a}tze for quantum machine learning, 2020.

\end{thebibliography}
